\documentclass[twocolumn, onecolappendix]{aastex631}

\usepackage{afterpage}
\usepackage{booktabs} 
 
\usepackage{siunitx} 
\usepackage{amsmath,amstext,color}
\usepackage{gensymb}
\usepackage{natbib}
\usepackage{hyperref}
\usepackage{longtable}
\usepackage{threeparttablex}
\usepackage{booktabs}
\usepackage{color}
\usepackage{graphicx} 
\usepackage{amsmath}
\usepackage{amssymb}
\usepackage[]{appendix}
\usepackage{threeparttable}
\usepackage{tabularx}
\usepackage{xcolor} 
\usepackage{pifont}

\title{24uj}

\begin{document}
\title{JWST Observations of Calcium-Strong Transients: I. Complex Nebular He Emission in SN\,2024uj}

\correspondingauthor{Saarah Hall}
\email{saarah@u.northwestern.edu}

\shorttitle{SN\,2024uj}
\shortauthors{Hall et al.}
\newcommand{\NUCIERA}{\affiliation{Center for Interdisciplinary Exploration and Research in Astrophysics and Department of Physics and Astronomy, \\Northwestern University, 1800 Sherman Avenue, Floor 8, Evanston, IL 60201-3497, USA}}

\newcommand{\NU}{\affiliation{Department of Physics and Astronomy, Northwestern University, 2145 Sheridan Road, Evanston, IL 60208, USA}}

\newcommand{\CIERA}{\affiliation{Center for Interdisciplinary Exploration and Research in Astrophysics (CIERA), 1800 Sherman Ave., Evanston, IL 60201, USA}}

\newcommand{\SKAI}{\affiliation{NSF-Simons AI Institute for the Sky (SkAI), 172 E. Chestnut St., Chicago, IL 60611, USA}}

\newcommand{\Rutgers}{\affiliation{Department of Physics and Astronomy, Rutgers, the State University of New Jersey, 136 Frelinghuysen Road, Piscataway, NJ 08854-8019, USA}}

\newcommand{\Berkeley}{\affiliation{Department of Astronomy, University of California, Berkeley, CA 94720-3411, USA}}

\newcommand{\Trinity}{\affiliation{School of Physics, Trinity College Dublin, the University of Dublin, College Green, Dublin 2, Ireland}}
\author[0000-0002-3841-380X]{Saarah~Hall}
\NU
\CIERA

\author[0000-0003-3108-1328]{Lindsey~A.~Kwok}
\thanks{NASA Hubble Fellow}
\CIERA

\author[0000-0002-7352-7845]{Aravind~P.~Ravi} 
\affiliation{Department of Physics and Astronomy, University of California, Davis, 1 Shields Avenue, Davis, CA 95616-5270, USA}

\author[0000-0001-9515-478X]{Adam~A.~Miller}
\NU
\CIERA
\SKAI

\author[0000-0003-0599-8407]{Luc~Dessart}
\affiliation{Institut d'Astrophysique de Paris, CNRS-Sorbonne Universit\'e, 98 bis boulevard Arago, F-75014 Paris, France}
\affiliation{French-Chilean Laboratory for Astronomy, IRL 3386, CNRS, Instituto de Astrofísica, Pontificia Universidad Católica de Chile, Casilla 306, Santiago, Chile}

\author[0000-0002-3934-2644]{W.~V.~Jacobson-Gal\'{a}n}
\thanks{NASA Hubble Fellow}
\affiliation{Cahill Center for Astrophysics, California Institute of Technology, MC 249-17, 1216 E California Boulevard, Pasadena, CA, 91125, USA}

\author[0000-0001-8023-4912]{Huei~Sears}
\Rutgers



\author[0000-0002-1895-6639]{Moira~Andrews}
\affiliation{Las Cumbres Observatory, Goleta, CA, 93117, USA}
\affiliation{University of California Santa Barbara, Santa Barbara, CA, 93111, USA}

\author[0000-0002-4924-444X]{K.~Azalee~Bostroem}
\affiliation{Steward Observatory, University of Arizona, 933 North Cherry Avenue, Tucson, AZ 85721-0065, USA}

\author[0000-0001-5955-2502]{Thomas~G.~Brink}
\Berkeley

\author[0000-0003-4914-5625]{Joseph~R.~Farah}
\affiliation{Center for Astrophysics $|$ Harvard \& Smithsonian, 60 Garden St, Cambridge, MA 02138, USA}
\affiliation{Black Hole Initiative, Harvard University, 20 Garden St, Cambridge, MA 02138, USA}

\author[0000-0003-3460-0103]{Alexei~V.~Filippenko}
\Berkeley

\author[0000-0002-2445-5275]{Ryan~J.~Foley}
\affiliation{Department of Astronomy and Astrophysics, University of California, Santa Cruz, CA 95064, USA}

\author[0000-0002-0832-2974]{Griffin~Hosseinzadeh}
\affiliation{Department of Astronomy \& Astrophysics, University of California, San Diego, 9500 Gilman Drive, MC 0424, La Jolla, CA 92093-0424, USA}

\author[0000-0003-4253-656X]{D.~Andrew~Howell}
\affiliation{Las Cumbres Observatory, Goleta, CA, 93117, USA}
\affiliation{University of California Santa Barbara, Santa Barbara, CA, 93111, USA}

\author[0000-0001-8738-6011]{Saurabh~W.~Jha}
\Rutgers

\author[0000-0003-2037-4619]{Conor~Larison}
\affiliation{Space Telescope Science Institute, 3700 San Martin Drive, Baltimore, MD 21218, USA}

\author[0000-0002-7866-4531]{Chang~Liu}
\NU
\CIERA
\SKAI

\author[0000-0002-9209-2787]{Colin~W.~Macrie}
\affiliation{Department of Physics and Astronomy, Purdue University, 525 Northwestern Avenue, West Lafayette, IN 47907-2036, USA}

\author[0000-0002-9770-3508]{Kate~Maguire}
\Trinity

\author[0000-0001-5807-7893]{Curtis~McCully}
\affiliation{Las Cumbres Observatory, Goleta, CA 93117, USA}

\author[0000-0002-7015-3446]{Nicolas~E.~Meza-Retamal}
\affiliation{Department of Physics and Astronomy, University of California, Davis, 1 Shields Avenue, Davis, CA 95616-5270, USA}

\author[0000-0001-7132-0333]{Maryam~Modjaz}
\affiliation{Department of Astronomy, University of Virginia, Charlottesville, VA 22904-4325, USA}

\author[0000-0001-9570-0584]{Megan~Newsome}
\affiliation{University of Texas at Austin, 1 University Station C1400, Austin, TX 78712-0259, USA}

\author[0000-0003-0209-9246]{Estefania~Padilla~Gonzalez}
\affiliation{Department of Physics and Astronomy, Johns Hopkins University, Baltimore, MD 21218, USA}

\author[0000-0003-4102-380X]{David~J.~Sand}
\affiliation{Steward Observatory, University of Arizona, 933 North Cherry Avenue, Tucson, AZ 85721-0065, USA}

\author[0000-0001-6797-1889]{Steve~Schulze}
\affiliation{Department of Particle Physics and Astrophysics, Weizmann Institute of Science, 234 Herzl St, 76100 Rehovot, Israel}

\author[0000-0003-0794-5982]{Giacomo~Terreran}
\affiliation{Adler Planetarium, 1300 S. DuSable Lake Shore Drive, Chicago, IL 60605, USA}

\author[0009-0000-0022-909X]{C.-G.~Touchard-Paxton}
\Trinity

\author[0000-0001-8818-0795]{Stefano~Valenti}
\affiliation{Department of Physics and Astronomy, University of California, Davis, 1 Shields Avenue, Davis, CA 95616-5270, USA}

\author[0000-0002-6535-8500]{Yi~Yang}
\affiliation{Physics Department, Tsinghua University, Beijing, 100084, China}

\author[0000-0002-2636-6508]{WeiKang~Zheng}
\Berkeley

\received{2026 June 30}

\newcommand{\heIone}{\ion{He}{1}\ 1.083\;$\mu$m}
\newcommand{\heItwo}{\ion{He}{1}\ 2.058\;$\mu$m} 
\newcommand{\oIdouble}{[\ion{O}{1}]~$\lambda\lambda6300,\,6364$}
\newcommand{\caIIdouble}{[\ion{Ca}{2}]~$\lambda\lambda7291,\,7324$}
\newcommand{\caIIdoubleleft}{[\ion{Ca}{2}]\,$\lambda7291$}
\newcommand{\caIIdoubleright}{[\ion{Ca}{2}]\,$\lambda7324$}
\newcommand{\cIIsix}{\ion{C}{2}\,$\lambda6570$}
\newcommand{\cIIseven}{\ion{C}{2}\,$\lambda7234$}
\newcommand{\sIone}{[\ion{S}{1}]\ 1.082\;$\mu$m}
\newcommand{\sItwo}{[\ion{S}{1}]\ 1.131\;$\mu$m}
\newcommand{\citepaperII}{S.\,Hall 2026b (in prep.)}
\newcommand{\citeCGsample}{C.G.\,Touchard-Paxton 2026 (in prep.)}
\newcommand{\kms}{\,km\,s$^{-1}$}
\newcommand{\microns}{\,$\mu$m}

\begin{abstract}
We present the first JWST observations of a Calcium-Strong Transient (CaST), SN\,2024uj, a rare class of supernovae (SNe) with observable properties that are consistent with both thermonuclear explosions of white dwarfs (WDs) and the core collapse of massive stars.  SN\,2024uj is offset by $\sim6.6$\;kpc from its host and exhibits a double-peaked light curve consistent with shock cooling of nearby circumstellar material. At early times, its optical spectra resemble those of normal SNe\,Ib, but strong \caIIdouble\ emission emerges between $+$2 and $+$17 days after maximum light. Radiative-transfer models of a massive stripped He star cannot reproduce this early forbidden Ca emission, even with artificially enhanced surface Ca, whereas it arises naturally in thermonuclear scenarios. The $+$150\;d JWST/NIRSpec spectrum reveals highly asymmetric, multicomponent \ion{He}{1} at both 1.083 and 2.058\,$\mu$m. The He extends to $\gtrsim+$5000\kms, with a strong, narrow peak at $+$1500\kms, indicating that He is distributed throughout the ejecta with a concentration offset from center. This He distribution overlaps central [\ion{Ca}{2}] and [\ion{O}{1}], implying a degree of mixing difficult to produce in a massive star explosion. The He peak might further trace interaction with a shocked, ejected companion in a thermonuclear system. The NIRSpec spectrum also shows molecular CO emission and a rising continuum that, together with a 10\,$\mu$m photometric detection, indicates dust emission extending into the mid-infrared. Given the remote environment, early forbidden Ca, mixed He/Ca/O ejecta, and possible companion signature, we favor a thermonuclear origin for SN\,2024uj involving at least one low-mass, partially He-rich WD.

\end{abstract}


\section{Introduction}

Calcium-Strong Transients (CaSTs) are a subclass of supernovae (SNe) that have puzzled observers since their first detections. The first members of this class were identified by \citet{filippenko_supernovae_2003} on the basis of unusually strong (``and perhaps suspect") calcium emission lines in their spectra. In early analyses \citep{perets_faint_2010, kasliwal_calcium-rich_2012}, the defining characteristics of this class came to be (i) peak magnitudes fainter than $\sim-16.5$\;mag, (ii) quick evolution to the nebular phase ($\sim 1-3$ months), and (iii) ``strong'' calcium emission, which was later quantified as an integrated flux ratio [\ion{Ca}{2}] / [\ion{O}{1}] $> 2$ \citep{milisavljevic_iPTF15eqv_2017}. Although they were originally thought to synthesize large amounts of calcium (``Ca-rich'' SNe), Ca is a very efficient coolant, so high abundances may not be necessary to reproduce the Ca line strengths (see models from \citealt{polin_nebular_2021}), and/or these systems may be weak in oxygen. Thus, we use the label ``Ca-strong'' \citep{shen_progenitors_2019} in this work.

CaSTs are often found at considerably large offsets from their host galaxies. 
In a sample of all CaSTs observed before 2025, \citeCGsample\ find an average host-SN offset of $5.76\pm0.75$\;kpc,\footnote{This average offset is smaller than previously thought for \\CaSTs (e.g., the average offset of the ``gold" sample in \\\citealt{shen_progenitors_2019} is 39\;kpc).} where $\sim$25$\%$ are offset by more than 20\;kpc.
In combination with the lack of star formation at the sites of CaSTs \citep{lyman_progenitors_2014}, this suggests that CaSTs originate from old stellar populations, disfavoring massive star progenitors \citep{lunnan_two_2017}, or that they are not formed in situ \citep{lyman_progenitors_2014, foley_kinematics_2015}. Furthermore, CaSTs also typically have small ejecta and $^{56}$Ni masses inferred from their faint and quickly fading light curves \citep{kawabata_massive_2010}, which has been interpreted as evidence that these are thermonuclear explosions \citep{perets_faint_2010, kasliwal_calcium-rich_2012}. 

Despite some evidence for a thermonuclear origin, CaSTs also have properties that instead suggest a massive star origin. Most CaST optical spectra near maximum light resemble those of Type Ib/c SNe, which are tied to the core collapse of massive stars. These optical spectra have helium emission lines, as do the small number of published near-infrared (NIR) CaST spectra (SN 2025coe, \citealt{kumar_multiwavelength_2026}; SN\,2021gno, \citealt{jacobson-galan_circumstellar_2022}; SN\,2019ehk, \citealt{jacobson-galan_sn_2020}; iPTF\,15eqv, \citealt{milisavljevic_iPTF15eqv_2017}). The presence of He in these spectra could naturally be explained with a massive star progenitor, but there are also thermonuclear scenarios that could produce He emission (e.g., a He-rich WD, \citealt{waldman_helium_2011, dessart_one-dimensional_2015}, or double detonations with a WD and a He-rich companion, \citealt{boyle_helium_2017, callan_exploring_2025}). 
The low ejecta and $^{56}$Ni masses from CaSTs could also be consistent with low-mass progenitors undergoing core collapse \citep{kawabata_massive_2010, de_peculiar_2021}; however, these progenitor models are ruled out when the masses are computed from the very late-time bolometric light curve (e.g., \citealt{jacobson-galan_late-time_2021}).

It may be the case that several progenitor channels and/or explosion mechanisms contribute to the CaST population. In a study of CaST host galaxies, \citet{dong_physical_2022} find that at least some CaSTs must come from thermonuclear WD explosions, leaving multiple channels as a possibility. For example, CaSTs in remote environments may be WD explosions, while those at smaller host-SN offsets and/or in star-forming galaxies may originate from massive stars \citep{prentice_rise_2020, de_peculiar_2021}. In a CaST sample study, \citet{de_zwicky_2020} identified subclasses that resemble Type Ia SNe (Ca-Ia) and Type Ib/c SNe (Ca-Ib/c), with a relative occurrence rate of 1:6. They conclude that the two classes can be explained with the same progenitor system (a WD with a He shell accumulated from a He-rich companion) but different explosion mechanisms (detonations or deflagrations, depending on the efficiency of the He burning).

\citet{touchard-paxton_sn_2025} proposed that the CaST SN 2023xwi originated from a thermonuclear explosion in an AM Canum Venaticorum (AM CVn) system. They also find that this system can explain observables seen in all CaSTs (e.g., the presence of forbidden transitions at very early times). AM CVn systems belong to a class of variables where a WD accretes matter from a He-rich semi/fully degenerate donor \citep{roelofs_hubble_2007}. 
As explored by \citet{bildsten_faint_2007}, the helium accreted from the donor onto the WD forms a He shell, which can produce a He-shell thermonuclear flash (He-nova) in cases of unstable burning at high accretion rates. These flashes in AM CVn systems occur periodically and can ultimately result in a faint thermonuclear SN, where a ``last flash'' with sufficient mass becomes dynamical and ejects radioactive material and fully unbinds the WD \citep{bildsten_faint_2007}.

To disentangle proposed CaST progenitors, many groups have developed radiative-transfer models for both core-collapse and thermonuclear CaST channels (for reviews, see \citealt{shen_progenitors_2019} and \citealt{zenati_origins_2023}). 
\citet{perets_faint_2010} first identified the prototype CaST SN\,2005E and proposed it was the product of a He-shell detonation on a WD. 
\citet{waldman_helium_2011} modeled this scenario specifically for WDs with low-mass carbon-oxygen (C/O) cores, and in cases where there is strong carbon enrichment, found good agreement between the simulated and early-time observed CaST spectra. 
\citet{dessart_one-dimensional_2015} adapted the \citet{waldman_helium_2011} model to a non-local-thermodynamic-equilibrium (NLTE) time-dependent framework with the \texttt{CMFGEN} code \citep{hillier_time-dependent_2012}. 
More recently, \citet{dessart_nebular_2021} simulated the core collapse of low-mass helium stars (He-stars).
Their NLTE radiative-transfer simulations replicate many spectral features seen in CaSTs, but also present strong \ion{Ti}{2} lines that have yet to be observed in a CaST. 
Lastly, we highlight the work of \citet{zenati_origins_2023}, who model the merger of a low-mass C/O WD and a hybrid helium-C/O (He/C/O) WD. In particular, through a mix of multidimensional hydrodynamical-thermonuclear (\texttt{FLASH}; \citealt{fryxell_flash_2000}) and one-dimensional (1D) NLTE radiative-transfer (\texttt{CMFGEN}) simulations, they model the full disruption of the C/O WD by the He/C/O WD and the following detonation of the material accreted onto the He/C/O. 
Notably, the progenitor CSM profile used in this model configuration is consistent with the CSM constraints made from X-ray CaST observations \citep{jacobson-galan_sn_2020,jacobson-galan_circumstellar_2022, kumar_multiwavelength_2026}. 
However, the spectra from these simulations lack helium emission lines seen in most CaSTs (Ca-Ib/c).

While the sample of observed CaSTs has grown to $\sim50$, only three have NIR spectra taken beyond $\sim+10$\;days post-peak (iPTF\,15eqv, \citealt{milisavljevic_iPTF15eqv_2017}; 
SN\,2019ehk, \citealt{jacobson-galan_sn_2020}; 
SN\,2025coe, \citealt{kumar_multiwavelength_2026}). Only iPTF\,15eqv has a late-time ($>$+50\;days) NIR spectrum, making nebular NIR observations of CaSTs a key underexplored frontier. In the late-time nebular phase, the SN ejecta becomes optically thin, revealing the innermost material that is obscured at earlier times. This enables inferences of the geometry and kinematics of the ejecta via the shapes of nebular emission lines \citep[for a review, see][]{jerkstrand_spectra_2017}. Nebular spectroscopy is especially fruitful in the NIR and mid-infrared (MIR) which host strong, relatively isolated lines tracing a range of ionization stages and nucleosynthetic groups. The distribution of different elements within the ejecta provides critical clues to the explosion mechanism and progenitor system \citep{derkacy_jwst_2023, derkacy_jwst_2024, derkacy_jwst_2026, kwok_jwst_2023, kwok_ground-based_2024, kwok_jwst_2025, kwok_jwst_2025-1, medler_flat-topped_2023, medler_jwst_2025, ashall_jwst_2024, dessart_optical_2025, jacobson-galan_panchromatic_2025, jacobson-galan_mapping_2026, mera_jwst_2026}.

We present an analysis of the first JWST observations of a CaST, SN\,2024uj. These data, taken at a phase of $+$150\;days after the second $r$-band peak, include a medium-resolution spectrum ($R \approx 1000$) from the Near-Infrared Spectrograph (NIRSpec; \citealt{jakobsen_near-infrared_2022}, \citealt{boker_-orbit_2023}) covering 1--5\microns\ and F1000W photometry from the Mid-Infrared Instrument (MIRI; \citealt{rieke_mid-infrared_2015}, \citealt{wright_mid-infrared_2015}). 
Paired with ground-based optical observations, these JWST data provide a more complete picture of the mechanisms behind this CaST and the class as whole.

\section{Discovery}

SN\,2024uj was first detected on 2024~January~9 (MJD~60318.45; all dates reported in UTC) by the Asteroid Terrestrial-impact Last Alert System (ATLAS; \citealt{tonry_atlas_2018, tonry_atlas_2024}). 
One day after its discovery (MJD~60319.21), \citet{petrushevska_epessto_2024} acquired a spectrum of SN\,2024uj through the ePESSTO+ program \citep{smartt_pessto_2015, nicholl_epessto_2019} with the EFOSC2 instrument \citep{buzzoni_eso_1984} on the 3.58\;m ESO New Technology Telescope (NTT) at the La Silla Observatory in Chile. On 2024 February 14 (MJD~60354.44), \citet{davis_ucsc_2024} acquired a spectrum of SN\,2024uj through the Young Supernova Experiment (YSE; \citealt{jones_young_2021}), utilizing the Low Resolution Imaging Spectrometer (LRIS; \citealt{oke_keck_1995}) on the Keck~I 10\;m telescope. With it, they classified SN\,2024uj as a Ca-rich SN~Ib (evoking the classification scheme from \citealt{de_zwicky_2020}), and noted a strong resemblance between SN\,2024uj and SN\,2019ehk \citep{jacobson-galan_sn_2020, de_peculiar_2021, nakaoka_calcium-rich_2021, jacobson-galan_late-time_2021}. 
We utilize the NTT and Keck spectra, both available via the Transient Name Server,\footnote{\href{https://www.wis-tns.org/object/2024uj}{wis-tns.org/object/2024uj}} in our analysis.

At the site of SN\,2024uj, there are two entries in the NGC catalog separated by $3.2''$: NGC 3565 and NGC 3566. There is also an entry for a galaxy pair positioned between the two. Visual inspection shows that this is one extended irregular galaxy (or two irregular overlapping galaxies). We adopt the redshift and distance of NGC 3566 from the NASA/IPAC Extragalactic Database (NED),\footnote{The NASA/IPAC Extragalactic Database (NED) is operated by the Jet Propulsion Laboratory, California Institute of Technology, under contract with the National Aeronautics and Space Administration.} $z=0.012505 \pm 0.00015$ and $D=60.08\pm4.34$ Mpc (includes Virgo + GA + Shapley peculiar-velocity correction, and assumes $H_0$ = 67.8$\pm$4.7$\;$km/s/Mpc; \citealt{mould_hubble_2000}). 
Like most CaSTs, SN\,2024uj exploded in a remote environment (see our LRS verification image, right panel of \autoref{fig:fullspec+phot}). With the coordinates of SN\,2024uj (\autoref{tab:snparams}) and NGC 3566 from NED,\footnote{\href{https://ned.ipac.caltech.edu/byname?objname=NGC+3566&hconst=67.8&omegam=0.308&omegav=0.692&wmap=4&corr_z=1}{ned.ipac.caltech.edu/byname?objname=NGC+3566}} we calculate a host--SN offset of 22$\farcs$7 ($6.6$\;kpc). There are bright regions distributed throughout the irregularly shaped host galaxy, which suggests the host may be actively star-forming.

\begin{deluxetable}{l l}
\tablecaption{Summary of SN 2024uj \label{tab:snparams}}
\startdata
&\\
Name & SN\,2024uj\\
Right Ascension & $11^{\rm hr}07^{\rm m}49^{\rm s}.06$\\
Declination & $-20^{\circ}01'34\farcs25$\\
Redshift & $0.012505 \pm 0.00015$\\
Host Galaxy & NGC 3566\\
Time of explosion$^{\dagger}$ & MJD~60315.10$^{+0.03}_{-0.01}$\\
Time of last non-detection & MJD~60317.29 \\
Time of discovery & MJD~60318.45 \\
Time of second $r$-band peak & MJD~60327.8$\pm$0.5\\
Host--SN offset & 22\farcs7 (6.6\;kpc)\\
Distance$^{\star}$ & $60.08\pm4.34$\;Mpc \\
\enddata
\tablecomments{SN coordinates are from Gaia via the Transient Name Server. \\${\dagger}$: Derived from bolometric light-curve modeling (see \autoref{sec:bolo_fit}).\\$\star$: Redshift-dependent distance to host galaxy; assumes $H_0$ = 67.8$\pm$4.7 km/s/Mpc.}
\end{deluxetable}

\section{Observations}

In this work, we utilize multiwavelength observations of SN\,2024uj, taken from early to late times ($>$150 days post-discovery). For all SN\,2024uj observations, we report phases with respect to the second $r$-band light-curve peak (MJD~60327.8$\pm$0.5, see \autoref{sec:bolo_fit}) unless stated otherwise. For all CaST spectra in this paper, we report phases in the frame of the SN (correcting for redshift). We list all SN\,2024uj spectra analyzed in this work in \autoref{tab:speclistuj}.

\subsection{Infrared Observations}\label{sec:IR-obs}

\begin{figure*}[htbp]
    \centering
    \includegraphics[width=0.99\textwidth]{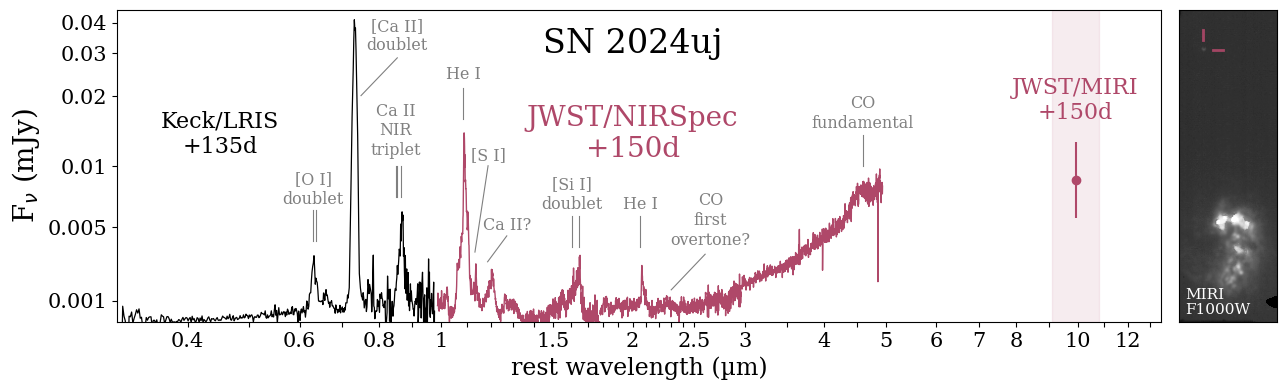}
    \caption{\textit{Left:} Nebular spectra of the CaST SN\,2024uj taken at $+$135\;d with Keck/LRIS and $+$150\;d with \textit{JWST/}NIRSpec (phases are with respect to second $r$-band peak). We also show the photometry measured with JWST/MIRI in the F1000W filter. The rebinned Keck spectrum is plotted in black  (bin size = 15 \AA). Prominent emission lines are labeled. Most lines are typical of the CaST class (e.g., strong [\ion{Ca}{2}] and weaker [\ion{O}{1}]), but some detections are a first for CaSTs (CO fundamental band, tentatively CO first overtone). There is also a clear flux continuum beyond 3 $\mu$m that we attribute to carbonaceous dust in \citepaperII\ \textit{Right:} LRS verification image for SN\,2024uj taken in the F1000W filter. The host galaxy, NGC 3565 is irregular and may be star-forming.}
    \label{fig:fullspec+phot}
\end{figure*}

We obtained NIR spectroscopy and MIR photometry of SN\,2024uj in its nebular phase (MJD~60479.88, $+$150\;d) through JWST Cycle 2 Director's Discretionary (DD) program 6591 \citep{kwok_cracking_2024}. These observations are presented
in the left panel of \autoref{fig:fullspec+phot}. The data are also available on the Mikulski Archive for Space Telescopes (MAST) at DOI \href{https://archive.stsci.edu/doi/resolve/resolve.html?doi=10.17909/fw08-m251}{10.17909/fw08-m251}.

We observed with NIRSpec in the fixed slits (FS) spectroscopy mode (\citealt{birkmann_-flight_2022}; \citealt{jakobsen_near-infrared_2022}; \citealt{rigby_science_2023}) with the S200A1 slit ($0\farcs2~\times~3\farcs3$). We used the following grating/filter combinations: G140M/F100LP, G235M/F170LP, and G395M/F290LP. The combined NIRSpec observations cover 0.96--5.1\,$\mu$m and the total exposure time, split between gratings, was 3838\,s (see \autoref{tab:speclistuj}). The NIRSpec spectra (G140M, G235M, and G395M) were reduced using the standard automatic JWST pipeline \citep{bushouse_jwst_2022}, obtained via MAST.

Unfortunately, owing to a guide-star acquisition failure, the planned MIRI/LRS spectrum was not obtained. We instead performed background-subtracted circular aperture photometry on the the LRS verification image, taken in the F1000W filter (right panel in \autoref{fig:fullspec+phot}).

\subsection{Optical Spectroscopy}

\begin{figure*}[htbp]
    \centering
    \includegraphics[width=0.98\textwidth]{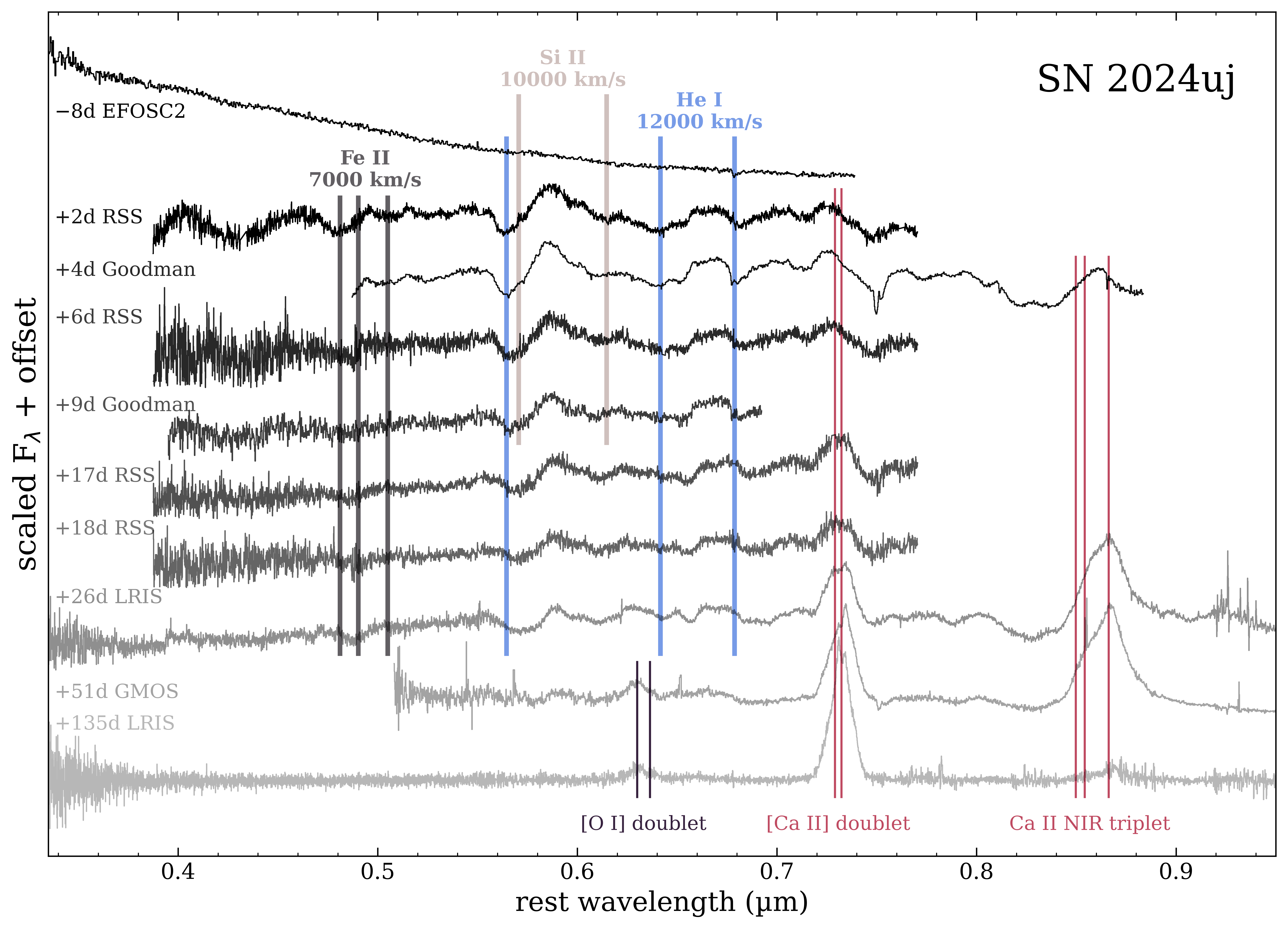}
    \caption{Optical spectral series of SN\,2024uj, from the discovery spectrum ($-$8\;d) to the latest nebular spectrum ($+$135\;d), see spectra details in \autoref{tab:speclistuj}. Phases are reported with respect to the second r-band peak. Wavelengths of line IDs are reported in \autoref{tab:lineIDs}. Photospheric line IDs and their approximate velocities (in bold) include \ion{Fe}{2} at 7000 km/s, \ion{Si}{2} at 10000 km/s, and \ion{He}{1} at 12000 km/s. Nebular line IDs (thin lines) include [\ion{O}{1}], [\ion{Ca}{2}], and the \ion{Ca}{2} NIR triplet.}
    \label{fig:optical-24uj}
\end{figure*}

In \autoref{fig:optical-24uj} we present the optical spectral series of SN\,2024uj. Between 2024~January~20 (MJD~60329.96, $+$2\;d) and February~5 (60345.92, $+$18\;d) we obtained six optical spectra of SN\,2024uj. Four were taken with the Robert Stobie Spectrograph (RSS; \citealt{burgh_prime_2003, kobulnicky_prime_2003}) on the Southern African Large Telescope (SALT; \citealt{buckley_completion_2006}), at phases $+$2\;d, $+$6\;d, $+$17\;d, and $+$18\;d. At phases $+$4\;d and $+$9\;d, we observed SN\,2024uj with the Goodman spectrograph \citep{clemens_goodman_2004} on the Southern Astrophysical Research telescope (SOAR). From these, the $+$2\;d, $+$4\;d, and $+$9\;d spectra were acquired through the Global Supernova Project \citep[GSP;][]{Howell19}. 

We also observed SN\,2024uj with the Gemini Multi-Object Spectrograph (GMOS; \citealt{hook_gemini-north_2004}) on the Gemini South telescope on MJD~60379.12 ($+$51\;d; \citealt{jacobson-galan_yse_2025}). This spectrum was reduced using the Data Reduction for Astronomy from Gemini Observatory North and South (\texttt{DRAGONS}) pipeline \citep{labrie_dragonsquick_2023, simpson_dragons_2026}. Lastly, we obtained a $+$135\;d Keck/LRIS spectrum (MJD~60464.31), when SN\,2024uj was well into its nebular phase. LRIS is equipped with an atmospheric dispersion corrector (ADC), thereby minimizing differential slit losses caused by atmospheric dispersion \citep{filippenko_importance_1982}.  This spectrum was reduced with the \texttt{LPipe} pipeline \citep{perley_fully_2019}.

\subsection{Optical and Ultraviolet Photometry}\label{sec:phot}

We obtained high-cadence follow-up photometry of SN\,2024uj using a network of worldwide 1\,m and 2\,m robotic telescopes within the Las Cumbres Observatory \citep{brown_cumbres_2013}. The observations were triggered through the Global Supernova Project \citep[GSP;][]{Howell19} and photometric data reduction was done with the \texttt{lcogtsnpipe}\footnote{\href{https://github.com/LCOGT/lcogtsnpipe}{github.com/LCOGT/lcogtsnpipe}} \citep{Valenti16} pipeline. The package \texttt{lcogtsnpipe} uses standard point-spread-function (PSF) fitting techniques to calculate instrumental magnitudes. Observations in the $gri$ filters are calibrated to 
the AAVSO Photometric All Sky Survey (APASS) catalog \citep{Henden16}, while $UBV$ filter photometry is calibrated to a local Landolt catalog \citep{landolt_ubvri_1992, landolt_ubvri_2009} constructed using standard fields observed with the same telescope/instrument on the same night as the science observations. As SN\,2024uj is significantly offset from its host galaxy (6.6\;kpc) it has no significant background contamination, and thus we do not perform difference imaging for photometry. 

The Neil Gehrels Swift Observatory (Swift; \citealt{gehrels_swift_2004}) monitored SN\,2024uj with its UV-Optical Telescope \citep[UVOT;][]{roming_swift_2005} across ultraviolet (UV) filters $UVW2$, $UVM2$, and $UVW1$. These data were reduced using the High-Energy Astrophysics Software (\texttt{HEASoft}). We chose a source region centered at the position of the SN with an aperture radius of 3$''$ for photometry. A background was chosen from a source-free region with an aperture radius of 5$''$. We adopted zero-points for the photometry from \cite{Breeveld10} and used the latest updates to the time-dependent sensitivity corrections in 2020.

\begin{figure}[htbp]
    \centering
    \includegraphics[width=0.48\textwidth]{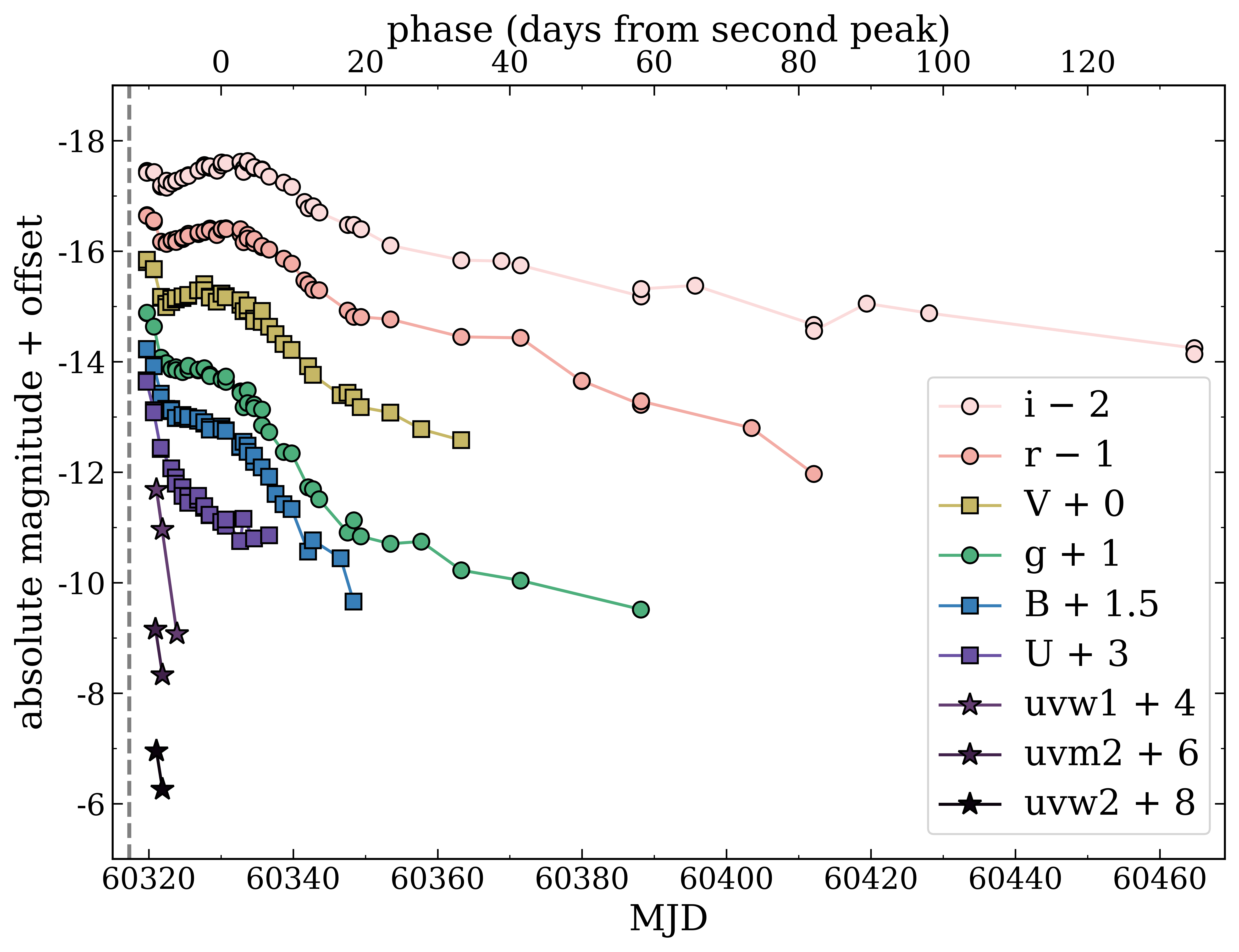}
    \caption{Multiband extinction-corrected optical and UV light curves of SN\,2024uj. The last non-detection is represented as a grey dashed line. This SN is part of a small but growing population of CaSTs with double-peaked light curves.}
    \label{fig:phot-obs}
\end{figure}

The multiband extinction-corrected UV and optical light curves of SN\,2024uj are shown in \autoref{fig:phot-obs} and \autoref{sec:appendixObsLog}.

\subsection{X-ray Observations} \label{sec:x-ray}

While observing with UVOT, Swift also monitored SN\,2024uj with its X-ray telescope (XRT) between 0.3 and 10\,keV in photon-counting mode \citep{burrows_swift_2005}. These observations took place from MJD~60320.575 to 60341.412, which corresponds to $-$7 to $+$13\;d post-second r-band peak (5 to 26 days post-explosion, see \autoref{sec:bolo_fit} for our explosion epoch estimate). We analyzed these data with the online-tools of the UK Swift team\footnote{\href{https://www.swift.ac.uk/user_objects/}{swift.ac.uk/user\_objects}} that use the software package \texttt{HEASoft} version 6.35.2 and methods described in \citet{evans_online_2007, evans_methods_2009}.

SN\,2024uj was not detected at any epoch. The median $3\sigma$ count-rate limit of all epochs is 0.008~s$^{-1}$ (0.3--10~keV). Combining all data pushes the limit to 0.001~s$^{-1}$. To convert count-rate into flux, we used WebPIMMS\footnote{\href{https://heasarc.gsfc.nasa.gov/cgi-bin/Tools/w3pimms/w3pimms.pl}{heasarc.gsfc.nasa.gov/cgi-bin/Tools/w3pimms/w3pimms.pl}} and assumed a power-law spectrum with a photon index\footnote{The photon index is defined as the power-law index of the photon flux density ($N(E)\propto E^{-\Gamma}$).} of $\Gamma=2$ and a Galactic neutral hydrogen column density of $4.1\times10^{20}$~cm$^{-2}$ \citep{hi4pi_collaboration_hi4pi_2016}. The energy conversion factor for the unabsorbed flux is $3.90\times10^{-11}\,\rm\left(erg\,s^{-1}\,cm^{-2}\right)/\left(s^{-1}\right)$.
The median count-rate limit of the individual observations corresponds to an unabsorbed flux of $<3.28\times10^{-13}~{\rm erg\,cm}^{-2}\,{\rm s}^{-1}$ between 0.3--10\,keV and a luminosity of $<1.42\times10^{41}~{\rm erg\,s}^{-1}$. 
The measurements of all individual epochs and the stacked data are shown in \autoref{tab:xray}.

With this luminosity limit, we cannot rule out that SN\,2024uj could have had X-ray emission similar to SN\,2019ehk, SN\,2021gno, and SN\,2025coe, the three CaSTs with Swift/XRT detections (all of which are Ca-Ib/c; \citealt{jacobson-galan_sn_2020,jacobson-galan_circumstellar_2022, kumar_multiwavelength_2026}). 
The brightest X-ray detection among these CaSTs is of SN\,2021gno at $\sim$1 day post-explosion, with $L_\mathrm{x}\approx5\times10^{41}~{\rm erg\,s}^{-1}$. At phases between $\sim$3 and 10 days post-explosion (consistent with our earliest X-ray observations of SN\,2024uj), SN\,2019ehk and SN\,2025coe were observed at $L_\mathrm{x}\approx10^{41}{\rm erg\,s}^{-1}$ and $L_\mathrm{x}\approx10^{40}~{\rm erg\,s}^{-1}$. 
These CaSTs detected in X-ray observations were all significantly closer than SN\,2024uj: SN\,2019ehk, SN\,2021gno, and SN\,2025coe are $\sim$16, 30, and 27\,Mpc away respectively, while SN\,2024uj is at $\sim$60\,Mpc. Thus, the XRT observations were not sufficiently deep to detect or rule out similar X-ray emission for SN\,2024uj.

\section{Photometric Analysis}\label{sec:bolo_fit}

\begin{figure}[h!]
    \centering
\includegraphics[width=0.48\textwidth]{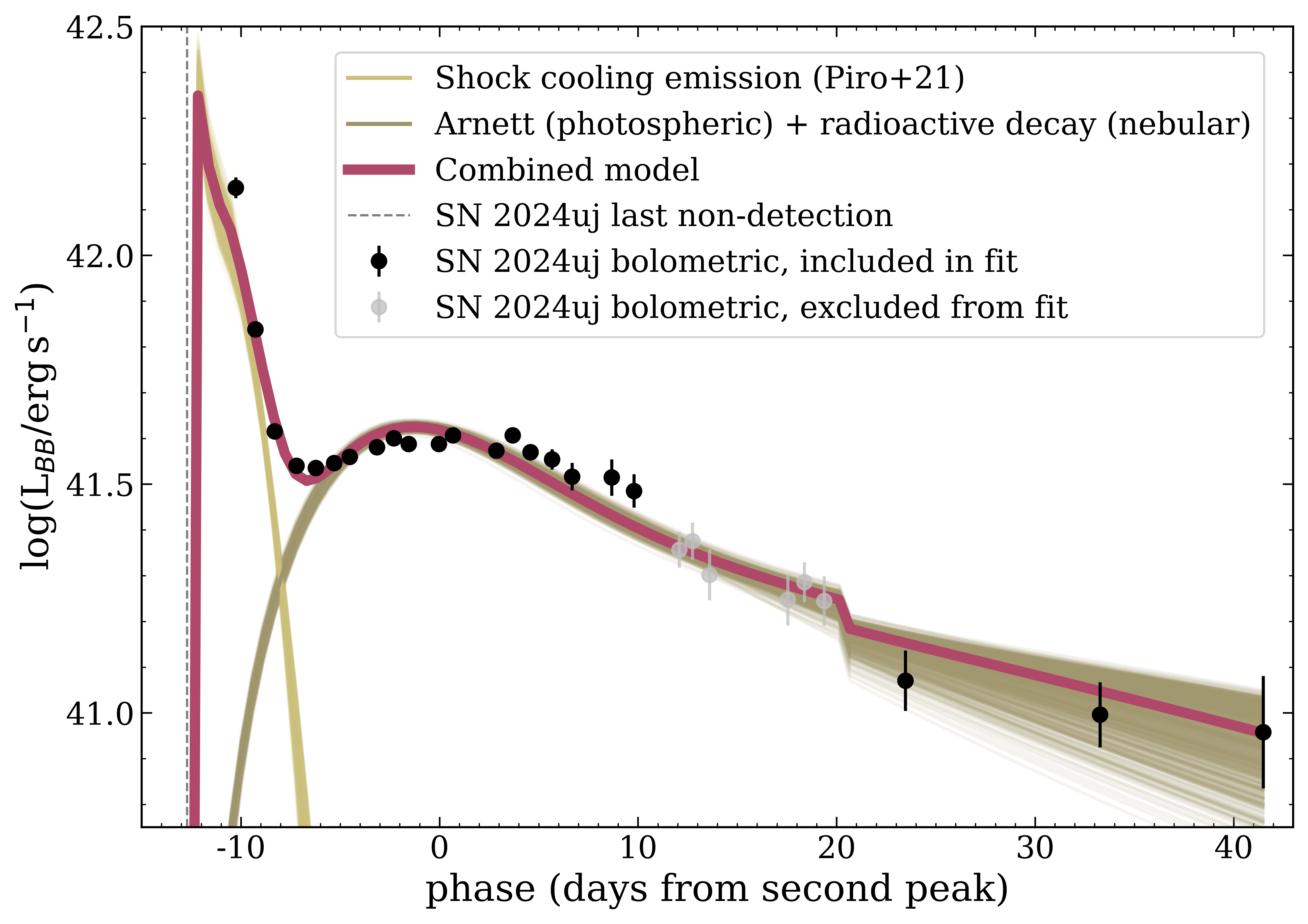}
    \caption{Bolometric light curve of SN\,2024uj based on optical and UV photometry. A combination of shock-heated envelope cooling emission and radioactivity (photospheric + nebular) can explain the observed luminosity as a function of time. The best-fit model and light curves randomly sampled from the posterior are overlaid over the observed luminosities. Points in black are used in the fitting and ones in grey (likely between the photospheric and nebular phases) are excluded in the fit. The last non-detection is marked as a grey dashed line and incorporated as a prior lower-limit to the explosion epoch for the combined model.}
    \label{fig:LC}
\end{figure}

To quantify the nature of the light curve evolution, we first construct a bolometric light curve of SN\,2024uj with the Light Curve Fitting package \citep{Hosseinzadeh23} using optical and UV photometry. For each telescope-night, if there are at least three photometry points, we fit a blackbody Planck function to the observed spectral energy distribution (SED; UV---optical) using a Markov Chain Monte Carlo (MCMC) routine to calculate the bolometric luminosity. We find that the bolometric luminosity estimated from SED fitting is reliable only until $\sim$40 days after the second peak in the optical light curve. Due to the rapid evolution from the photospheric to nebular phase in SN\,2024uj, the peak of the SED quickly shifts from the UV to the IR, and with no IR photometry, the blackbody temperatures are left unconstrained at epochs beyond 40 days. Moreover, as the SN evolves into the nebular phase, [\ion{Ca}{2}] and other emission lines strengthen, breaking down the blackbody assumption. The constructed bolometric light curve is presented in \autoref{fig:LC}.

The bolometric light curve shows an early rapid decline from a peak luminosity of $\sim$\,1.4 $\times$ 10$^{42}$\,erg\,s$^{-1}$ before rising to a second peak ($\sim$9 days after discovery) with a luminosity of $\sim$\,4 $\times$ 10$^{41}$\,erg\,s$^{-1}$. We note that there are no UV observations at the time of the first observed peak, so the corresponding bolometric luminosity is a lower limit. The double peaked nature of the light curve is similar to several other CaSTs discovered early, where the rapid first peak arises from the cooling of a shock-heated envelope and/or shock interaction with ambient circumstellar material (CSM), while the slower second peak is powered by the radioactive decay of $^{56}$Ni \citep{de_iptf_2018, jacobson-galan_ca_2020, jacobson-galan_circumstellar_2022, ertini_sn_2023, crawford_peaky_2025, ravi_double-peaked_2026}.

We adopt a combination of the shock-envelope cooling formalism of \cite{Piro21} with the radioactive decay of $^{56}$Ni to explain the double-peaked bolometric light curve of SN\,2024uj. Physical scenarios leading to the excess shock-cooling luminosity are degenerate, but must involve a gravitationally bound compact envelope, as opposed to, or in combination with, an ambient CSM around the progenitor. In the \citet{Piro21} envelope cooling model, extended material of mass $M_{e}$ within a radial extent of $R_{e}$ is imparted an energy of $E_{e}$ as the shock from the explosion passes through. The resulting cooling of the extended envelope adds extra luminosity to the expected contributions from radioactivity alone leading to the first peak. Homologous expansion is assumed and the luminosity due to a cooling envelope is found to be proportional to the initial envelope radius \citep{Piro21}. 

To determine the other physical parameters associated with the explosion of SN\,2024uj, we fit the bolometric LC with the luminosity expected from radioactivity at photospheric and nebular phases. In the photospheric phase ($t \lesssim$\,30 days from explosion), the luminosity is driven by photon diffusion time which is a function of the ejecta mass $M_{\mathrm{ej}}$, the photospheric ejecta velocity $v_{\mathrm{ph}}$, and the opacity, assuming spherical symmetry \citep{Arnett82}. If we assume that at the radioactively-powered second peak of SN\,2024uj the rise time is comparable to the photon diffusion time, we can infer $M_{\mathrm{ej}}$, the nickel mass $M_{\mathrm{Ni}}$, and the explosion kinetic energy $E_{\mathrm{k}}$. We fix the optical opacity ($\kappa_{\rm opt}$) at 0.1\,cm$^{2}$\,g$^{-1}$, a value consistent with electron scattering in singly ionized helium (adopted for SN ejecta mass estimates in e.g., \citealt{kleiser_rapidly_2014}). To break the degeneracy between $M_{\mathrm{ej}}$ and $E_{\mathrm{k}}$ \citep[assuming a homogeneous density as per][]{Arnett82}, we fix $v_{\mathrm{ph}}$ at 6,800\,km\,s$^{-1}$ from the measured absorption velocity of \ion{Fe}{2}\,5169\,\AA\ around the second peak from optical spectroscopy ($+$2\;d). For the free parameter $t_{\mathrm{exp}}$, the explosion epoch, we implement a hard limit as a prior based on the last-available reliable non-detection (MJD~60317.29). The radioactive light curve tail is mostly dominated by the decay of $^{56}$Co and we adopt the analytical formalism described by \cite{Valenti08} to self-consistently account for potentially incomplete trapping of $\gamma$-rays during this later phase where the ejecta are rapidly becoming optically thin.

In \autoref{fig:LC} we show the best-fit combined (shock-cooling $+$ radioactive powering) model and a few randomly sampled posterior model light curves. For our modeling, we implemented an ensemble sampling with \texttt{emcee}, a python-based affine invariant MCMC sampler \citep{foreman-mackey_emcee_2013}. 
First, we recover an explosion epoch $t_{\mathrm{exp}}=-$12.7$^{+0.03}_{-0.01}$\;days from the second peak (MJD~60315.10 $^{+0.03}_{-0.01}$).
From the first peak, our data is consistent with a model compact envelope of radius $R_\mathrm{env}=$ 28.7 $^{+8.6}_{-7.2}$\,$R_{\odot}$ and mass $M_\mathrm{env}=$ 0.14$^{+0.02}_{-0.01}$\,$M_{\odot}$.
From the second peak and subsequent evolution, our data is consistent with $M_{\mathrm{ej}}=$ 0.61$\pm$0.02\;$M_{\odot}$, 
$M_{\mathrm{Ni}}=$ $($1.36 $\pm$0.03)$\times$10$^{-2}$\;$M_{\odot}$, 
and $E_{\mathrm{k}}=$ (0.17 $\pm$ 0.01)$\times$10$^{51}$\;erg. 
We note that these uncertainties are purely statistical, and the systematic uncertainties are much larger. The uncertainty on $M_{\mathrm{Ni}}$, for example, is dominated by the systematic uncertainty of $\sim$\,0.002\;$M_{\odot}$ due to the error on the assumed distance to SN\,2024uj. There are also systematic uncertainties introduced by simplistic assumptions from our choice of models (e.g., assuming one constant opacity value for the entirety of the SN ejecta, and imposing a blackbody fit at post-peak epochs to construct the bolometric light curve). We thus utilize these derived values as order-of-magnitude estimates of the explosion properties of SN\,2024uj.

The shock-heated compact envelope properties ($R_\mathrm{env}, M_\mathrm{env}$) explaining the first peak in SN\,2024uj are consistent with other double-peaked CaSTs \citep{jacobson-galan_ca_2020, jacobson-galan_circumstellar_2022, ertini_sn_2023, ravi_double-peaked_2026}. Unlike SN\,2019ehk \citep{jacobson-galan_ca_2020}, SN\,2021gno \citep{jacobson-galan_circumstellar_2022}, and SN\,2025coe \citep{kumar_multiwavelength_2026}, no corresponding X-ray detections were observed in SN\,2024uj (see Section \ref{sec:x-ray}). The low ejecta, nickel masses, and a weak explosion all estimated from modeling the second peak agree exceptionally well with the properties of these and other CaSTs (e.g., SN\,2019ehk and SN\,2024uj mass and energy estimates agree to within a factor of $\sim$2; \autoref{sec:obs-othercasts}, \citealt{jacobson-galan_ca_2020}). We note that since we could construct a bolometric light curve out to only $\sim$40\;days after the second peak, 
there are only a few points on the radioactive decay tail, introducing a larger scatter on the posterior sample (\autoref{fig:LC}). Nevertheless, our combined model and posterior reasonably estimates the observed luminosity and derived explosion properties of SN\,2024uj.

SN\,2024uj is the seventh member of an emerging subclass of CaSTs with confirmed double-peaked light curves: iPTF\,16hgs \citep{de_iptf_2018}, SN\,2018lqo \citep{de_zwicky_2020}, SN\,2019ehk \citep{jacobson-galan_sn_2020}, SN\,2021gno \citep{jacobson-galan_circumstellar_2022, ertini_sn_2023}, SN\,2021inl \citep{jacobson-galan_circumstellar_2022}, and SN\,2025coe \citep{chen_sn_2025, ravi_double-peaked_2026}. We suggest that this double-peaked CaST subclass is not inherently small -- the first peak can be easily missed without high-cadence, ideally sub-daily, follow-up observations. Even in the case of SN\,2024uj, our earliest photometric observation was taken during the fall from the first peak, $\sim$1.3 days after discovery and $\sim$2.4 days after the last non-detection. 

\section{Spectral analysis}

\subsection{Line Identification}
\begin{deluxetable}{c c c c}
\tablecaption{SN\,2024uj Line Identifications \label{tab:lineIDs}}
\newcommand{\rowhead}[1]{\multicolumn{4}{c}{\bfseries #1}}
\tablehead{\rowhead{Photospheric IDs}
}
\startdata
$\lambda$ ($\mu$m) & Species & $\lambda$ ($\mu$m) & Species\\
\cmidrule(lr){1-2}\cmidrule(lr){3-4}
0.4924&\ion{Fe}{2}& 0.6355&\ion{Si}{2}\\
0.5018&\ion{Fe}{2}& 0.6570&\ion{C}{2}?\\
0.5169&\ion{Fe}{2}& 0.6678&\ion{He}{1}\\
0.5876&\ion{He}{1}& 0.7065&\ion{He}{1}\\
0.5900&\ion{Si}{2}& 0.7234&\ion{C}{2}\\
        \midrule
        \rowhead{Nebular IDs} \\
        \midrule
        $\lambda$ ($\mu$m) & Species & $\lambda$ ($\mu$m) & Species\\
        \cmidrule(lr){1-2}\cmidrule(lr){3-4}
0.6300&$[$\ion{O}{1}$]^a$&   1.131&$[$\ion{S}{1}$]$\\
0.6364&$[$\ion{O}{1}$]^a$&   1.184&\ion{Ca}{2}? \\
0.7291&$[$\ion{Ca}{2}$]^b$&  1.607&$[$\ion{Si}{1}$]^d$\\
0.7324&$[$\ion{Ca}{2}$]^b$&  1.645&$[$\ion{Si}{1}$]^d$\\
0.8498&\ion{Ca}{2}$^c$&     2.058&\ion{He}{1}\\
0.8542&\ion{Ca}{2}$^c$&     2.2 -- 2.4&CO first overtone?\\
0.8662&\ion{Ca}{2}$^c$&     4.5 -- 5.1&CO fundamental band\\
1.083&\ion{He}{1}&&\\
\enddata
\tablecomments{Wavelengths are given in air. Transitions within the same doublet / triplet are denoted with shared superscripts.
}
\end{deluxetable}

\subsubsection{Photospheric phase}
We list all line identifications for SN\,2024uj in \autoref{tab:lineIDs}. 
\ion{Fe}{2} and \ion{Si}{2} absorption lines have been identified in photospheric spectra of other CaSTs (e.g., SN\,2021gno, \citealt{ertini_sn_2023}). In the $+$2\;d spectrum of SN\,2024uj (\autoref{fig:optical-24uj}), there is strong \ion{Fe}{2} absorption at 4924 \AA\, and 5169 \AA, and weaker absorption at 5018 \AA. As for \ion{Si}{2}, there is weak absorption at 5900 \AA\, and 6355 \AA\, between $+$2\;d and $+$9\;d. There are also \ion{He}{1} features at 5876, 6678, and 7065 \AA\, between $+$2\;d and $+$26\;d, consistent with the CaST-Ib classification.

In the nebular SN\,2024uj spectrum we identify molecular CO (as well as carbonaceous dust, 
additional analysis to be given in S.\,Hall 2026b, in prep.), motivating a search for carbon features in the early spectra. We find that the \cIIsix\ 
absorption feature may be present in the $+$[2 -- 9]\;d spectra of SN\,2024uj at a velocity of 14,000\kms, but due to line blending -- e.g., with [\ion{Co}{3}] or [\ion{Co}{1}] at $\lambda$6578, $\lambda$6586 -- we are unable to confirm this feature or \cIIseven. 

\subsubsection{Nebular phase}

The $+$135\;d nebular Keck/LRIS spectrum of SN\,2024uj (\autoref{fig:fullspec+phot}) exhibits strong [\ion{Ca}{2}] $\lambda\lambda7291,$ $7324$\ emission and weaker \oIdouble. In line with the CaST definition from \cite{milisavljevic_iPTF15eqv_2017}, the integrated flux ratio [\ion{Ca}{2}]/[\ion{O}{1}]$>2$ for SN\,2024uj: from direct integration we measure [\ion{Ca}{2}] / [\ion{O}{1}] $=$ 3.2, 5.6, and 8.1 
at $+$26\;d, $+$51\;d, and $+$135\;d respectively. The \ion{Ca}{2} NIR triplet is also prominent, and decreases in strength over time relative to the [\ion{Ca}{2}] emission.

While the \oIdouble\ feature is a standard identification for emission at this wavelength in CCSNe and CaSTs, \citet{dessart_one-dimensional_2015} instead attribute this feature to blended \ion{Sc}{2} and \ion{Ti}{2} lines. In the absence of other strong Ti or Sc lines in our SN\,2024uj observations, we keep the [\ion{O}{1}] identification but encourage further exploration of this feature in future CaSTs.

The $+$150\;d JWST/NIRSpec spectrum of SN\,2024uj clearly detects both \heIone\ and the weaker \heItwo\ line. Following the NIR line identifications from \cite{milisavljevic_iPTF15eqv_2017}, we also detect \sItwo, indicating that \sIone\ must contribute to the \heIone\ profile. The \sIone\ line shares the same upper level as \sItwo, with the intensity ratio set by atomic physics to be 3.4:1 (1.082$\;\mu$m:1.131$\;\mu$m). These lines are discussed in detail in \autoref{sec:O-C-He}.

There is a prominent emission feature around 1.184 $\mu$m, not seen in iPTF\,15eqv \citep{milisavljevic_iPTF15eqv_2017}, which we attribute to \ion{Ca}{2}, following the identification of this line in the earlier phase NIR spectra of SN\,2019ehk and SN\,2021gno \citep{jacobson-galan_sn_2020, jacobson-galan_late-time_2021}. Its velocity profile is consistent with the other confident Ca lines in the spectra of all three of these CaSTs. We also detect the [\ion{Si}{1}] doublet at $1.607$, $1.645~\mu$m, which was also present in iPTF\,15eqv \citep{milisavljevic_iPTF15eqv_2017}.

Finally, we detect CO fundamental band emission between 4.5--5.1\,$\mu$m. This feature is broader and smoother than other lines in the $+$150\;d JWST/NIRSpec spectrum, which points to molecular emission. The only alternative ID from previous works would be [\ion{Fe}{2}] at 4.888 $\mu$m, a weak emission line seen in nebular SN Ia radiative-transfer simulations \citep{blondin_nebular_2023}. However, we do not identify any other, typically stronger, Fe lines in SN\,2024uj, and are thus confident that this feature is molecular CO.
The small bump at 2.2--2.4 microns may be the CO first overtone, which is expected to be much weaker. This constitutes the first detection of molecular emission in a CaST. We also note that there is a clear dust continuum beyond 2.5 $\mu$m in \autoref{fig:fullspec+phot}. We model and analyze the CO and dust features in SN\,2024uj in \citepaperII

\subsection{SN 2024uj and other CaSTs}\label{sec:obs-othercasts}

\begin{figure*}[htb]
    \centering
    \includegraphics[width=0.95\textwidth]{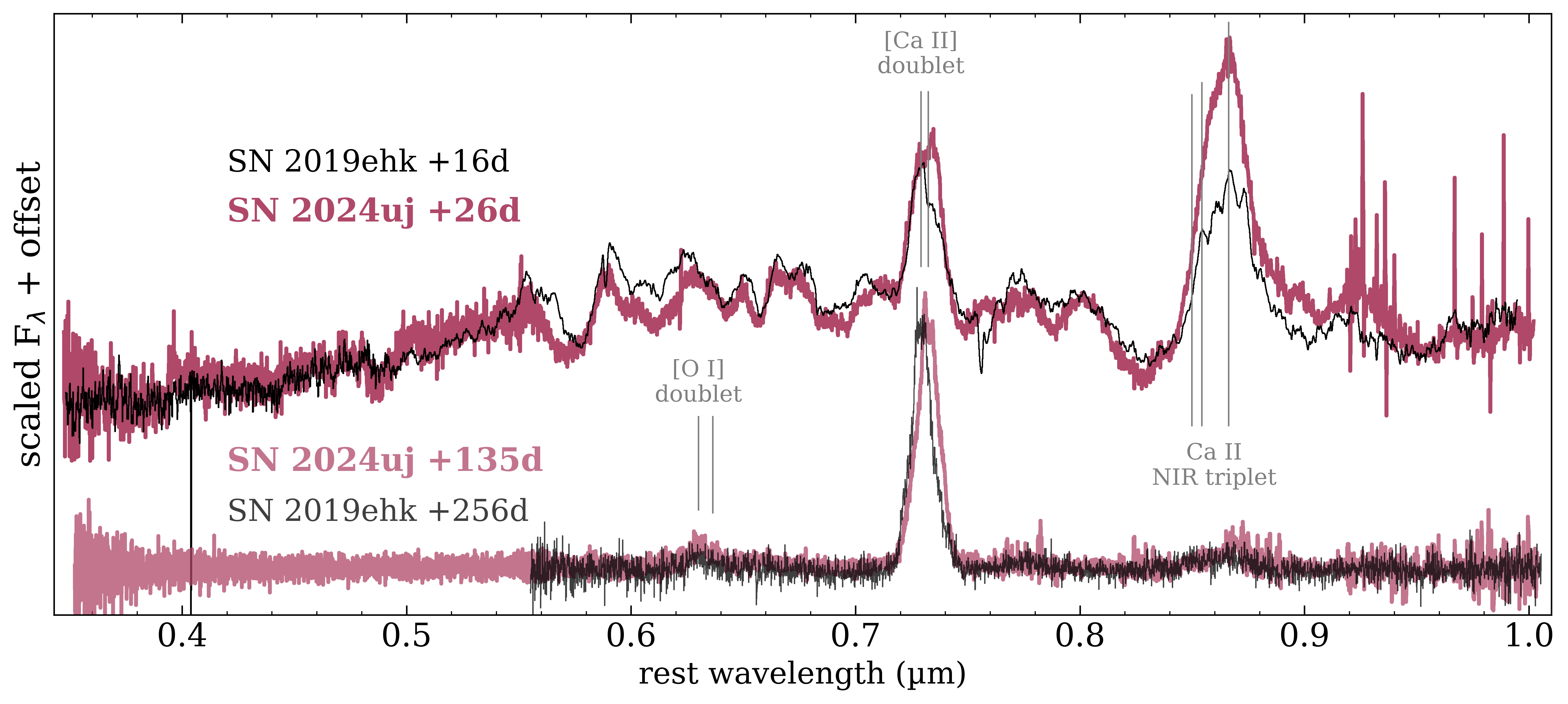}
    \caption{Spectral series of SN\,2024uj and SN\,2019ehk, with striking resemblances at both early and late phases. A reddening correction with $E(B-V)=0.34$ has been applied to the SN\,2019ehk spectra as in \citet{jacobson-galan_sn_2020}. Phases for both CaSTs are with respect to second light curve peak (r-band for SN\,2024uj, B-band for SN\,2019ehk)}
    \label{fig:uj-ehk}
\end{figure*}

\autoref{fig:uj-ehk} shows that SN\,2024uj is spectroscopically very similar to SN\,2019ehk at both early and late times. At earlier times, these two CaSTs exhibit similar relative strengths of emission and absorption features, including \caIIdouble\, and the \ion{Ca}{2} NIR triplet. At late times, both have similar [\ion{Ca}{2}] $/$ [\ion{O}{1}] ratios, and \caIIdouble\ features with similar widths. Both CaSTs also exhibit double-peaked bolometric light curves, from which the estimated $M_{\mathrm{ej}}$, $M_{\mathrm{Ni}}$, and $E_{\mathrm{k}}$ agree to within order unity ($M_{\mathrm{ej},2019\text{ehk}} / M_{\mathrm{ej},2024\text{uj}}=1.2$, $M_{\mathrm{Ni},2019\text{ehk}} / M_{\mathrm{Ni},2024\text{uj}}=2.1$ and $E_{\mathrm{k},2019\text{ehk}} / E_{\mathrm{k},2024\text{uj}}=1.1$; \autoref{sec:bolo_fit}, \citealt{jacobson-galan_sn_2020}). 
These similarities suggest that SN\,2019ehk and SN\,2024uj share a common origin.

Additionally, SN\,2024uj and SN\,2019ehk exploded in similar environments. NGC 3566 is a barred lenticular or irregular type galaxy which may be star-forming, from which SN\,2024uj is offset by 6.6\;kpc, in a region with no apparent star formation. M100, the host of SN\,2019ehk, is a star-forming barred spiral galaxy. SN\,2019ehk appears to be embedded in a spiral arm of M100 (offset by 2\;kpc), however, \citet{jacobson-galan_sn_2020} estimate a low local star formation rate of (9.2$\pm$2.9)$\times$10$^{-5}$\;$M_{\odot}$yr$^{-1}$. It could be the case that the small host-SN offset of SN\,2019ehk is a projection effect. Additionally, \citet{jacobson-galan_sn_2020} find that pre-explosion limits rule out a single massive star ($>$8$M_\odot$) progenitor for SN\,2019ehk.

\begin{figure}[h!]
    \centering
    \includegraphics[scale=0.4]{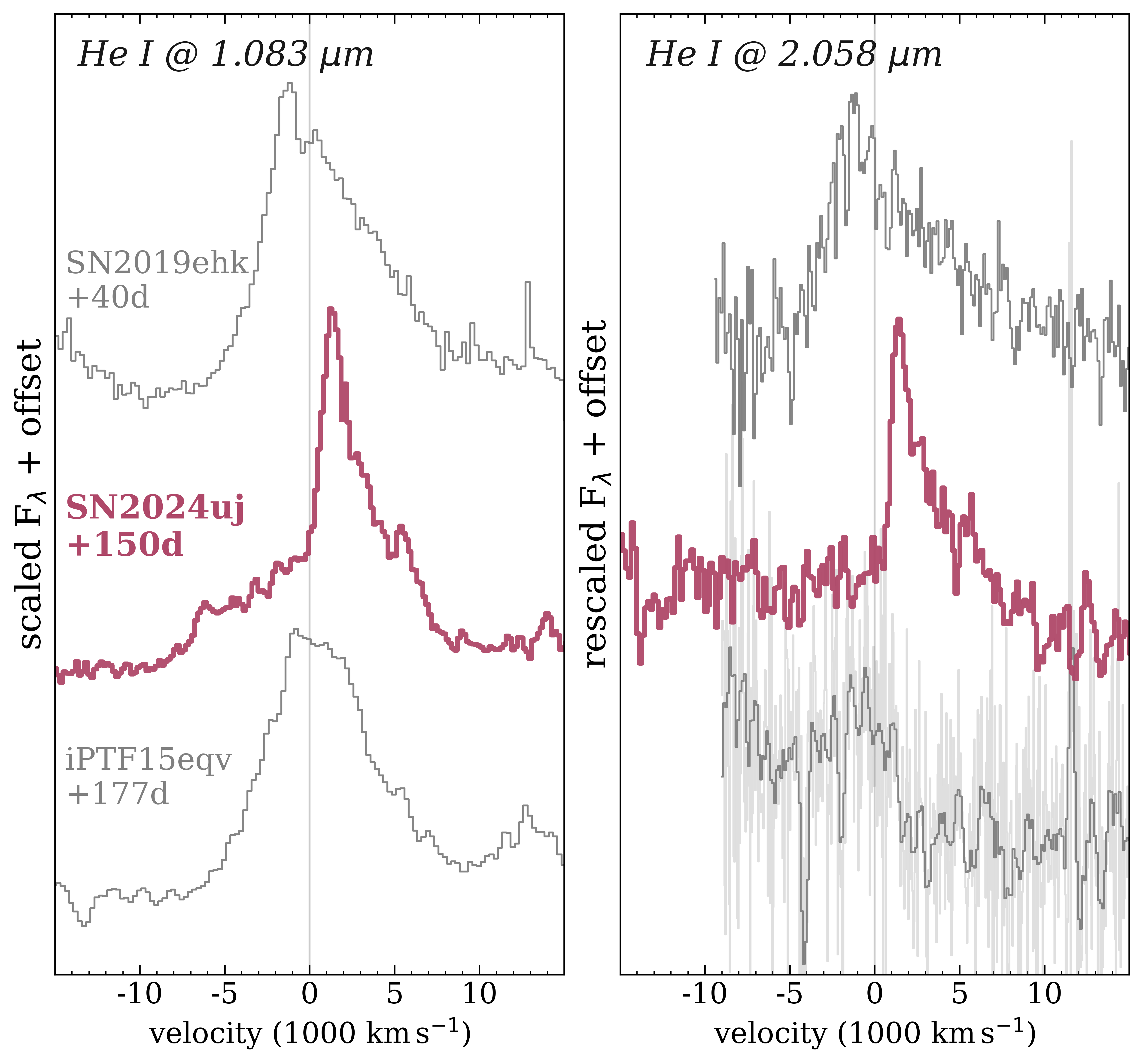}
    \caption{Comparison of the helium profiles present in SN\,2019ehk, SN\,2024uj, and iPTF\,15eqv, the only CaSTs with published near-infrared spectra taken at phases $\geq+$40\;d. The \heItwo\ line is clearest in SN\,2024uj and shows striking resemblance to the \heIone\ line. The spectra of SN\,2019ehk and iPTF\,15eqv have been rebinned  to the resolution of the SN\,2024uj spectrum. Phases are reported with respect to the second \textit{B}-band maximum for SN\,2019ehk \citep{jacobson-galan_sn_2020}, second $r$-band maximum for SN\,2024uj, and discovery for iPTF\,15eqv \citep{milisavljevic_iPTF15eqv_2017}.}
    \label{fig:cast-HeI}
\end{figure}

Including SN\,2024uj, six of the $\sim$50 CaSTs observed to date have published NIR spectra. Four of these CaSTs have spectral coverage at photospheric phases ($+$40\;d and earlier): SN\,2016hnk \citep{galbany_evidence_2019}, SN\,2019ehk \citep{jacobson-galan_sn_2020}, SN\,2021gno \citep{jacobson-galan_circumstellar_2022}, and SN\,2025coe \citep{kumar_multiwavelength_2026}. The only other CaST with published nebular NIR spectra is iPTF\,15eqv 
($+$177\;d post-discovery, \citealt{milisavljevic_iPTF15eqv_2017}). 
All of these CaSTs show strong \heIone\ emission lines with the exception of SN\,2016hnk (a Ia-like CaST), although there may be \heIone\ absorption in SN\,2016hnk spectra taken at earlier phases (\citealt{galbany_evidence_2019}; see Figure 5 and 6 in \citealt{liu_sn_2023-1}). 
We thus suggest that \ion{He}{1} emission is likely a common feature of the Ib-like CaST class, and possibly the Ia-like class as well. 
We compare the \heIone\, and \heItwo\, emission line profiles of iPTF\,15eqv, SN\,2024uj, and SN\,2019ehk, in \autoref{fig:cast-HeI}.

The \heIone\ line has a higher intensity than \heItwo\ but can be blended with \ion{C}{1}, \ion{O}{1}, and \ion{Mg}{2} (demonstrated in stripped-envelope SN models by \citealt{dessart_radiative-transfer_2015}), 
or with [\ion{S}{1}] (identified in iPTF\,15eqv by \citealt{milisavljevic_iPTF15eqv_2017}). SN\,2019ehk and SN\,2024uj have the clearest detections of \heItwo, with SN\,2024uj being the first with a high S/N detection in a fully nebular spectrum.
This allows for detailed modeling of \ion{He}{1} and its distribution in the SN ejecta (\autoref{sec:O-C-He}).

All three \heIone\, profiles in \autoref{fig:cast-HeI} are asymmetric: they appear steep on the blue side and relatively shallow on the red side (in SN\,2019ehk, this could be an optical depth effect since the spectrum is not yet fully nebular). In SN\,2024uj, there are at least three discernible peaks in both \ion{He}{1} lines between 0 and 10000\kms, which suggests that there are distinct emitting regions of \ion{He}{1}. SN\,2024uj differs from the others with a \ion{He}{1} profile that is redshifted by $\sim$1000\kms, rather than centered around 0\kms. This difference in profiles between CaSTs may not be surprising if these systems are inherently asymmetric and subject to viewing angle dependencies. Interestingly, \citet{kumar_multiwavelength_2026} instead identify a boxy \heItwo\ profile in their NIR spectrum of SN\,2025coe taken at +34\;d (which may not be a fully nebular phase). A boxy \ion{He}{1} profile is typically associated with a He envelope in massive star progenitor scenarios (e.g., SN\,2020acat \citealt{medler_flat-topped_2023}), but it does not preclude a WD progenitor system.

We searched the literature for nebular-phase NIR spectra with \ion{He}{1} profiles similar in structure to those in SN\,2024uj (multipeaked, highly asymmetric) and find that the closest match is the AM CVn system V445 Puppis (see Figure 4 in \citealt{woudt_expanding_2009}). The implications of this are further discussed in \autoref{sec:progenitors}.

\subsection{Line Profiles and Velocities}\label{sec:O-C-He}
\begin{figure*}[htbp]
    \centering
    \includegraphics[width=0.95\textwidth]{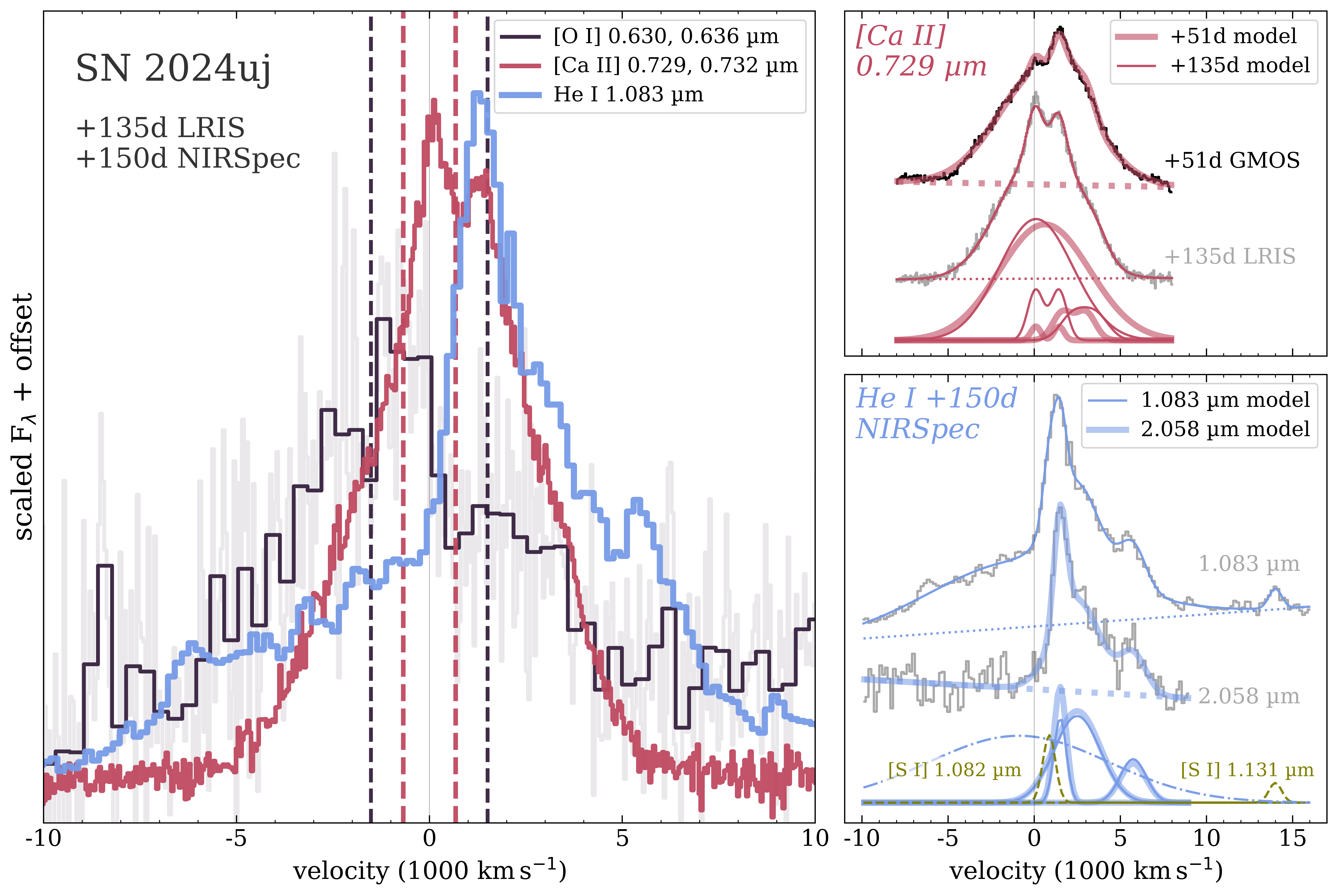}
    \caption{\textit{Left:} [\ion{O}{1}], [\ion{Ca}{2}], and \ion{He}{1} velocity profiles of SN\,2024uj at $+$135\;d and $+$150\;d after the second r-band peak. Doublets are in velocity-space with respect to the average wavelength, and actual line centers are marked with vertical dashed lines. The [\ion{O}{1}] profile (black) is rebinned, with raw data shown in low opacity underneath. There are regions where each ion distribution overlaps ($[0-2000]$\kms), and others where only one is present (e.g., \ion{He}{1} at $\sim5000$\kms). \textit{Top right:} multicomponent fits to the [\ion{Ca}{2}] line profile at both $+$51\;d (thick, lighter) and $+$135\;d epochs (thin, darker). Three sets of doublet Gaussians captures the asymmetry and evolution of this feature. \textit{Bottom right:} multicomponent fits to both the \ion{He}{1} 1.083\,$\mu$m (thin, darker) and 2.058\,$\mu$m (thick, lighter) profiles at $+$150\;d. Three Gaussians can replicate both profile shapes, with an additional ``shoulder" Gaussian (dashed-dotted line) only present in the 1.083\,$\mu$m feature. We also account for blending with [\ion{S}{1}] (green dashed lines) in our \heIone\ fit.}
    \label{fig:profile-vel-24uj}
\end{figure*}

The \oIdouble\ doublet, the [\ion{Ca}{2}] $\lambda\lambda7291,$$~7342$ doublet, and \ion{He}{1} at both 1.083\,$\mu$m and 2.058\,$\mu$m are relatively strong and isolated emission lines in SN\,2024uj (\autoref{fig:fullspec+phot}).
In the left panel of  \autoref{fig:profile-vel-24uj}, we show the velocity-space line profiles for each of these lines. These nebular-phase emission lines directly probe the innermost ejecta, which has become optically thin, giving an integrated flux distribution as a function of projected velocity along the line of sight.
The shape of these line profiles therefore encodes how a given ion is distributed throughout the ejecta \citep{jerkstrand_nebular_2014}, providing constraints on their spatial distribution that can differentiate between SN progenitor systems and explosion mechanisms (see \autoref{sec:discussOCaHe} for our physical interpretations).

Although the \oIdouble\ feature has low S/N, the rebinned data show two distinct peaks at velocities roughly corresponding to each line in the doublet (black dashed lines). The doublet profiles in the left panel of \autoref{fig:profile-vel-24uj} are transformed into velocity space using the average wavelength of the doublet, 
so the actual distribution of  [\ion{O}{1}] peaks around $\sim0$\kms, indicating it is located roughly centrally in the SN. The [\ion{O}{1}] profile shape is also relatively peaked (i.e., not flat-topped), and not highly asymmetric around $v\approx$0\kms. The theoretical 6300:6364 [\ion{O}{1}] flux ratio (based on transition probabilities) is $\sim$3:1, and ratios closer to 1:1 have been attributed to optically thick emission \citep{li_o_1992, williams_extinction_1994}.
In our case, the $+$135d \oIdouble\ emission has an intensity ratio that deviates from 3:1 (1.9:1.0), which may be evidence for clumping in the ejecta \citep{modjaz_double-peaked_2008, milisavljevic_doublets_2010}. 

The \caIIdouble\ doublet is also clearly double-peaked in the left panel of \autoref{fig:profile-vel-24uj}. The overall emission profile is relatively centered around $v\approx0$\kms, but the peaks appear redshifted by $\sim$750\kms, with the separation between the observed peaks being similar to the separation between the lines in the doublet (red dashed lines). 
In the top right panel of \autoref{fig:profile-vel-24uj}, the double-peaked structure differs between the $+$51\;d and $+$135\;d [\ion{Ca}{2}] profiles. At $+$51\;d, the \caIIdoubleright\ emission has a larger amplitude than \caIIdoubleleft, while the opposite is true at $+$135\;d. Both epochs deviate from the theoretical $\sim$1:1 probability ratio of the lines in this doublet. We find that this asymmetric profile evolution can only be modeled with multiple contributing components of [\ion{Ca}{2}] (see \autoref{sec:modelCa}).

In contrast to the broadly central [\ion{Ca}{2}] and [\ion{O}{1}], the \heIone\ emission profile is complex and asymmetric. There is a strong, narrow peak close to $v\approx1000$\kms, another clear peak at $v\approx5000$\kms, and a broad, shallow, blueshifted shoulder between $[-5000,0]$\kms.

There is some overlap between the [\ion{O}{1}], [\ion{Ca}{2}], and \ion{He}{1} profiles. For example, there are strong peaks of all three ions that coincide in the [$0-2000$]\kms\ range. There are also regions with only one, however, such as the peak at $\sim$5000\kms\ and the shallow blueshifted feature in \ion{He}{1} that are not coincident with components of [\ion{Ca}{2}] or [\ion{O}{1}]. The high S/N of the [\ion{Ca}{2}] and \ion{He}{1} (at both $1.083$ and $2.058\,\mu$m) emission enables us to fit them with multiple components and infer properties of their kinematics and distribution within the ejecta.

\subsubsection{Gaussian Fits to Calcium}\label{sec:modelCa}

We present our fits to the \caIIdouble\ line profile as it evolves in the right top panel of \autoref{fig:profile-vel-24uj}. Multiple components are required to capture the asymmetry in the two clear [\ion{Ca}{2}] peaks and the way in which they evolve. We model this feature as three sets of Gaussian pairs representing independently evolving concentrations of [\ion{Ca}{2}] in the ejecta, and a linear component for the local pseudo-continuum.

We use a simple weighted least-squares algorithm to find a three-component model of [\ion{Ca}{2}] with maximized likelihood. We start by fitting the fully nebular $+$135\;d emission line with Gaussians pairs at fixed $\Delta\lambda,$ allowing any value for the means (velocity offset of the emitting material $v$) of each pair. To ensure our fit captures physical features in the emission profile rather than narrow low S/N features, we set a lower bound of $\sim$600\kms\ for the Full Width at Half Maximum (FWHM), or, the velocity dispersion of the emitting material. Using the best-fit parameters from the $+$135\;d \caIIdouble\ line profile (setting $v_{+49d} = v_{+134d}\pm1500$\kms\ and FWHM$_{+49d}=\rm{FWHM}_{+134d}\pm$1800\kms, and leaving amplitudes unconstrained), we fit the $+$51\;d profile. 

In our resulting fits, the broad $\sim0$\kms\ Gaussian pair and the Gaussian pair near $\sim2500$\kms\ are nearly static (similar amplitudes) between epochs. The other narrower Gaussian pair near $\sim700$\kms, becomes more pronounced relative to the other components between epochs, giving rise to the double peaked structure. Our multicomponent fit parameters are given in \autoref{tab:velocities} and we discuss physical interpretations in \autoref{sec:discussOCaHe}.

\begin{deluxetable*}{c c c c c c}
\tablecaption{SN\,2024uj Line Velocities \label{tab:velocities}}
\tablehead{
\colhead{Feature} & \colhead{Phase} & \colhead{Gaussian label} & \colhead{mean $\mu$ (velocity)} & \colhead{FWHM (velocity dispersion)}\\ 
    \colhead{} & \colhead{} & \colhead{} & \colhead{(1000 km/s)} & 
    \colhead{(1000 km/s)}
}
\startdata
        $[$\ion{Ca}{2}$]$$^\star$ & $+$51\;d & narrow, strengthening&0.77$\pm$0.03&0.68$\pm$0.09\\
        $[$\ion{Ca}{2}$]$$^\star$ & $+$51\;d & broad, strong&0.61$\pm$0.03&5.93$\pm$0.07\\
        $[$\ion{Ca}{2}$]$$^\star$ & $+$51\;d & redshifted, weak&2.32$\pm$0.03&1.34$\pm$0.07\\
        \midrule
        $[$\ion{Ca}{2}$]$$^\star$ & $+$135\;d & narrow, strengthening&0.76$\pm$0.01&1.08$\pm$0.05\\
        $[$\ion{Ca}{2}$]$$^\star$ & $+$135\;d & broad, strong&0.11$\pm$0.11&4.76$\pm$0.12\\
        $[$\ion{Ca}{2}$]$$^\star$ & $+$135\;d & redshifted, weak&2.91$\pm$0.04&1.74$\pm$0.19\\
        \midrule
     \ion{He}{1}$^\dagger$  & $+$150\;d & narrow, blended & 1.51 [1.45, 1.51] & 0.85 [0.85, 1.01]\\
     \ion{He}{1}$^\dagger$  & $+$150\;d & broad, blended & 2.49 [2.45, 2.64] & 3.17 [2.88, 3.34]\\
     \ion{He}{1}$^\dagger$ & $+$150\;d & high-velocity & 5.74 [5.64, 5.77] & 1.78 [1.70, 1.95]\\
     \heIone & $+$150\;d & shoulder & -0.89 [-1.21, -0.82]& 12.30 [11.75, 12.71]\\
\enddata
\tablecomments{For [\ion{Ca}{2}], quoted $\mu$ and FWHM values are  those with minimized weighted-$\chi^2$ values and their uncertainties are the standard deviations. For \ion{He}{1}, quoted $\mu$ and FWHM values are the maximum a posteriori value, and the 16th and the 84th percentile credible interval are in brackets.\\ $^\star$: Fits are with respect to the average wavelength of the doublet (7307.5 \AA).\\ $^\dagger$: Fits are with respect to 1.083 and 2.058 $\mu$m simultaneously.}
\end{deluxetable*}

\subsubsection{Gaussian Fits to Helium}\label{sec:modelHe}

The highly redshifted and asymmetric \ion{He}{1} profiles in SN\,2024uj, along with our multicomponent fits, are shown in the bottom right panel of \autoref{fig:profile-vel-24uj}. The \ion{He}{1} profile is more complex than that of [\ion{Ca}{2}], and with only one epoch, we can run the risk of overfitting this feature with an overly complicated model. For this reason, we utilize nested sampling in our fits to \ion{He}{1}. The primary goal of nested sampling is to estimate the Bayesian evidence $\mathcal{Z}$ (the marginal likelihood), which represents the probability of the data given a specific model. These evidence values quantify how well a given model explains the data while penalizing unnecessary complexity. When comparing different models, selecting the one with higher evidence guards against overfitting.

In our fits to \ion{He}{1}, we model the 1.083$\mu$m\, and 2.058$\mu$m\, features simultaneously. Our \heIone\, model is comprised of a linear pseudo-continuum and three Gaussian components that represent distinct regions of neutral helium within the SN ejecta. The choice of three Gaussians, rather than two, is a result of model comparison: the ratio of evidences, or, the log Bayes factor = 114, in favor of the triple-Gaussian model. We include an additional broad Gaussian for the blueshifted ``shoulder'' (non-zero flux between [-8000, 0]\kms). Lastly, because \heIone\, is blended with \sIone, and there is also an isolated neighboring \sItwo\, line, we include 2 Gaussian components in our model for these [S I] lines. The fit to the isolated \sItwo\ line sets the amplitude, $v$, and FWHM of the \sIone\ Gaussian, based on the theoretical intensity ratio of these two transitions.

Our \heItwo\, fit is the same as our \heIone\, fit, but without the broad blueshifted ``shoulder'' Gaussian component or the nearby [\ion{S}{1}] components.
When fitting both \heIone\, and \heItwo\, profiles simultaneously, the three shared Gaussians have the same velocity parameters, and we allow their amplitudes to vary slightly between each profile. We use the nested sampling package \texttt{dynesty} \citep{speagle_dynesty_2020} to find the solution with a maximized posterior probability. For more details on prior selection and the model setup, see \autoref{sec:appendixHe}.
\autoref{fig:profile-vel-24uj} (bottom right panel) shows the fit using maximum a posteriori (MAP) values of the samples for each parameter. MAP values and credible ranges for the fit parameters are given in \autoref{tab:velocities}. 

\subsubsection{Distinct O, Ca, and He distributions}\label{sec:discussOCaHe}

With our best-fitting models of [\ion{Ca}{2}] we identify a central, narrow component which gives rise to the doublet peak and strengthens over time, a redshifted component that shifts towards higher velocities over time. These dynamic components cause the red peak seen in the $+$51 day spectrum to fade with respect to the blue side of the profile. On top of these components, there is a strong and broad centrally-located component which stays relatively stagnant over the 84\;d phase difference.

In our fits to \ion{He}{1}, all 3 Gaussians shared between $1.083$ and $2.058\,\mu$m are redshifted (positive velocities). Two reside at similar line of sight velocities and are blended with one another (narrow feature at $v=1510$\kms, broad feature at $v=2490$\kms) and one is significantly offset ($v=5740$\kms). This offset, high-velocity component could be explained by a He-rich pre-SN outburst. The other components can reasonably be connected to the SN alone, although this is hard to constrain without several epochs of He detected. The broad ``shoulder'' term only present in the \heIone\ feature is centered at $v=-890$\kms, with a very broad dispersion (FWHM$=12300$\kms). This feature could be attributed to an asymmetric shell of He sitting at high velocities, or neighboring weak, blended lines.

Our multicomponent fits have elucidated the physical layout of SN\,2024uj in its nebular phase. The [\ion{Ca}{2}] and \ion{He}{1} profiles are complex, asymmetric, and distinct from one another and [\ion{O}{1}]. This suggests distinct emitting regions and distributions of these three elements within the ejecta. We find that the [\ion{Ca}{2}] and [\ion{O}{1}] appear to come from similar, more central regions, whereas the \ion{He}{1} is very redshifted and extends to higher velocities. We discuss the implications for the progenitor system and explosion mechanism in \autoref{sec:progenitors}.

\subsection{Early [\ion{Ca}{2}] emission}

Forbidden transitions, with their extremely small probabilities (long lifetimes), are only expected to occur in low-density environments. In a high-density environment, it would be more likely for an atom to be collisionally de-excited than to emit a photon from a forbidden transition. With the exception of some extreme super-luminous supernovae (SLSNe), where observed photospheric [\ion{Ca}{2}], [\ion{O}{2}], and [\ion{O}{3}] emission is connected to CSM interaction \citep{lunnan_ps1-14bj_2016, inserra_complexity_2017, aamer_precursor_2024, schulze_1100_2024}, forbidden emission lines are only expected to emerge in the late-time nebular phase, when SN ejecta become optically thin.

We observed forbidden \caIIdouble\ emission emerging in the early optical spectra of SN\,2024uj (\autoref{fig:optical-24uj}). This line has clearly emerged by $+$17\;d ($\sim$30\;d post-explosion), and may be present as early as $+$2\;d ($\sim$15\;d post-explosion). The rapid onset of this forbidden emission is a common characteristic of CaSTs: in their sample study, \citeCGsample\ find that 90$\%$ of CaSTs with near-peak spectra exhibit [\ion{Ca}{2}] before $+$10\;d.

A possible explanation for early [\ion{Ca}{2}] emission is the presence of Ca in the outermost layers of the ejecta, which would become optically thin faster than deeper material within the ejecta. In thermonuclear explosions triggered via burning in an outer helium-rich shell (e.g., \citealt{liu_sn_2023-1}), newly synthesized Ca is formed in what becomes the outermost layers of the ejecta, followed by a detonation of the WD \citep{polin_observational_2019}. 
For massive stars that produce helium-rich SNe, however, there is no mechanism to naturally mix significant amounts of Ca into the helium-rich stellar envelope that ultimately forms the outermost layers of the ejecta. To test the possibility of early [\ion{Ca}{2}] emission in both thermonuclear and core collapse explosions we run radiative-transfer simulations of both scenarios and compare to the observations.

\begin{figure*}[htbp]
    \centering
    \includegraphics[width=0.98\textwidth]{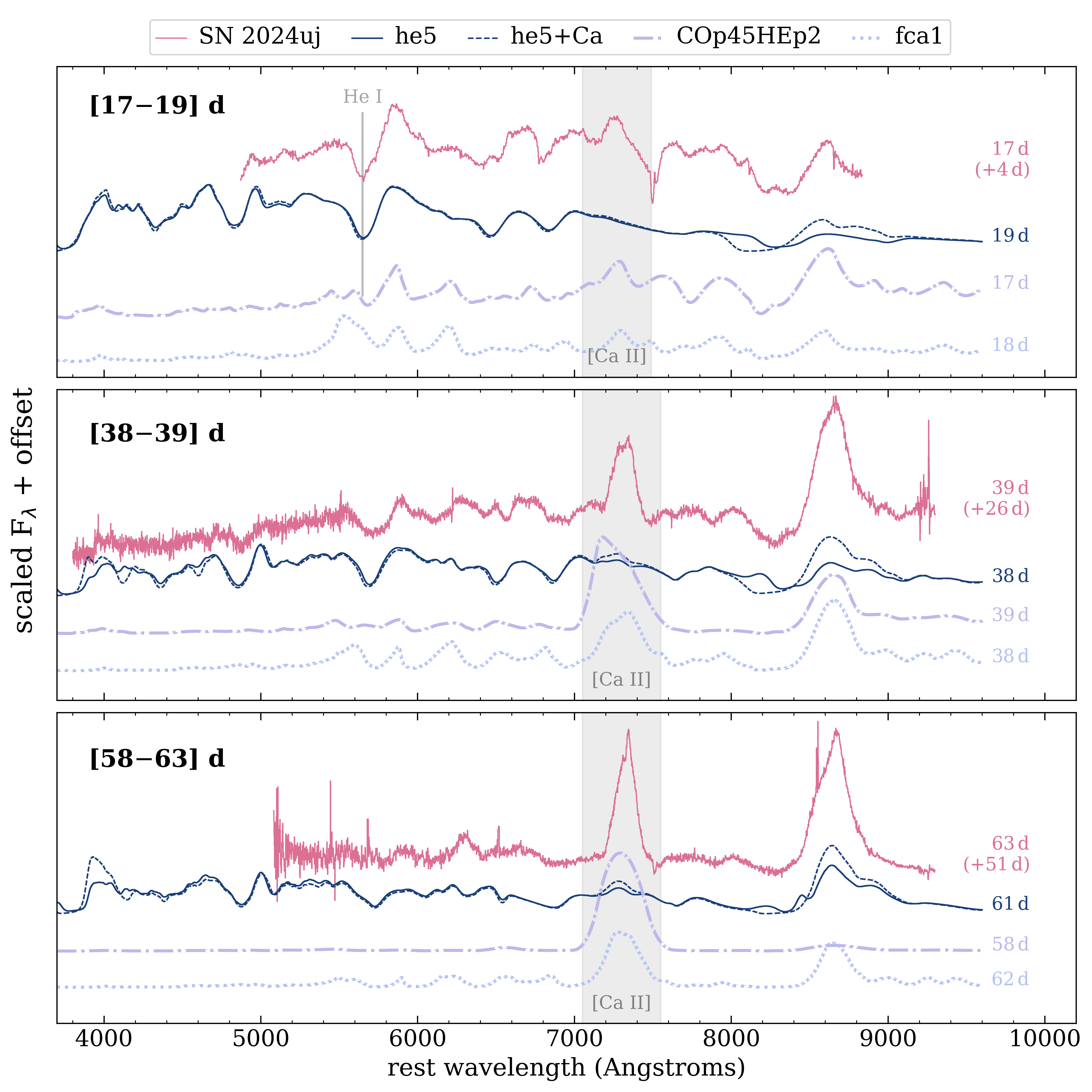}
    \caption{Multi-epoch comparison of SN\,2024uj observations (pink, thin solid lines) and three radiative-transfer models. The overlaid solid and dashed navy lines are two versions of a He-star core-collapse model (\texttt{he5}; \citealt{dessart_supernovae_2020}), where the dashed line has extra Ca in the outermost layers of the star. Even with this unphysical Ca addition, this model does not reproduce the strong [\ion{Ca}{2}] emission in our observations (grey shaded regions). The thermonuclear models (dotted-dashed lines \texttt{COp45HEp2} from \citealt{waldman_helium_2011} and \citealt{dessart_one-dimensional_2015}; dotted lines \texttt{fca1} from \citealt{zenati_origins_2023}) do produce strong [\ion{Ca}{2}] at all three epochs. \texttt{COp45HEp2} simulates a He-shell detonation on a low-mass WD, and \texttt{fca1} simulates a double-WD merger. \texttt{fca1} is the only model unable to reproduce the strong \ion{He}{1} absorption in the top panel (earliest phase). Phases are given with respect to explosion, those in parentheses are with respect to the second r-band peak.}
    \label{fig:ca-tweak}
\end{figure*}

In \autoref{fig:ca-tweak}, we compare optical observations of SN\,2024uj with model SN spectra from non-local thermodynamic equilibrium (non-LTE) \texttt{CMFGEN} radiative-transfer simulations \citep{hillier_time-dependent_2012}. We focus on potential progenitor scenarios to explain CaSTs, including: a helium star, significantly stripped from binary mass transfer, that undergoes core collapse \citep[the \texttt{he5} model from][]{dessart_supernovae_2020}; a helium shell ($M_{\text{He~shell}}=0.2M_\odot$) detonation on a low-mass WD with C/O core mass $M_{\rm C/O}=0.45M_\odot$ \citep[the \texttt{COp45HEp2} model from][]{waldman_helium_2011, dessart_one-dimensional_2015}; and the full disruption of a low-mass C/O WD by a more massive hybrid He/C/O WD, which undergoes a ``weak'' detonation following the accretion of C/O onto the He/C/O-WD that ultimately does not fully unbind the He/C/O-WD \cite[the \texttt{fca1} model from][]{zenati_origins_2023}. In \autoref{fig:ca-tweak}, for direct comparison to the simulated spectra, all phases are reported with respect to explosion (MJD~60315.10$^{+0.03}_{-0.01}$) instead of the second r-band peak.

The two thermonuclear models in \autoref{fig:ca-tweak} (\texttt{COp45HEp2} and \texttt{fca1}) show strong [\ion{Ca}{2}] emission identifiable $\sim$20\;d post-explosion (around the second peak in SN\,2024uj).
In the core-collapse model (\texttt{he5}), the [\ion{Ca}{2}] emission remains weak even two months after explosion. In an attempt to identify a core-collapse scenario where strong [\ion{Ca}{2}] emission emerges early, we artificially modify the \texttt{he5} model, adding Ca to the outer He envelope of the star. This modified \texttt{he5+Ca} model has increased Ca abundance, where 1$\%$ of the He mass fraction is replaced with Ca. This represents a very small change for He, but a huge change for Ca. The resulting models are plotted in \autoref{fig:ca-tweak} as navy dashed lines.

At early times (17\;d post-explosion and $+$4\;d post-second r-band peak, top panel \autoref{fig:ca-tweak}), SN\,2024uj most closely resembles the core-collapse \texttt{he5} and \texttt{he5+Ca} models. 
The strong \ion{He}{1} absorption in SN\,2024uj is well matched to the \texttt{he5} and \texttt{COp45HEp2} models, but not the \texttt{fca1} model. \citet{zenati_origins_2023} find that \texttt{fca1} is not fully consistent with Ib-like CaSTs spectroscopically or photometrically, in part because of this missing \ion{He}{1} absorption.

At intermediate times (39\;d post-explosion, middle panel), no single model matches the observations particularly well in terms of relative line strengths.
By 63\;d post-explosion, SN\,2024uj most closely resembles the thermonuclear models, largely due to the strong [\ion{Ca}{2}]. Even with an unphysically large addition of Ca in the outer layers, the \texttt{he5+Ca} model does not reproduce strong enough [\ion{Ca}{2}] lines to resemble the Ca in SN\,2024uj at any phase. This demonstrates that the Ca in \texttt{he5+Ca} is not reaching optically thin states early enough, and an even larger amount of Ca not expected in massive stars, would be required to match the observations of SN\,2024uj. We therefore suggest that the massive star scenario is improbable for SN\,2024uj, and since early [\ion{Ca}{2}] is a common feature for the class, improbable for all CaSTs.

We also note that spectra in \autoref{fig:ca-tweak} are not presented in absolute flux; rather, they are rescaled to better compare relative line strengths. \texttt{COp45HEp2} has the lowest overall flux and is rescaled by a factor of $\sim$100 to match the strengths of the other models and observations. This is an important discrepancy -- even though \texttt{COp45HEp2} demonstrates strong [\ion{Ca}{2}] and overall similar relative line strengths to SN\,2024uj at phases beyond $\sim38$\;d post explosion, there is a clear mismatch in luminosity which encourages further exploration and/or revisions to the model scenario.

\section{Discussion}\label{sec:discuss}

\subsection{Progenitor Scenarios}\label{sec:progenitors}

\subsubsection{SN\,2024uj: CCSN?}

Some properties of CaSTs resemble those of stripped envelope SNe (Ib/c), suggesting a possible massive star origin. In SN\,2024uj, the early time spectra, particularly near the second r-band peak (17\;d post-explosion), closely match the stripped He star \texttt{he5} model (top panel of \autoref{fig:ca-tweak}), largely due to the presence of He absorption. At late times, CCSNe show prominent He emission, as does SN\,2024uj. Binary systems that give rise to many stripped-envelope supernovae (SESNe) can be asymmetric, viewing-angle dependent, and shrouded by aspherical CSM (e.g., \citealt{chandra_supernova_2020, baer-way_multiwavelength_2025}).
This is consistent with the significantly asymmetric nebular lines in the nebular SN\,2024uj spectra; however, the complex \ion{He}{1} line extending to $\gtrsim5000$\kms and peaking at $+$1500\kms\ does not closely resemble the He profiles seen in CCSNe, even asymmetric SESN. Finally, SN\,2024uj displays a shock-cooling first light curve peak, and molecular and dust emission, both of which are typically associated with CCSNe due to pre-explosion mass loss in massive stars. 
There are, however, rare SN Ia subtypes that also exhibit early light curve bumps: 02es-like \citep{cao_iptf_2015, miller_spectacular_2020, xi_sn_2024} and 03fg-like \citep{jiang_discovery_2021, srivastav_luminous_2023, dimitriadis_sn_2023}, with the latter potentially arising from shock-cooling.

There are three key observables in SN\,2024uj and other CaSTs that are difficult to reconcile with a massive star progenitor system: the early emergence of [\ion{Ca}{2}], the redshifted \ion{He}{1} component (v$=$5740\kms), and a remote environment. In an experiment with the massive He-star radiative-transfer model \texttt{he5}, we are unable to reproduce early [\ion{Ca}{2}] emission even after adding unphysically large amounts of Ca ($\chi_{\rm Ca}$=0.01) to the star's outer layers, which would require a level of mixing that is not expected in a massive star scenario. We encourage future exploration of this point with a broader range of massive star progenitor models and configurations.
Next, while the highly redshifted peak velocity of \ion{He}{1} at 5740\kms\ could be tied to previous mass-loss episodes or a high bulk velocity, it is unclear how a CCSN scenario could produce the peculiar shape of the \ion{He}{1} line, with bulk redshift and wings extending to the broadest velocities, while [\ion{Ca}{2}] and [\ion{O}{1}] remain centrally located (\autoref{fig:profile-vel-24uj}). Finally, the lack of local star formation and the large host-galaxy offset for SN\,2024uj and other CaSTs makes a young, massive star progenitor less likely. However, the remote location alone does not preclude a massive star origin (e.g., the SN Ic-BL SN\,2002ap was observed 10\;kpc away from its host, at a site with no local star formation; \citealt{gal-yam_supernova_2002, modjaz_measured_2008}).

\subsubsection{SN\,2024uj: AM CVn-like?}
An AM CVn-like progenitor system modeled after V445 Puppis could explain a number of features in the SN\,2024uj observations. AM CVn systems can undergo successive He novae, episodes of helium burning and mass ejection, which have also been proposed to explain the helium observed in some Type Iax SNe (e.g., \citealt{jacobson-galan_detection_2019}). These He novae can give rise to lobes with He, Ca, and O at extended radii \citep{woudt_expanding_2009}. Within an AM CVn-like CaST progenitor scenario, these low-density Ca-rich regions formed pre-SN could help explain the early emergence of [\ion{Ca}{2}] emission \citep{touchard-paxton_sn_2025}. The bipolar lobes (or similarly complex structure) of a V445 Puppis-like system would also likely be able to produce the multicomponent Ca and He profiles, distinct regions of Ca and O, He moving at $\gtrsim$5600\kms, and overall asymmetry of the nebular emission lines seen in SN\,2024uj. The presence of dust and CO in SN\,2024uj (\autoref{fig:fullspec+phot}) is also broadly consistent with V445 Puppis-like system, as \citet{ashok_enigmatic_2003} identify a dust shell in V445 Puppis that was formed $\sim$1 month after the He-nova outburst. Lastly, the remote location of SN\,2024uj is compatible with an AM CVn-like progenitor composed of a double degenerate WD binary system (e.g., the V445 Puppis-like CaST model in \citealt{touchard-paxton_sn_2025}), although it is unclear whether this channel can reproduce the distribution of large host-SN offsets that define the CaST class.

Based on the bipolar lobe geometry of V445 Puppis, the AM CVn-like CaST progenitor in \citet{touchard-paxton_sn_2025} also predicts spectral `transition features’: emission lines dominated by the external regions at early times (profiles with strong high-velocity components) and by the central SN at later times (profiles with strong components near $v\approx$0\kms). In our multicomponent fits to [\ion{Ca}{2}] in SN\,2024uj we identify a central component ($v\approx750$\kms\ between $+$51 and $+$135\;d) which grows over time, which may be tied to the SN overpowering external material from pre-SN activity. We encourage further exploration into `transition features' in the CaST class at large, especially with high resolution spectra taken with high cadence.

There is also a compelling similarity between the \ion{He}{1} profiles of SN\,2024uj and V445 Puppis. \citet{woudt_expanding_2009} identify four distinct peaks in the \ion{He}{1} lines at $7065~$\AA\, and $2.058~\mu$m, two with negative velocities and two with positive, which they attribute to blue- and redshifted material distributed asymmetrically in each lobe. In SN\,2024uj we see similarly peaked asymmetric structure (\autoref{fig:profile-vel-24uj}), but all peaks have positive velocities $v>1400$ km/s, besides the blue ``shoulder'' only present in \heIone. While we cannot directly attribute the \ion{He}{1} in SN\,2024uj to a bipolar shell structure like the one in V445 Puppis due to a lack of similar spatial information, this possibility should be investigated in future high-resolution NIR samples of CaSTs and three-dimensional models. We note that of the other CaSTs with $\geq+$35\;d NIR spectra (SN\,2019ehk and iPTF\,15eqv, \autoref{fig:cast-HeI}) only SN\,2024uj has \ion{He}{1} profiles that resemble those in V445 Puppis. A larger sample would help determine whether this is a viewing angle effect or a shortcoming of the model. 

Although aspects of this model are promising, it has yet to be modeled with radiative-transfer simulations, and may struggle to explain other CaST properties. In particular, it is unclear whether the lobes in the V445 Puppis-like CaST model \citep{touchard-paxton_sn_2025} can reach high enough densities to generate the shock-cooling first light curve peak and associated X-ray emission seen in several well-observed CaSTs \citep{jacobson-galan_sn_2020, jacobson-galan_circumstellar_2022, kumar_multiwavelength_2026}. Additionally, \citet{de_zwicky_2020} note that the rate of AM CVn systems is well below their estimate of the CaST volumetric rate, but they do not rule out that some CaSTs---particularly some members of the Ca-Ib subclass---could originate from AM CVn systems.

\subsubsection{SN\,2024uj: Low-mass binary WD merger?}
Finally, we consider a WD merger scenario with at least one He WD or hybrid He/C/O WD, which we refer to as a low-mass WD binary. We note that the AM CVn-like and low-mass WD binary scenarios are likely not mutually exclusive. It may be that an AM CVn-like progenitor explodes terminally in a low-mass WD merger, and we encourage modeling to explore this possibility.

The late-time optical spectra of SN\,2024uj and other CaSTs reasonably resemble model spectra of binary WD systems (see \texttt{COp45HEp2} and \texttt{fca1} in \autoref{fig:ca-tweak}), which predict strong late-time [\ion{Ca}{2}] and also the early emergence of [\ion{Ca}{2}]. There are several possible low-mass WD merger models which explode in various ways. The \texttt{fca1} model \citep{zenati_origins_2023} follows the accretion of C/O material onto an He/C/O-WD that causes a detonation of the He shell, burning away most of the He while leaving a hot remnant C/O core that may radiate X-rays as it cools. This causes the \texttt{fca1} model not to produce significant He features \citep{zenati_origins_2023}. \cite{moran-fraile_faint_2024} simulated the merger of a C/O WD and He WD where the He WD is disrupted into a massive He shell around the C/O WD and the He ignites, triggering a double detonation. \cite{callan_nlte_2025} showed that full NLTE radiative transfer of this model does produce He absorption at early times and He emission at later times in both the optical and NIR, validating that He lines can be produced in thermonuclear explosions. 

Merger systems can be quite asymmetric and viewing angle dependent \citep[][]{pakmor_normal_2012, kwok_ground-based_2024, pakmor_violent_2026}, which might provide an explanation for the asymmetric \ion{He}{1} and [\ion{Ca}{2}] profiles in the $+$150\;d NIRSpec spectrum of SN\,2024uj. The nebular \ion{He}{1} also has a strong redshifted peak at $\sim$$+$1500 \kms, which could potentially be attributed to a shocked helium-rich companion. \cite{wong_mass_2025} predict narrow nebular He emission lines (1000--4000\kms) for double-degenerate SNe\,Ia involving a low-mass He WD companion that is flung off at its orbital velocity of ~1000--3000\kms. The higher-velocity (5740\kms) \ion{He}{1} in SN\,2024uj might also be connected to material that was stripped prior to merger.

Several other characteristics of WD mergers of low-mass WDs can naturally account for the observed low energy and ejecta masses of SN\,2024uj and other CaSTs. Thermonuclear explosions produce higher kinetic energy to ejecta mass ratios than CCSNe because they produce $^{56}$Ni more efficiently and have less additional envelope mass to lift off, enabling the SN ejecta to cool and become optically thin quickly. This is consistent with the observation of early forbidden Ca emission and rapid transition to the nebular phase in CaSTs. In a WD merger scenario, material can be stripped off during the inspiral, providing a potential avenue for nearby CSM; however, it is unclear whether such material would be of sufficient density to produce the shock cooling and X-rays associated with CaSTs. Unburned C and O from a merger system, either stripped pre-explosion or left over if the explosion does not fully disrupt the WD (as in the \texttt{fca1} model; \citealt{zenati_origins_2023}), could potentially form CO and seed dust formation, in line with the CO and dust detections in SN\,2024uj (see \autoref{fig:fullspec+phot} and S.\,Hall 2026b, in prep.) Finally, the remote locations of most CaSTs are suggestive of WD progenitors.

Some characteristics of SN\,2024uj are not fully consistent with our current set of low-mass binary WD merger models. Future hydrodynamical simulations and radiative-transfer models should investigate the pre-merger mass loss and include physics for molecule and dust formation. Multidimensional models should also investigate whether such systems give rise to ejecta distributions that are asymmetric enough to reproduce the distributions of He, Ca, and O seen in SN\,2024uj (\autoref{fig:profile-vel-24uj}). Although still difficult computationally, such models should follow the pre-merger inspiral, such as was done by \cite{pakmor_violent_2026}, to help determine whether these systems can produce dense enough material prior to explosion. Lastly, we note that a WD progenitor system, one which could exist at many locations throughout a given host galaxy, does not fully explain why CaSTs consistently explode at large host-SN offsets.

\section{Summary and Conclusions}

\newcommand{\cmark}{\ding{51}}
\newcommand{\xmark}{\ding{55}}

\begin{deluxetable}{l c c}[h!]
\centering
\small
\setlength{\tabcolsep}{4pt}
\tablecaption{Observables of SN\,2024uj and their compatibility with different explosion mechanism/progenitor origins \label{tab:obs_interp}}
\tablehead{
\colhead{Observable} & \colhead{Thermo./} & \colhead{CC/}\\[-0.2cm]
\colhead{} & \colhead{WD} & \colhead{massive star}
}
\startdata
Large host-SN offset$^{\dagger}$  &  \cmark  & ? \\
Early [\ion{Ca}{2}] emission$^{\dagger}$  &  \cmark    & \xmark \\
Large $E_\mathrm{k}/M_{ej}$ & \cmark & \cmark \\
Strong \ion{He}{1} emission$^{\dagger}$  &  ?  &
\cmark\\
Multicomponent \ion{He}{1}  &  \cmark  & ? \\
\ion{He}{1}, [\ion{Ca}{2}], and [\ion{O}{1}] overlap  &  ?     & \xmark \\
Molecular CO  &  ?   & \cmark \\
\enddata
\tablecomments{${\dagger}:$ Feature is common to all He-rich (Ib-like) CaSTs.}
\end{deluxetable}

We present the first JWST observations of a CaST, SN\,2024uj, taken $+$150\;d post-second r-band peak. From our analysis of the nebular JWST/NIRSpec spectrum, in combination with photometry and optical spectra taken from $-$8\;d to $+$135\;d, we favor a thermonuclear origin for SN\,2024uj, but cannot entirely rule out a CCSN origin.

We summarize our findings for SN\,2024uj as follows:

\begin{enumerate}
    \item The large host-SN offset of $\sim$ 6.6\;kpc and lack of local star formation disfavor a young, massive star progenitor for SN\,2024uj.
    \item The double-peaked bolometric light curve of SN\,2024uj is best fit with a shock-cooling model (first peak) and the radioactive decay of $^{56}$Ni (second peak). 
    The shock-cooling emission supports an explosion scenario with pre-explosion mass loss that provides a dense ambient CSM and/or a gravitationally bound envelope.
    The large ratio of kinetic energy to ejecta mass ($E_{\mathrm{k}}\approx$ 0.2 $\times$ 10$^{51}$ erg, $M_{\mathrm{ej}}\approx0.6$\,$M_{\odot}$) inferred from the radioactive decay model (second peak) is also consistent with a thermonuclear explosion. 
    \item There is a rising flux continuum beyond 3\,$\mu$m in the JWST/NIRSpec spectrum and a bright 10 $\mu$m MIR detection. We also detect CO emission in the fundamental band, between 4.5--5.1 $\mu$m.
    \item In the nebular phase, \ion{He}{1} and [\ion{Ca}{2}] have asymmetric and complex line profiles distinct from one another and [\ion{O}{1}]. $[$\ion{Ca}{2}$]$ and [\ion{O}{1}] are centrally located in SN\,2024uj while \ion{He}{1} has a redshifted peak ($\sim$1500\kms) and extends to high velocities ($\sim$6000\kms). This suggests that these ions have distinct emission distributions within the ejecta. The asymmetry and multicomponent structure of these key elements is suggestive of a highly anisotropic explosion.
    \item $[$\ion{Ca}{2}$]$ is observed in the early optical spectra, with a tentative \caIIdouble\ detection near maximum light that becomes clear by $+$17\;d. This early forbidden emission suggests that Ca exists in the outermost layers of the SN ejecta. We explored greatly enhancing the abundance of Ca in the outer layers of a stripped He star explosion model and found that it was unable to replicate the early Ca line strengths observed in SN\,2024uj, whereas thermonuclear models naturally produce this feature.
\end{enumerate}

The complex \heIone\ and \heItwo\ line profiles suggest an asymmetric explosion, with He present throughout the ejecta. This is difficult to reconcile with a core-collapse scenario, as there is not a straightforward mechanism to produce a multicomponent He outflow in the outermost layers of the unburnt ejecta.
In our search for nebular NIR SN spectra in the literature, we could not find examples of stripped-envelope CCSNe with similar \ion{He}{1} profiles to SN\,2024uj. Instead, the profile shape is highly reminiscent of the helium detections (\ion{He}{1} at 0.7065 and 2.0581 microns) in V445 Puppis, an AM CVn system \citep{woudt_expanding_2009}. The AM CVn-like progenitor model proposed for CaSTs \citep{touchard-paxton_sn_2025} is currently untested from a radiative-transfer modeling standpoint; however, it appears consistent with our observations, with \ion{He}{1} at $v=5740$\kms\ potentially connected to pre-explosion ejecta. 

To date, He-rich (Ib-like) CaSTs have been interpreted as both core-collapse and thermonuclear SNe. The CaST-Ib SN\,2019ehk, which is spectroscopically and photometrically very similar to SN\,2024uj, has been tied to the core collapse of a stripped massive star progenitor \citep{de_peculiar_2021} and the tidal disruption of a low-mass C/O (or hybrid) white dwarf by a more massive hybrid white dwarf  \citep{jacobson-galan_sn_2020, jacobson-galan_late-time_2021}.
In this work, based on a remote explosion site, early [\ion{Ca}{2}] emission, small $^{56}$Ni and ejecta masses, and complex asymmetric nebular lines, particularly the He I emission, we favor the interpretation that SN\,2024uj is the product of a thermonuclear explosion. Specifically, we find that a binary WD merger, with at least one He WD or hybrid He/C/O WD, potentially within an AM CVn-like system is the most likely progenitor system. However, we cannot rule out a CCSN origin and encourage future modeling to further explore both avenues. Other CaST-Ib SNe also exist at large host-SN offsets, exhibit early [\ion{Ca}{2}] emission, show double-peak light curves, and feature strong \ion{He}{1} emission, similar to SN\,2024uj, so we suggest that our interpretation may extend to the larger subclass. CaSTs as thermonuclear explosions would reinforce that low-mass WDs with He can produce strong spectroscopic He signatures, and likely explode via a double detonation or He shell burning mechanism.

Finally, our detection of molecular CO emission in the $+$150\;d NIRSpec and MIRI observations of SN\,2024uj is the first such detection for a CaST and thermonuclear SN in general. In \citepaperII, we analyze the dust and CO in SN\,2024uj.
These results may suggest that faint thermonuclear SN can be dust producers. 
We encourage future hydrodynamical and radiative-transfer models of low-mass WD binaries with He, as well as investigating pre-explosion mass loss, and molecule and dust formation in these systems. Larger samples and multi-epoch late-time NIR and MIR observations of CaSTs will be essential to improving our understanding of this class of SN and their diversity.\\

\section{Data Availability}
All spectral observations analyzed in this work, and the software written to model them, are available at the following repository: \hyperlink{https://github.com/saarahhall/SN24uj}{https://github.com/saarahhall/SN24uj}.\\

S.H. and C.W.M. thank the LSST-DA Data Science Fellowship Program, which is funded by LSST-DA, the Brinson Foundation, the WoodNext Foundation, and the Research Corporation for Science Advancement Foundation; their participation in the program has benefited this work.
L.A.K. and W.J.-G.\ are supported by NASA through Hubble Fellowship grants (HF2-51579.001-A and HSTHF2-51558.001-A) awarded by the Space Telescope Science Institute, which is operated for NASA by the Association of Universities for Research in Astronomy, Inc., under contract NAS5-26555.
S.V., A.P.R., and the UC Davis time-domain research team acknowledge support from National Science Foundation (NSF) grant AST-240756
A.V.F.’s research group at UC Berkeley acknowledges financial assistance from the Christopher R. Redlich Fund, as well as donations from Gary and Cynthia Bengier, Clark and Sharon Winslow, Alan Eustace and Kathy Kwan, Timothy and Melissa Draper, Briggs and Kathleen Wood, Sanford Robertson, and Ellyn and Alan Seelenfreund (W.Z. is a Bengier-Winslow-Eustace Specialist in Astronomy, T.G.B. is a Draper-Wood-Seelenfreund Specialist in Astronomy, Y.Y. was a Bengier-Winslow-Robertson Fellow in Astronomy), and numerous other donors. Y.Y.'s research is partially supported by the Tsinghua University Dushi Program.

S.W.J. gratefully acknowledges support from a Guggenheim Fellowship.
C. Liu is supported by DoE award \#\,DE-SC0025599.
K.M. acknowledges funding from Horizon Europe ERC grant 101125877
C.W.M. acknowledges support by the NSF Graduate Research Fellowship Program under grant 2444108.
Time-domain research by the University of Arizona team and D.J.S. is supported by  NSF grants 2308181, 2407566, and 2432036.
C.-G.T.-P. acknowledges funding from Horizon Europe ERC grant 101125877.
M.M. acknowledge support in part from ADAP program grant 80NSSC22K0486, from  NSF grant AST-2206657, and from the NSF under Cooperative Agreement 2421782 and the Simons Foundation grant MPS-AI-00010515 awarded to the NSF-Simons AI Institute for Cosmic Origins — CosmicAI, https://www.cosmicai.org/.

This work was granted access to the HPC resources of TGCC under the allocation 2025 -- A0190416871 on Irene-Rome made by GENCI, France.
Some of the data presented herein were obtained at the W.\~M. Keck Observatory, which is operated as a scientific partnership among the California Institute of Technology, the University of California, and NASA. The Observatory was made possible by the generous financial support of the W.\~M. Keck Foundation. The authors wish to recognize and acknowledge the very significant cultural role and reverence that the summit of Maunakea has always had within the indigenous Hawaiian community. We are most fortunate to have the opportunity to conduct observations from this mountain.

This work makes use of observations from the Las Cumbres Observatory network. The LCO team is supported by NSF grant AST-2308113.
Data presented here from the Southern African Large Telescope were obtained via Rutgers University program 2023-1-MLT-008 (PI S. Jha). 
Based in part on observations obtained at the Southern Astrophysical Research (SOAR) telescope, which is a joint project of the Ministério da Ciência, Tecnologia, Inovações e Comunicações (MCTIC) do Brasil, the U.S. National Optical Astronomy Observatory (NOAO), the University of North Carolina at Chapel Hill (UNC), and Michigan State University (MSU).

\textit{Facilities:} JWST, Gemini (GMOS-S), Keck I (LRIS), Las Cumbres (Sinistro, MuSCAT), NTT (EFOSC2), SALT (RSS), SOAR (Goodman), Swift (UVOT; XRT)

\textit{Software:} 
\texttt{astropy} \citep{astropy_collaboration_astropy_2013, astropy_collaboration_astropy_2018}, 
\texttt{CMFGEN} \citep{hillier_time-dependent_2012},
\texttt{DRAGONS} \citep{labrie_dragonsquick_2023},
\texttt{dynesty} \citep{speagle_dynesty_2020},
\texttt{emcee} \citep{foreman-mackey_emcee_2013},
\texttt{HEASoft} \citep{heasarc_heasoft_2014},
\texttt{lcogtsnpipe} \citep{Valenti16},
\texttt{LPipe} \citep{perley_fully_2019},
\texttt{Matplotlib} \citep{hunter_matplotlib_2007}, 
\texttt{NumPy} \citep{harris_array_2020}, 
\texttt{SciPy} \citep{virtanen_scipy_2020}.

\restartappendixnumbering
\appendix

\section{Observations}\label{sec:appendixObsLog}

\begin{deluxetable}{c c c c c c}[h!]
\tablecaption{Log of SN\,2024uj spectral observations \label{tab:speclistuj}}
\tablehead{
\colhead{MJD} & \colhead{Phase} & \colhead{Telescope/Instrument} & \colhead{Wavelength range} & \colhead{Exp. time} & \colhead{Source}\\ 
    \colhead{} & \colhead{(day)} & \colhead{} & \colhead{($\mu$m)} & \colhead{(s)} & \colhead{}
}
\startdata
60319.21  &  $-$8  & NTT/EFOSC2 & 0.332 -- 0.739 & 1800 & \citet{petrushevska_epessto_2024}\\
60329.96  &  $+$2     & SALT/RSS & 0.388 -- 0.770 & 1533 & This work\\
60332.34  &  $+$4     & SOAR/Goodman & 0.487 -- 0.884 & 2700 & This work\\
60333.95  &  $+$6     & SALT/RSS & 0.388 -- 0.770 & 1533 & This work\\
60337.28  &  $+$9     & SOAR/Goodman & 0.395 -- 0.692 & 3600 & This work\\
60344.92  &  $+$17     & SALT/RSS & 0.388 -- 0.770 & 1533 & This work\\
60345.92  &  $+$18     & SALT/RSS & 0.388 -- 0.770 & 1533 & This work\\
60354.44  &  $+$26   &  Keck I/LRIS  & 0.312 -- 1.002 & 1200 & \citet{davis_ucsc_2024} \\
60379.12  &  $+$51   &  Gemini/GMOS  & 0.508 -- 0.976 & 2400$^{b}$ & This work \\
60464.31  &  $+$135  &  Keck I/LRIS  & 0.315 -- 1.012 & 2400 & This work \\
60479.88  &  $+$150  &  JWST/NIRSpec & 0.962 -- 5.100 & 3838$^{c}$ & This work \\
\enddata
\tablecomments{Phases are with respect to second r-band peak (MJD 60327.8). \\$^{b}$ Total exposure over 2 gratings, 1200\,s per grating\\$^{c}$ Total exposure over 3 gratings (1255\,s, 1255\,s, and 1328\,s for gratings G140M, G235M, and G395M, respectively).}
\end{deluxetable}

\begin{deluxetable}{l c c c r}
\tablecaption{Optical and UV photometry of SN\,2024uj \label{tab:photometry}}
\tablehead{
\colhead{MJD} & \colhead{$m$} & \colhead{$\sigma_{\rm mag}$} & \colhead{Filter} & \colhead{Telescope/Instrument} \\
\colhead{} & \colhead{(mag)} & \colhead{(mag)} & \colhead{} & \colhead{}
}
\startdata
60319.722 & 17.229 & 0.024 & U & Sinistro / Las Cumbres \\
60319.726 & 17.256 & 0.025 & U & Sinistro / Las Cumbres \\
60319.730 & 18.172 & 0.015 & B & Sinistro / Las Cumbres \\
60319.733 & 18.164 & 0.016 & B & Sinistro / Las Cumbres \\
60319.735 & 18.090 & 0.016 & V & Sinistro / Las Cumbres \\
60319.737 & 18.046 & 0.016 & V & Sinistro / Las Cumbres \\
60319.739 & 18.012 & 0.018 & g & Sinistro / Las Cumbres \\
60319.742 & 18.006 & 0.018 & g & Sinistro / Las Cumbres \\
60319.745 & 18.236 & 0.030 & r & Sinistro / Las Cumbres \\
... & ... & ... & ... & ... \\
\enddata
\tablecomments{This table is available in its entirety in machine-readable form.}
\end{deluxetable}

\begin{deluxetable}{c c c c c}[h!]
\newcommand{\rowhead}[1]{\multicolumn{1}{l}{\bfseries #1}}
\tablecaption{X-ray Non-detections of SN\,2024uj\label{tab:xray}}
\tablehead{
\colhead{MJD} & \colhead{MJD$_{\rm START}$} & \colhead{MJD$_{\rm STOP}$} & \colhead{Count rate} & \colhead{Flux (0.3-10 keV)} \\ 
    \colhead{} & \colhead{} & \colhead{} & \colhead{} & 
    \colhead{(erg/cm$^2$/s)}
}
\startdata
        \rowhead{Individual epochs} \\
        \midrule
60320.575 & 0.134 & -0.134 & $<$0.011 & $<$4.14E-13 \\
60321.410 & 0.300 & -0.300 & $<$0.007 & $<$2.71E-13 \\
60322.529 & 0.430 & -0.430 & $<$0.005 & $<$2.03E-13 \\
60323.414 & 0.395 & -0.395 & $<$0.011 & $<$4.13E-13 \\
60324.795 & 0.003 & -0.003 & $<$0.016 & $<$6.20E-13 \\
60328.822 & 0.002 & -0.002 & $<$0.023 & $<$9.04E-13 \\
60331.334 & 0.007 & -0.007 & $<$0.007 & $<$2.62E-13 \\
60334.228 & 0.134 & -0.134 & $<$0.006 & $<$2.35E-13 \\
60337.794 & 0.007 & -0.007 & $<$0.010 & $<$3.85E-13 \\
60341.412 & 0.008 & -0.008 & $<$0.006 & $<$2.46E-13 \\
    \addlinespace
    \midrule
    \rowhead{Combining all data} \\
    \midrule
60325.009 & 16.410 & -4.569 & $<$0.001 & $<$3.88E-14  
\enddata
\tablecomments{The modified Julian date reports the mid-exposure time. The phase error indicates when the observation started and ended. (Swift can split up an observation into chunks. All limits are reported at $3\sigma$ confidence. The measurements are corrected for Milky Way absorption and reported for the bandpass from 0.3 to 10~keV.}
\end{deluxetable}

\section{Model setup for He I}\label{sec:appendixHe}
Here, we outline our modeling procedure for the \heIone\ and \heItwo\ profiles in our nebular SN\,2024uj observations. This information, and the software used to perform this modeling, can also be found in the GitHub repository associated with this paper: \hyperlink{https://github.com/saarahhall/SN24uj?tab=readme-ov-file}{github.com/saarahhall/SN24uj}

\begin{enumerate}
    \item Define the data to be fit: 
    \begin{enumerate}
        \item \ion{He}{1} 2.058 $\mu$m: we model the profile from -10000 to 9000 km/s. We extend to -10000 km/s to verify the absence of a blue ``shoulder'' (justify the modified model for \heIone) and to help define the baseline for this feature. 
        \item \ion{He}{1} 1.083 $\mu$m: we include the spectrum between -10000 and 16000 km/s, capturing the ``shoulder'' on the blue end and also the [S I] \sItwo\ line on the red end. The choice of where to cut off this spectrum, and how to represent the baseline, has a non-negligible impact on the amplitudes in resulting fits. This does not affect our interpretations, however, because we only gleam physical information from the means and standard deviations, not the amplitudes.
    \end{enumerate}
    \item Define the model: We model the \heIone\ and \heItwo\ features simultaneously. This requires a complex model with 21 parameters in total. 
    \begin{enumerate}
        \item 12 of the parameters define the three Gaussians that are common to both \ion{He}{1} features: $\{\mu_{1}$,  FWHM$_{1}$, $\mu_2$, FWHM$_{2}$,  $\mu_3$, FWHM$_3$, $A_{1,1.083\mu m}$, $A_{1,2.058\mu m}$, $A_{2,1.083\mu m}$, $A_{2,2.058\mu m}$, $A_{3,1.083\mu m}$, $A_{3,2.058\mu m}\}$. Note: the means (velocities) and FWHM (velocity dispersions) of these three Gaussians are the same in our models for \heIone\ and \heItwo, but the amplitudes for each Gaussian are free to vary, since these values can differ due to differences in the linear baseline.
        \item 4 of the 21 parameters define the linear baselines in each feature: $\{m_{1.083\mu m}$ $b_{1.083\mu m}$, $m_{2.058\mu m}$, $b_{2.058\mu m}\}$. 
        \item 3 model parameters define the Gaussian ``shoulder'' present only in \heIone: $\{\mu_{\text{shoulder}}$, FWHM$_{\text{shoulder}}$, $A_{\text{shoulder}}\}$. 
        \item The final 2 parameters define the Gaussian contributions from [\ion{S}{1}]: $\{$FWHM$_{\text{[S I]}~1.131\mu m}$, $A_{\text{[S I]}~1.131\mu m}\}$. The 1.131$\mu$m [\ion{S}{1}] line is isolated, thus we know the value of $\mu_{\text{[S I]}~1.131\mu m}$. The Gaussian fit to this isolated feature then defines the \sIone\ Gaussian term (which has the same FWHM value, and the amplitude is scaled based on the transition probability ratio of these two lines).
    \end{enumerate}
    \item Define the priors: We set uniform priors on all model parameters, and define their bounds carefully to be physically-informed without introducing bias Priors on all amplitudes are wide, and we ensure positive values with a lower limit of 0. We also set wide, uninformative priors for the slopes and intercepts of the linear baselines in each profile ($m_1$, $b_1$, $m_2$, $b_2$).
    \begin{enumerate}
        \item Velocities $\mu$: The velocities defining the 3 Gaussians shared between \heIone\ and \heItwo, $\mu_1$, $\mu_2$, and $\mu_3$, have uniform priors between $[0,9000]$ km/s. The lower bound is set to avoid doubly-modeling the shoulder feature, and the upper bound is defined by the right edge of the \heItwo\ data we are fitting. The prior on $\mu_{\text{shoulder}}$ is $[-6000,4000]$ km/s. 
        \item Velocity dispersions FWHM: We set a prior of $2\sqrt{2\ln2}\cdot[100, 2000]$ km/s for FWHM$_1$, FWHM$_2$, and FWHM$_3$. The lower bound is set to avoid fitting Gaussians to small, non-physical peaks from the noise. Because the \sItwo\ feature is isolated, we set a narrower FWHM$_{\text{[S I]}~1.131}$: $2\sqrt{2\ln2}\cdot[200,600]$ km/s. The shoulder feature is inherently broad, so we set a large upper limit for FWHM$_{\text{shoulder}}$: $2\sqrt{2\ln2}\cdot[200, 6000]$ km/s.
    \end{enumerate}
    \end{enumerate}

Finally, we run nested sampling with \texttt{dynesty}. The resulting run has 36724 steps. After accounting for burn-in, we have 10000 samples, from which we grab the maximum a posteriori (MAP) values of each model parameter to define the fits shown in \autoref{fig:profile-vel-24uj}.


\begin{thebibliography}{}
\expandafter\ifx\csname natexlab\endcsname\relax\def\natexlab#1{#1}\fi
\providecommand{\url}[1]{\href{#1}{#1}}
\providecommand{\dodoi}[1]{doi:~\href{http://doi.org/#1}{\nolinkurl{#1}}}
\providecommand{\doeprint}[1]{\href{http://ascl.net/#1}{\nolinkurl{http://ascl.net/#1}}}
\providecommand{\doarXiv}[1]{\href{https://arxiv.org/abs/#1}{\nolinkurl{https://arxiv.org/abs/#1}}}

\bibitem[{Aamer {et~al.}(2024)Aamer, Nicholl, Jerkstrand, Gomez, Oates, Smartt, Srivastav, Leloudas, Anderson, Berger, de~Boer, Chambers, Chen, Galbany, Gao, Gompertz, González-Bañuelos, Gromadzki, Gutiérrez, Inserra, Lowe, Magnier, Mazzali, Moore, Müller-Bravo, Pursiainen, Rest, Schulze, Smith, Terwel, Wainscoat, \& Young}]{aamer_precursor_2024}
Aamer, A., Nicholl, M., Jerkstrand, A., {et~al.} 2024, Monthly Notices of the Royal Astronomical Society, 527, 11970, \dodoi{10.1093/mnras/stad3776}

\bibitem[{{Arnett}(1982)}]{Arnett82}
{Arnett}, W.~D. 1982, \apj, 253, 785, \dodoi{10.1086/159681}

\bibitem[{Ashall {et~al.}(2024)Ashall, Hoeflich, Baron, Shahbandeh, DerKacy, Medler, Shappee, Tucker, Fereidouni, Mera, Andrews, Baade, Bostroem, Brown, Burns, Burrow, Cikota, de~Jaeger, Do, Dong, Dominguez, Fox, Galbany, Hsiao, Krisciunas, Khaghani, Kumar, Lu, Maund, Mazzali, Morrell, Patat, Pfeffer, Phillips, Schmidt, Stangl, Stevens, Stritzinger, Suntzeff, Telesco, Wang, \& Yang}]{ashall_jwst_2024}
Ashall, C., Hoeflich, P., Baron, E., {et~al.} 2024, The Astrophysical Journal, 975, 203, \dodoi{10.3847/1538-4357/ad6608}

\bibitem[{Ashok \& Banerjee(2003)}]{ashok_enigmatic_2003}
Ashok, N.~M., \& Banerjee, D. P.~K. 2003, Astronomy \& Astrophysics, 409, 1007.
\newblock \url{https://www.aanda.org/articles/aa/abs/2003/39/aah4263/aah4263.html}

\bibitem[{{Astropy Collaboration} {et~al.}(2013){Astropy Collaboration}, Robitaille, Tollerud, Greenfield, Droettboom, Bray, Aldcroft, Davis, Ginsburg, Price-Whelan, Kerzendorf, Conley, Crighton, Barbary, Muna, Ferguson, Grollier, Parikh, Nair, Unther, Deil, Woillez, Conseil, Kramer, Turner, Singer, Fox, Weaver, Zabalza, Edwards, Azalee~Bostroem, Burke, Casey, Crawford, Dencheva, Ely, Jenness, Labrie, Lim, Pierfederici, Pontzen, Ptak, Refsdal, Servillat, \& Streicher}]{astropy_collaboration_astropy_2013}
{Astropy Collaboration}, Robitaille, T.~P., Tollerud, E.~J., {et~al.} 2013, A\&A, 558, A33, \dodoi{10.1051/0004-6361/201322068}

\bibitem[{{Astropy Collaboration} {et~al.}(2018){Astropy Collaboration}, Price-Whelan, Sip{\textbackslash}Hocz, Günther, Lim, Crawford, Conseil, Shupe, Craig, Dencheva, Ginsburg, VanderPlas, Bradley, Pérez-Suárez, de~Val-Borro, Aldcroft, Cruz, Robitaille, Tollerud, Ardelean, Babej, Bach, Bachetti, Bakanov, Bamford, Barentsen, Barmby, Baumbach, Berry, Biscani, Boquien, Bostroem, Bouma, Brammer, Bray, Breytenbach, Buddelmeijer, Burke, Calderone, Cano~Rodríguez, Cara, Cardoso, Cheedella, Copin, Corrales, Crichton, D'Avella, Deil, Depagne, Dietrich, Donath, Droettboom, Earl, Erben, Fabbro, Ferreira, Finethy, Fox, Garrison, Gibbons, Goldstein, Gommers, Greco, Greenfield, Groener, Grollier, Hagen, Hirst, Homeier, Horton, Hosseinzadeh, Hu, Hunkeler, Ivezić, Jain, Jenness, Kanarek, Kendrew, Kern, Kerzendorf, Khvalko, King, Kirkby, Kulkarni, Kumar, Lee, Lenz, Littlefair, Ma, Macleod, Mastropietro, McCully, Montagnac, Morris, Mueller, Mumford, Muna, Murphy, Nelson, Nguyen, Ninan, Nöthe, Ogaz, Oh, Parejko, Parley,
  Pascual, Patil, Patil, Plunkett, Prochaska, Rastogi, Reddy~Janga, Sabater, Sakurikar, Seifert, Sherbert, Sherwood-Taylor, Shih, Sick, Silbiger, Singanamalla, Singer, Sladen, Sooley, Sornarajah, Streicher, Teuben, Thomas, Tremblay, Turner, Terrón, van Kerkwijk, de~la Vega, Watkins, Weaver, Whitmore, Woillez, Zabalza, \& {Astropy Contributors}}]{astropy_collaboration_astropy_2018}
{Astropy Collaboration}, Price-Whelan, A.~M., Sip{\textbackslash}Hocz, B.~M., {et~al.} 2018, {\textbackslash}aj, 156, 123, \dodoi{10.3847/1538-3881/aabc4f}

\bibitem[{Baer-Way {et~al.}(2025)Baer-Way, Chandra, Modjaz, Kumar, Pellegrino, Chevalier, Crawford, Sarangi, Smith, Maeda, Nayana, Filippenko, Andrews, Arcavi, Bostroem, Brink, Dong, Dwarkadas, Farah, Howell, Hiramatsu, Hosseinzadeh, McCully, Meza, Newsome, Padilla~Gonzalez, Pearson, Sand, Shrestha, Terreran, Valenti, Wyatt, Yang, \& Zheng}]{baer-way_multiwavelength_2025}
Baer-Way, R., Chandra, P., Modjaz, M., {et~al.} 2025, The Astrophysical Journal, 983, 101, \dodoi{10.3847/1538-4357/adc00a}

\bibitem[{Bildsten {et~al.}(2007)Bildsten, Shen, Weinberg, \& Nelemans}]{bildsten_faint_2007}
Bildsten, L., Shen, K.~J., Weinberg, N.~N., \& Nelemans, G. 2007, The Astrophysical Journal, 662, L95, \dodoi{10.1086/519489}

\bibitem[{Birkmann {et~al.}(2022)Birkmann, Giardino, Sirianni, Ferruit, Rauscher, Alves~de Oliveira, Böker, Kumari, Lützgendorf, Manjavacas, Proffitt, Rawle, te~Plate, \& Zeidler}]{birkmann_-flight_2022}
Birkmann, S.~M., Giardino, G., Sirianni, M., {et~al.} 2022, in Proceedings of {SPIE}, {Volume} 12180, Vol. 12180, eprint: arXiv:2208.12686, 121802P, \dodoi{10.1117/12.2629545}

\bibitem[{Blondin {et~al.}(2023)Blondin, Dessart, Hillier, Ramsbottom, \& Storey}]{blondin_nebular_2023}
Blondin, S., Dessart, L., Hillier, D.~J., Ramsbottom, C.~A., \& Storey, P.~J. 2023, Astronomy \& Astrophysics, 678, A170, \dodoi{10.1051/0004-6361/202347147}

\bibitem[{Boyle {et~al.}(2017)Boyle, Sim, Hachinger, \& Kerzendorf}]{boyle_helium_2017}
Boyle, A., Sim, S.~A., Hachinger, S., \& Kerzendorf, W. 2017, Astronomy \& Astrophysics, 599, A46.
\newblock \url{https://www.aanda.org/articles/aa/full_html/2017/03/aa29712-16/aa29712-16.html}

\bibitem[{{Breeveld} {et~al.}(2010){Breeveld}, {Curran}, {Hoversten}, {Koch}, {Landsman}, {Marshall}, {Page}, {Poole}, {Roming}, {Smith}, \& et~al.}]{Breeveld10}
{Breeveld}, A.~A., {Curran}, P.~A., {Hoversten}, E.~A., {et~al.} 2010, \mnras, 406, 1687, \dodoi{10.1111/j.1365-2966.2010.16832.x}

\bibitem[{Brown {et~al.}(2013)Brown, Baliber, Bianco, Bowman, Burleson, Conway, Crellin, Depagne, De~Vera, Dilday, Dragomir, Dubberley, Eastman, Elphick, Falarski, Foale, Ford, Fulton, Garza, Gomez, Graham, Greene, Haldeman, Hawkins, Haworth, Haynes, Hidas, Hjelstrom, Howell, Hygelund, Lister, Lobdill, Martinez, Mullins, Norbury, Parrent, Paulson, Petry, Pickles, Posner, Rosing, Ross, Sand, Saunders, Shobbrook, Shporer, Street, Thomas, Tsapras, Tufts, Valenti, Vander~Horst, Walker, White, \& Willis}]{brown_cumbres_2013}
Brown, T.~M., Baliber, N., Bianco, F.~B., {et~al.} 2013, Publications of the Astronomical Society of the Pacific, 125, 1031, \dodoi{10.1086/673168}

\bibitem[{Buckley {et~al.}(2006)Buckley, Swart, \& Meiring}]{buckley_completion_2006}
Buckley, D. A.~H., Swart, G.~P., \& Meiring, J.~G. 2006, in Proceedings of {SPIE}, {Volume} 6267, Vol. 6267, 62670Z, \dodoi{10.1117/12.673750}

\bibitem[{Burgh {et~al.}(2003)Burgh, Nordsieck, Kobulnicky, Williams, O'Donoghue, Smith, \& Percival}]{burgh_prime_2003}
Burgh, E.~B., Nordsieck, K.~H., Kobulnicky, H.~A., {et~al.} 2003, in Proceedings of {SPIE}, {Volume} 4841, Vol. 4841, 1463--1471, \dodoi{10.1117/12.460312}

\bibitem[{Burrows {et~al.}(2005)Burrows, Hill, Nousek, Kennea, Wells, Osborne, Abbey, Beardmore, Mukerjee, Short, Chincarini, Campana, Citterio, Moretti, Pagani, Tagliaferri, Giommi, Capalbi, Tamburelli, Angelini, Cusumano, Bräuninger, Burkert, \& Hartner}]{burrows_swift_2005}
Burrows, D.~N., Hill, J.~E., Nousek, J.~A., {et~al.} 2005, Space Science Reviews, 120, 165, \dodoi{10.1007/s11214-005-5097-2}

\bibitem[{Bushouse {et~al.}(2022)Bushouse, Eisenhamer, Dencheva, Davies, Greenfield, Morrison, Hodge, Simon, Grumm, Droettboom, Slavich, Sosey, Pauly, Miller, Jedrzejewski, Hack, Davis, Crawford, Law, Gordon, Regan, Cara, MacDonald, Bradley, Shanahan, Jamieson, Teodoro, \& Williams}]{bushouse_jwst_2022}
Bushouse, H., Eisenhamer, J., Dencheva, N., {et~al.} 2022, {JWST} {Calibration} {Pipeline},  Zenodo, \dodoi{10.5281/zenodo.7041998}

\bibitem[{Buzzoni {et~al.}(1984)Buzzoni, Delabre, Dekker, Dodorico, Enard, Focardi, Gustafsson, Nees, Paureau, \& Reiss}]{buzzoni_eso_1984}
Buzzoni, B., Delabre, B., Dekker, H., {et~al.} 1984, The Messenger, 38, 9.
\newblock \url{https://ui.adsabs.harvard.edu/abs/1984Msngr..38....9B}

\bibitem[{Böker {et~al.}(2023)Böker, Beck, Birkmann, Giardino, Keyes, Kumari, Muzerolle, Rawle, Zeidler, Abul-Huda, de~Oliveira, Arribas, Bechtold, Bhatawdekar, Bonaventura, Bunker, Cameron, Carniani, Charlot, Curti, Espinoza, Ferruit, Franx, Jakobsen, Karakla, López-Caniego, Lützgendorf, Maiolino, Manjavacas, Marston, Moseley, Ogle, Perna, Peña-Guerrero, Pirzkal, Plesha, Proffitt, Rauscher, Rix, del Pino, Rustamkulov, Sabbi, Sing, Sirianni, Plate, Úbeda, Wahlgren, Wislowski, Wu, \& Willott}]{boker_-orbit_2023}
Böker, T., Beck, T.~L., Birkmann, S.~M., {et~al.} 2023, Publications of the Astronomical Society of the Pacific, 135, 038001, \dodoi{10.1088/1538-3873/acb846}

\bibitem[{Callan {et~al.}(2025{\natexlab{a}})Callan, Collins, Sim, Shingles, Pakmor, Srivastav, Pollin, Gronow, Röpke, \& Seitenzahl}]{callan_exploring_2025}
Callan, F.~P., Collins, C.~E., Sim, S.~A., {et~al.} 2025{\natexlab{a}}, Monthly Notices of the Royal Astronomical Society, 539, 1404, \dodoi{10.1093/mnras/staf539}

\bibitem[{Callan {et~al.}(2025{\natexlab{b}})Callan, Holas, Morán-Fraile, Sim, Collins, Shingles, Pollin, Röpke, Pakmor, \& Schneider}]{callan_nlte_2025}
Callan, F.~P., Holas, A., Morán-Fraile, J., {et~al.} 2025{\natexlab{b}}, {NLTE} spectral modelling for a carbon-oxygen and helium white-dwarf merger as a {Ca}-rich transient candidate,  arXiv, \dodoi{10.48550/arXiv.2503.12105}

\bibitem[{Cao {et~al.}(2015)Cao, Kulkarni, Cook, \& Vreeswijk}]{cao_iptf_2015}
Cao, Y., Kulkarni, S.~R., Cook, D., \& Vreeswijk, P. 2015, The Astronomer's Telegram, 8428, 1.
\newblock \url{https://ui.adsabs.harvard.edu/abs/2015ATel.8428....1C}

\bibitem[{Chandra {et~al.}(2020)Chandra, Chevalier, Chugai, Milisavljevic, \& Fransson}]{chandra_supernova_2020}
Chandra, P., Chevalier, R.~A., Chugai, N., Milisavljevic, D., \& Fransson, C. 2020, The Astrophysical Journal, 902, 55, \dodoi{10.3847/1538-4357/abb460}

\bibitem[{Chen {et~al.}(2025)Chen, Sun, Xi, Tinyanont, Aguado, Pérez-Fournon, Poidevin, Maund, Kumar, Jin, Mao, Wang, Zhang, Guo, Li, Rojas-Bravo, Shen, Wang, Wang, Zhao, Zheng, Zhu, Fernández-Nespral, López-Oramas, Niu, Wang, Wiersema, \& Liu}]{chen_sn_2025}
Chen, C., Sun, N.-C., Xi, Q., {et~al.} 2025, {SN} 2025coe: {A} {Triple}-{Peaked} {Calcium}-{Strong} {Transient} from {A} {White}-{Dwarf} {Progenitor},  arXiv, \dodoi{10.48550/arXiv.2510.00135}

\bibitem[{Clemens {et~al.}(2004)Clemens, Crain, \& Anderson}]{clemens_goodman_2004}
Clemens, J.~C., Crain, J.~A., \& Anderson, R. 2004, in Proceedings of {SPIE}, {Volume} 5492, Vol. 5492, 331--340, \dodoi{10.1117/12.550069}

\bibitem[{Crawford {et~al.}(2025)Crawford, Pritchard, Modjaz, Pellegrino, Kumar, \& Baer-Way}]{crawford_peaky_2025}
Crawford, A., Pritchard, T.~A., Modjaz, M., {et~al.} 2025, The Astrophysical Journal, 989, 192, \dodoi{10.3847/1538-4357/adea3a}

\bibitem[{Davis {et~al.}(2024)Davis, Arunachalam, Foley, \& Kwok}]{davis_ucsc_2024}
Davis, K., Arunachalam, P., Foley, R., \& Kwok, L. 2024, Transient Name Server Classification Report, 2024-521, 1.
\newblock \url{https://ui.adsabs.harvard.edu/abs/2024TNSCR.521....1D}

\bibitem[{De {et~al.}(2021)De, Fremling, Gal-Yam, Yaron, Kasliwal, \& Kulkarni}]{de_peculiar_2021}
De, K., Fremling, U.~C., Gal-Yam, A., {et~al.} 2021, The Astrophysical Journal, 907, L18, \dodoi{10.3847/2041-8213/abd627}

\bibitem[{De {et~al.}(2018)De, Kasliwal, Cantwell, Cao, Cenko, Gal-Yam, Johansson, Kong, Kulkarni, Lunnan, Masci, Matuszewski, Mooley, Neill, Nugent, Ofek, Perrott, Rebbapragada, Rubin, Sullivan, \& Yaron}]{de_iptf_2018}
De, K., Kasliwal, M.~M., Cantwell, T., {et~al.} 2018, The Astrophysical Journal, 866, 72, \dodoi{10.3847/1538-4357/aadf8e}

\bibitem[{De {et~al.}(2020)De, Kasliwal, Tzanidakis, Fremling, Adams, Aloisi, Andreoni, Bagdasaryan, Bellm, Bildsten, Cannella, Cook, Delacroix, Drake, Duev, Dugas, Frederick, Gal-Yam, Goldstein, Golkhou, Graham, Hale, Hankins, Helou, Ho, Irani, Jencson, Kaplan, Kaye, Kulkarni, Kupfer, Laher, Leadbeater, Lunnan, Masci, Miller, Neill, Ofek, Perley, Polin, Prince, Quataert, Reiley, Riddle, Rusholme, Sharma, Shupe, Sollerman, Tartaglia, Walters, Yan, \& Yao}]{de_zwicky_2020}
De, K., Kasliwal, M.~M., Tzanidakis, A., {et~al.} 2020, The Astrophysical Journal, 905, 58, \dodoi{10.3847/1538-4357/abb45c}

\bibitem[{DerKacy {et~al.}(2023)DerKacy, Ashall, Hoeflich, Baron, Shappee, Baade, Andrews, Bostroem, Brown, Burns, Burrow, Cikota, de~Jaeger, Do, Dong, Dominguez, Galbany, Hsiao, Karamehmetoglu, Krisciunas, Kumar, Lu, Evans, Maund, Mazzali, Medler, Morrell, Patat, Phillips, Shahbandeh, Stangl, Stevens, Stritzinger, Suntzeff, Telesco, Tucker, Valenti, Wang, Yang, Jha, \& Kwok}]{derkacy_jwst_2023}
DerKacy, J.~M., Ashall, C., Hoeflich, P., {et~al.} 2023, The Astrophysical Journal Letters, 945, L2, \dodoi{10.3847/2041-8213/acb8a8}

\bibitem[{DerKacy {et~al.}(2024)DerKacy, Ashall, Hoeflich, Baron, Shahbandeh, Shappee, Andrews, Baade, Balangan, Bostroem, Brown, Burns, Burrow, Cikota, de~Jaeger, Do, Dong, Dominguez, Fox, Galbany, Hoang, Hsiao, Janzen, Jencson, Krisciunas, Kumar, Lu, Lundquist, Mera~Evans, Maund, Mazzali, Medler, Meza~Retamal, Morrell, Patat, Pearson, Phillips, Shrestha, Stangl, Stevens, Stritzinger, Suntzeff, Telesco, Tucker, Valenti, Wang, \& Yang}]{derkacy_jwst_2024}
---. 2024, The Astrophysical Journal, 961, 187, \dodoi{10.3847/1538-4357/ad0b7b}

\bibitem[{DerKacy {et~al.}(2026)DerKacy, Ashall, Baron, Medler, Mera, Hoeflich, Shahbandeh, Burns, Stritzinger, Tucker, Shappee, Auchettl, Angus, Desai, Do, Hinkle, Hoogendam, Huber, Payne, Jones, Shi, Kong, Romagnoli, Syncatto, Moran, Fereidouni, Brown, Engesser, Fox, Galbany, Hsiao, de~Jaeger, Kumar, Lu, Matsuura, Mazzali, Morrell, Pfeffer, Phillips, Rest, Shiber, Strolger, Suntzeff, Temim, Tinyanont, Wang, Wesson, Park, \& Rho}]{derkacy_jwst_2026}
DerKacy, J.~M., Ashall, C., Baron, E., {et~al.} 2026, The Astrophysical Journal, 997, 179, \dodoi{10.3847/1538-4357/ae1f87}

\bibitem[{Dessart \& Hillier(2015)}]{dessart_one-dimensional_2015}
Dessart, L., \& Hillier, D.~J. 2015, Monthly Notices of the Royal Astronomical Society, 447, 1370, \dodoi{10.1093/mnras/stu2520}

\bibitem[{Dessart {et~al.}(2021)Dessart, Hillier, Sukhbold, Woosley, \& Janka}]{dessart_nebular_2021}
Dessart, L., Hillier, D.~J., Sukhbold, T., Woosley, S.~E., \& Janka, H.-T. 2021, Astronomy and Astrophysics, 656, A61, \dodoi{10.1051/0004-6361/202141927}

\bibitem[{Dessart {et~al.}(2015)Dessart, Hillier, Woosley, Livne, Waldman, Yoon, \& Langer}]{dessart_radiative-transfer_2015}
Dessart, L., Hillier, D.~J., Woosley, S., {et~al.} 2015.
\newblock \url{https://ui.adsabs.harvard.edu/abs/2015MNRAS.453.2189D/abstract}

\bibitem[{Dessart {et~al.}(2025)Dessart, Kotak, Jacobson-Galan, Das, Fremling, Kasliwal, Qin, \& Rose}]{dessart_optical_2025}
Dessart, L., Kotak, R., Jacobson-Galan, W., {et~al.} 2025, An optical to infrared study of type {II} {SN2024ggi} at nebular times,  arXiv, \dodoi{10.48550/arXiv.2507.05803}

\bibitem[{Dessart {et~al.}(2020)Dessart, Yoon, Aguilera-Dena, \& Langer}]{dessart_supernovae_2020}
Dessart, L., Yoon, S.-C., Aguilera-Dena, D.~R., \& Langer, N. 2020, Astronomy and Astrophysics, 642, A106, \dodoi{10.1051/0004-6361/202038763}

\bibitem[{Dimitriadis {et~al.}(2023)Dimitriadis, Maguire, Karambelkar, Lebron, Liu (刘~畅), Kozyreva, Miller, Ridden-Harper, Anderson, Chen, Coughlin, Valle, Drake, Galbany, Gromadzki, Groom, Gutiérrez, Ihanec, Inserra, Johansson, Müller-Bravo, Nicholl, Polin, Rusholme, Schulze, Sollerman, Srivastav, Taggart, Wang, Yang (杨~轶), \& Young}]{dimitriadis_sn_2023}
Dimitriadis, G., Maguire, K., Karambelkar, V.~R., {et~al.} 2023, Monthly Notices of the Royal Astronomical Society, 521, 1162, \dodoi{10.1093/mnras/stad536}

\bibitem[{Dong {et~al.}(2022)Dong, Milisavljevic, Leja, Sarbadhicary, Nugent, Margutti, Jacobson-Galán, Polin, Banovetz, Reynolds, \& Subrayan}]{dong_physical_2022}
Dong, Y., Milisavljevic, D., Leja, J., {et~al.} 2022, The Astrophysical Journal, 927, 199, \dodoi{10.3847/1538-4357/ac5257}

\bibitem[{Ertini {et~al.}(2023)Ertini, Folatelli, Martinez, Bersten, Anderson, Ashall, Baron, Bose, Brown, Burns, DerKacy, Ferrari, Galbany, Hsiao, Kumar, Lu, Mazzali, Morrell, Orellana, Pessi, Phillips, Piro, Polin, Shahbandeh, Shappee, Stritzinger, Suntzeff, Tucker, Elias-Rosa, Kuncarayakti, Gutiérrez, Kozyreva, Müller-Bravo, Chen, Hinkle, Payne, Székely, Szalai, Barna, Könyves-Tóth, Bánhidi, Bíró, Csányi, Kriskovits, Pál, Szabó, Szakáts, Vida, Vinkó, Gromadzki, Harvey, Nicholl, Paraskeva, Young, \& Englert}]{ertini_sn_2023}
Ertini, K., Folatelli, G., Martinez, L., {et~al.} 2023, Monthly Notices of the Royal Astronomical Society, 526, 279, \dodoi{10.1093/mnras/stad2705}

\bibitem[{Evans {et~al.}(2007)Evans, Beardmore, Page, Tyler, Osborne, Goad, O'Brien, Vetere, Racusin, Morris, Burrows, Capalbi, Perri, Gehrels, \& Romano}]{evans_online_2007}
Evans, P.~A., Beardmore, A.~P., Page, K.~L., {et~al.} 2007, Astronomy and Astrophysics, 469, 379, \dodoi{10.1051/0004-6361:20077530}

\bibitem[{Evans {et~al.}(2009)Evans, Beardmore, Page, Osborne, O'Brien, Willingale, Starling, Burrows, Godet, Vetere, Racusin, Goad, Wiersema, Angelini, Capalbi, Chincarini, Gehrels, Kennea, Margutti, Morris, Mountford, Pagani, Perri, Romano, \& Tanvir}]{evans_methods_2009}
---. 2009, Monthly Notices of the Royal Astronomical Society, 397, 1177, \dodoi{10.1111/j.1365-2966.2009.14913.x}

\bibitem[{Filippenko(1982)}]{filippenko_importance_1982}
Filippenko, A.~V. 1982, Publications of the Astronomical Society of the Pacific, 94, 715, \dodoi{10.1086/131052}

\bibitem[{Filippenko {et~al.}(2003)Filippenko, Chornock, Swift, Modjaz, Simcoe, \& Rauch}]{filippenko_supernovae_2003}
Filippenko, A.~V., Chornock, R., Swift, B., {et~al.} 2003, International Astronomical Union Circular, 8159, 2.
\newblock \url{https://ui.adsabs.harvard.edu/abs/2003IAUC.8159....2F}

\bibitem[{Foley(2015)}]{foley_kinematics_2015}
Foley, R.~J. 2015, Monthly Notices of the Royal Astronomical Society, 452, 2463, \dodoi{10.1093/mnras/stv789}

\bibitem[{Foreman-Mackey {et~al.}(2013)Foreman-Mackey, Hogg, Lang, \& Goodman}]{foreman-mackey_emcee_2013}
Foreman-Mackey, D., Hogg, D.~W., Lang, D., \& Goodman, J. 2013, {\textbackslash}pasp, 125, 306, \dodoi{10.1086/670067}

\bibitem[{Fryxell {et~al.}(2000)Fryxell, Olson, Ricker, Timmes, Zingale, Lamb, MacNeice, Rosner, Truran, \& Tufo}]{fryxell_flash_2000}
Fryxell, B., Olson, K., Ricker, P., {et~al.} 2000, \dodoi{10.1086/317361}

\bibitem[{Gal-Yam {et~al.}(2002)Gal-Yam, Ofek, \& Shemmer}]{gal-yam_supernova_2002}
Gal-Yam, A., Ofek, E.~O., \& Shemmer, O. 2002, Monthly Notices of the Royal Astronomical Society, 332, L73, \dodoi{10.1046/j.1365-8711.2002.05535.x}

\bibitem[{Galbany {et~al.}(2019)Galbany, Ashall, Höflich, González-Gaitán, Taubenberger, Stritzinger, Hsiao, Mazzali, Baron, Blondin, Bose, Bulla, Burke, Burns, Cartier, Chen, Della~Valle, Diamond, Gutiérrez, Harmanen, Hiramatsu, Holoien, Hosseinzadeh, Howell, Huang, Inserra, de~Jaeger, Jha, Kangas, Kromer, Lyman, Maguire, Marion, Milisavljevic, Prentice, Razza, Reynolds, Sand, Shappee, Shekhar, Smartt, Stassun, Sullivan, Valenti, Villanueva, Wang, Wheeler, Zhai, \& Zhang}]{galbany_evidence_2019}
Galbany, L., Ashall, C., Höflich, P., {et~al.} 2019, Astronomy and Astrophysics, 630, A76, \dodoi{10.1051/0004-6361/201935537}

\bibitem[{Gehrels {et~al.}(2004)Gehrels, Chincarini, Giommi, Mason, Nousek, Wells, White, Barthelmy, Burrows, Cominsky, Hurley, Marshall, Mészáros, Roming, Angelini, Barbier, Belloni, Campana, Caraveo, Chester, Citterio, Cline, Cropper, Cummings, Dean, Feigelson, Fenimore, Frail, Fruchter, Garmire, Gendreau, Ghisellini, Greiner, Hill, Hunsberger, Krimm, Kulkarni, Kumar, Lebrun, Lloyd-Ronning, Markwardt, Mattson, Mushotzky, Norris, Osborne, Paczynski, Palmer, Park, Parsons, Paul, Rees, Reynolds, Rhoads, Sasseen, Schaefer, Short, Smale, Smith, Stella, Tagliaferri, Takahashi, Tashiro, Townsley, Tueller, Turner, Vietri, Voges, Ward, Willingale, Zerbi, \& Zhang}]{gehrels_swift_2004}
Gehrels, N., Chincarini, G., Giommi, P., {et~al.} 2004, \dodoi{10.1086/422091}

\bibitem[{Harris {et~al.}(2020)Harris, Millman, van~der Walt, Gommers, Virtanen, Cournapeau, Wieser, Taylor, Berg, Smith, Kern, Picus, Hoyer, van Kerkwijk, Brett, Haldane, del Río, Wiebe, Peterson, Gérard-Marchant, Sheppard, Reddy, Weckesser, Abbasi, Gohlke, \& Oliphant}]{harris_array_2020}
Harris, C.~R., Millman, K.~J., van~der Walt, S.~J., {et~al.} 2020, {\textbackslash}nat, 585, 357, \dodoi{10.1038/s41586-020-2649-2}

\bibitem[{{HEASARC}(2014)}]{heasarc_heasoft_2014}
{HEASARC}. 2014, {HEAsoft}: {Unified} {Release} of {FTOOLS} and {XANADU}

\bibitem[{Henden {et~al.}(2016)Henden, Templeton, Terrell, Smith, Levine, \& Welch}]{Henden16}
Henden, A.~A., Templeton, M., Terrell, D., {et~al.} 2016, VizieR Online Data Catalog: AAVSO Photometric All Sky Survey (APASS) DR9 (Henden+, 2016), VizieR On-line Data Catalog: II/336. Originally published in: 2015AAS...22533616H.
\newblock \url{https://ui.adsabs.harvard.edu/abs/2016yCat.2336....0H}

\bibitem[{{HI4PI Collaboration} {et~al.}(2016){HI4PI Collaboration}, Ben~Bekhti, Flöer, Keller, Kerp, Lenz, Winkel, Bailin, Calabretta, Dedes, Ford, Gibson, Haud, Janowiecki, Kalberla, Lockman, McClure-Griffiths, Murphy, Nakanishi, Pisano, \& Staveley-Smith}]{hi4pi_collaboration_hi4pi_2016}
{HI4PI Collaboration}, Ben~Bekhti, N., Flöer, L., {et~al.} 2016, Astronomy and Astrophysics, 594, A116, \dodoi{10.1051/0004-6361/201629178}

\bibitem[{Hillier \& Dessart(2012)}]{hillier_time-dependent_2012}
Hillier, D.~J., \& Dessart, L. 2012, Monthly Notices of the Royal Astronomical Society, 424, 252, \dodoi{10.1111/j.1365-2966.2012.21192.x}

\bibitem[{Hook {et~al.}(2004)Hook, Jørgensen, Allington-Smith, Davies, Metcalfe, Murowinski, \& Crampton}]{hook_gemini-north_2004}
Hook, I.~M., Jørgensen, I., Allington-Smith, J.~R., {et~al.} 2004, Publications of the Astronomical Society of the Pacific, 116, 425, \dodoi{10.1086/383624}

\bibitem[{Hosseinzadeh {et~al.}(2023)Hosseinzadeh, Bostroem, \& Gomez}]{Hosseinzadeh23}
Hosseinzadeh, G., Bostroem, K.~A., \& Gomez, S. 2023, Light Curve Fitting, v0.8.0,  Zenodo, \dodoi{10.5281/zenodo.7872772}

\bibitem[{{Howell}(2019)}]{Howell19}
{Howell}, D. 2019, in American Astronomical Society Meeting Abstracts, Vol. 233, American Astronomical Society Meeting Abstracts \#233, 258.16

\bibitem[{Hunter(2007)}]{hunter_matplotlib_2007}
Hunter, J.~D. 2007, Computing in Science and Engineering, 9, 90, \dodoi{10.1109/MCSE.2007.55}

\bibitem[{Inserra {et~al.}(2017)Inserra, Nicholl, Chen, Jerkstrand, Smartt, Krühler, Anderson, Baltay, Della~Valle, Fraser, Gal-Yam, Galbany, Kankare, Maguire, Rabinowitz, Smith, Valenti, \& Young}]{inserra_complexity_2017}
Inserra, C., Nicholl, M., Chen, T.-W., {et~al.} 2017, Monthly Notices of the Royal Astronomical Society, 468, 4642, \dodoi{10.1093/mnras/stx834}

\bibitem[{Jacobson-Galán(2025)}]{jacobson-galan_yse_2025}
Jacobson-Galán, W. 2025, Transient Name Server AstroNote, 21, 1.
\newblock \url{https://ui.adsabs.harvard.edu/abs/2025TNSAN..21....1J}

\bibitem[{Jacobson-Galán {et~al.}(2026)Jacobson-Galán, Dessart, \& Vartanyan}]{jacobson-galan_mapping_2026}
Jacobson-Galán, W.~V., Dessart, L., \& Vartanyan, D. 2026, Mapping 3-{D} {Explosive} {Nucleosynthesis} with {Type} {II} {Supernova} {Infrared} {Emission} {Lines},  arXiv, \dodoi{10.48550/arXiv.2605.18951}

\bibitem[{Jacobson-Galán {et~al.}(2019)Jacobson-Galán, Foley, Schwab, Dimitriadis, Dong, Jha, Kasen, Kilpatrick, \& Thomas}]{jacobson-galan_detection_2019}
Jacobson-Galán, W.~V., Foley, R.~J., Schwab, J., {et~al.} 2019, Monthly Notices of the Royal Astronomical Society, 487, 2538, \dodoi{10.1093/mnras/stz1305}

\bibitem[{Jacobson-Galán {et~al.}(2020{\natexlab{a}})Jacobson-Galán, Margutti, Kilpatrick, Hiramatsu, Perets, Khatami, Foley, Raymond, Yoon, Bobrick, Zenati, Galbany, Andrews, Brown, Cartier, Coppejans, Dimitriadis, Dobson, Hajela, Howell, Kuncarayakti, Milisavljevic, Rahman, Rojas-Bravo, Sand, Shepherd, Smartt, Stacey, Stroh, Swift, Terreran, Vinko, Wang, Anderson, Baron, Berger, Blanchard, Burke, Coulter, DeMarchi, DerKacy, Fremling, Gomez, Gromadzki, Hosseinzadeh, Kasen, Kriskovics, McCully, Müller-Bravo, Nicholl, Ordasi, Pellegrino, Piro, Pál, Ren, Rest, Rich, Sai, Sárneczky, Shen, Short, Siebert, Stauffer, Szakáts, Zhang, Zhang, \& Zhang}]{jacobson-galan_sn_2020}
Jacobson-Galán, W.~V., Margutti, R., Kilpatrick, C.~D., {et~al.} 2020{\natexlab{a}}, The Astrophysical Journal, 898, 166, \dodoi{10.3847/1538-4357/ab9e66}

\bibitem[{Jacobson-Galán {et~al.}(2020{\natexlab{b}})Jacobson-Galán, Polin, Foley, Dimitriadis, Kilpatrick, Margutti, Coulter, Jha, Jones, Kirshner, Pan, Piro, Rest, \& Rojas-Bravo}]{jacobson-galan_ca_2020}
Jacobson-Galán, W.~V., Polin, A., Foley, R.~J., {et~al.} 2020{\natexlab{b}}, The Astrophysical Journal, 896, 165, \dodoi{10.3847/1538-4357/ab94b8}

\bibitem[{Jacobson-Galán {et~al.}(2021)Jacobson-Galán, Margutti, Kilpatrick, Raymond, Berger, Blanchard, Bobrick, Foley, Gomez, Hosseinzadeh, Milisavljevic, Perets, Terreran, \& Zenati}]{jacobson-galan_late-time_2021}
Jacobson-Galán, W.~V., Margutti, R., Kilpatrick, C.~D., {et~al.} 2021, The Astrophysical Journal, 908, L32, \dodoi{10.3847/2041-8213/abdebc}

\bibitem[{Jacobson-Galán {et~al.}(2022)Jacobson-Galán, Venkatraman, Margutti, Khatami, Terreran, Foley, Angulo, Angus, Auchettl, Blanchard, Bobrick, Bright, Brout, Chambers, Couch, Coulter, Clever, Davis, de~Boer, DeMarchi, Dodd, Jones, Johnson, Kilpatrick, Khetan, Lai, Langeroodi, Lin, Magnier, Milisavljevic, Perets, Pierel, Raymond, Rest, Rest, Ridden-Harper, Shen, Siebert, Smith, Taggart, Tinyanont, Valdes, Villar, Wang, Yadavalli, Zenati, \& Zenteno}]{jacobson-galan_circumstellar_2022}
Jacobson-Galán, W.~V., Venkatraman, P., Margutti, R., {et~al.} 2022, The Astrophysical Journal, 932, 58, \dodoi{10.3847/1538-4357/ac67dc}

\bibitem[{Jacobson-Galán {et~al.}(2025)Jacobson-Galán, Dessart, Kilpatrick, Patel, Auchettl, Tinyanont, Margutti, Dwarkadas, Bostroem, Chornock, Foley, Abunemeh, Ahumada, Arunachalam, Bustamante-Rosell, Coulter, Gall, Gao, Guo, Jones, Hjorth, Kaewmookda, Kasliwal, Kaur, Larison, LeBaron, Miao, Narayan, Pan, Park, Patra, Qin, Ransome, Rest, Rho, Rose, Sears, Swift, Taggart, Villar, Wang, Zenati, \& Zhou}]{jacobson-galan_panchromatic_2025}
Jacobson-Galán, W.~V., Dessart, L., Kilpatrick, C.~D., {et~al.} 2025, The Astrophysical Journal, 994, L14, \dodoi{10.3847/2041-8213/ae157a}

\bibitem[{Jakobsen {et~al.}(2022)Jakobsen, Ferruit, Oliveira, Arribas, Bagnasco, Barho, Beck, Birkmann, Böker, Bunker, Charlot, Jong, Marchi, Ehrenwinkler, Falcolini, Fels, Franx, Franz, Funke, Giardino, Gnata, Holota, Honnen, Jensen, Jentsch, Johnson, Jollet, Karl, Kling, Köhler, Kolm, Kumari, Lander, Lemke, López-Caniego, Lützgendorf, Maiolino, Manjavacas, Marston, Maschmann, Maurer, Messerschmidt, Moseley, Mosner, Mott, Muzerolle, Pirzkal, Pittet, Plitzke, Posselt, Rapp, Rauscher, Rawle, Rix, Rödel, Rumler, Sabbi, Salvignol, Schmid, Sirianni, Smith, Strada, Plate, Valenti, Wettemann, Wiehe, Wiesmayer, Willott, Wright, Zeidler, \& Zincke}]{jakobsen_near-infrared_2022}
Jakobsen, P., Ferruit, P., Oliveira, C. A.~d., {et~al.} 2022, Astronomy \& Astrophysics, 661, A80.
\newblock \url{https://www.aanda.org/articles/aa/full_html/2022/05/aa42663-21/aa42663-21.html}

\bibitem[{Jerkstrand(2017)}]{jerkstrand_spectra_2017}
Jerkstrand, A. 2017, in Handbook of {Supernovae} (Springer), 795--842, \dodoi{10.1007/978-3-319-21846-5_29}

\bibitem[{Jerkstrand {et~al.}(2014)Jerkstrand, Smartt, Fraser, Fransson, Sollerman, Taddia, \& Kotak}]{jerkstrand_nebular_2014}
Jerkstrand, A., Smartt, S.~J., Fraser, M., {et~al.} 2014, Monthly Notices of the Royal Astronomical Society, 439, 3694, \dodoi{10.1093/mnras/stu221}

\bibitem[{Jiang {et~al.}(2021)Jiang, Maeda, Kawabata, Doi, Shigeyama, Tanaka, Tominaga, Nomoto, Niino, Sako, Ohsawa, Schramm, Yamanaka, Kobayashi, Takahashi, Nakaoka, Kawabata, Isogai, Aoki, Kondo, Mori, Arimatsu, Kasuga, Okumura, Urakawa, Reichart, Taguchi, Arima, Beniyama, Uno, \& Hamada}]{jiang_discovery_2021}
Jiang, J.-a., Maeda, K., Kawabata, M., {et~al.} 2021, The Astrophysical Journal Letters, 923, L8, \dodoi{10.3847/2041-8213/ac375f}

\bibitem[{Jones {et~al.}(2021)Jones, Foley, Narayan, Hjorth, Huber, Aleo, Alexander, Angus, Auchettl, Baldassare, Bruun, Chambers, Chatterjee, Coppejans, Coulter, DeMarchi, Dimitriadis, Drout, Engel, French, Gagliano, Gall, Hung, Izzo, Jacobson-Galán, Kilpatrick, Korhonen, Margutti, Raimundo, Ramirez-Ruiz, Rest, Rojas-Bravo, Siebert, Smartt, Smith, Terreran, Wang, Wojtak, Agnello, Ansari, Arendse, Baldeschi, Blanchard, Brethauer, Bright, Brown, de~Boer, Dodd, Fairlamb, Grillo, Hajela, Hede, Kolborg, Law-Smith, Lin, Magnier, Malanchev, Matthews, Mockler, Muthukrishna, Pan, Pfister, Ramanah, Rest, Sarangi, Schrøder, Stauffer, Stroh, Taggart, Tinyanont, Wainscoat, \& {Young Supernova Experiment}}]{jones_young_2021}
Jones, D.~O., Foley, R.~J., Narayan, G., {et~al.} 2021, The Astrophysical Journal, 908, 143, \dodoi{10.3847/1538-4357/abd7f5}

\bibitem[{Kasliwal {et~al.}(2012)Kasliwal, Kulkarni, Gal-Yam, Nugent, Sullivan, Bildsten, Yaron, Perets, Arcavi, Ben-Ami, Bhalerao, Bloom, Cenko, Filippenko, Frail, Ganeshalingam, Horesh, Howell, Law, Leonard, Li, Ofek, Polishook, Poznanski, Quimby, Silverman, Sternberg, \& Xu}]{kasliwal_calcium-rich_2012}
Kasliwal, M.~M., Kulkarni, S.~R., Gal-Yam, A., {et~al.} 2012, The Astrophysical Journal, 755, 161, \dodoi{10.1088/0004-637X/755/2/161}

\bibitem[{Kawabata {et~al.}(2010)Kawabata, Maeda, Nomoto, Taubenberger, Tanaka, Deng, Pian, Hattori, \& Itagaki}]{kawabata_massive_2010}
Kawabata, K.~S., Maeda, K., Nomoto, K., {et~al.} 2010, Nature, 465, 326, \dodoi{10.1038/nature09055}

\bibitem[{Kleiser \& Kasen(2014)}]{kleiser_rapidly_2014}
Kleiser, I. K.~W., \& Kasen, D. 2014, Monthly Notices of the Royal Astronomical Society, 438, 318, \dodoi{10.1093/mnras/stt2191}

\bibitem[{Kobulnicky {et~al.}(2003)Kobulnicky, Nordsieck, Burgh, Smith, Percival, Williams, \& O'Donoghue}]{kobulnicky_prime_2003}
Kobulnicky, H.~A., Nordsieck, K.~H., Burgh, E.~B., {et~al.} 2003, in Proceedings of {SPIE}, {Volume} 4841, Vol. 4841, 1634--1644, \dodoi{10.1117/12.460315}

\bibitem[{Kumar {et~al.}(2026)Kumar, Baer-Way, Ravi, Modjaz, Chandra, Valenti, Kwok, Tinyanont, Foley, Howell, Hiramatsu, Andrews, Bostroem, Christy, Franz, Hsu, Pearson, Sand, Shrestha, Smith, \& Subrayan}]{kumar_multiwavelength_2026}
Kumar, S., Baer-Way, R., Ravi, A.~P., {et~al.} 2026, A multiwavelength view of the nearby {Calcium}-{Strong} {Transient} {SN} 2025coe in the {X}-{Ray}, {Near}-{Infrared}, and {Radio} {Wavebands},  arXiv, \dodoi{10.48550/arXiv.2601.19018}

\bibitem[{Kwok {et~al.}(2024{\natexlab{a}})Kwok, Dessart, Filippenko, Foley, Howell, Jacobson-Galan, Jha, Larison, Modjaz, Pellegrino, Sand, \& Valenti}]{kwok_cracking_2024}
Kwok, L., Dessart, L., Filippenko, A.~V., {et~al.} 2024{\natexlab{a}}, JWST Proposal. Cycle 2, 6591.
\newblock \url{https://ui.adsabs.harvard.edu/abs/2024jwst.prop.6591K}

\bibitem[{Kwok {et~al.}(2023)Kwok, Jha, Temim, Fox, Larison, Camacho-Neves, Brenner~Newman, Pierel, Foley, Andrews, Badenes, Barna, Bostroem, Deckers, Flörs, Garnavich, Graham, Graur, Hosseinzadeh, Howell, Hughes, Johansson, Kendrew, Kerzendorf, Maeda, Maguire, McCully, O'Brien, Rest, Sand, Shahbandeh, Strolger, Szalai, Ashall, Baron, Burns, DerKacy, Evans, Fisher, Galbany, Hoeflich, Hsiao, de~Jaeger, Karamehmetoglu, Krisciunas, Kumar, Lu, Maund, Mazzali, Medler, Morrell, Phillips, Shappee, Stritzinger, Suntzeff, Telesco, Tucker, \& Wang}]{kwok_jwst_2023}
Kwok, L.~A., Jha, S.~W., Temim, T., {et~al.} 2023, The Astrophysical Journal, 944, L3, \dodoi{10.3847/2041-8213/acb4ec}

\bibitem[{Kwok {et~al.}(2024{\natexlab{b}})Kwok, Siebert, Johansson, Jha, Blondin, Dessart, Foley, Hillier, Larison, Pakmor, Temim, Andrews, Auchettl, Badenes, Barnabas, Bostroem, Brenner~Newman, Brink, Bustamante-Rosell, Camacho-Neves, Clocchiatti, Coulter, Davis, Deckers, Dimitriadis, Dong, Farah, Filippenko, Flörs, Fox, Garnavich, Padilla~Gonzalez, Graur, Hambsch, Hosseinzadeh, Howell, Hughes, Kerzendorf, Saux, Maeda, Maguire, McCully, Mihalenko, Newsome, O’Brien, Pearson, Pellegrino, Pierel, Polin, Rest, Rojas-Bravo, Sand, Schwab, Shahbandeh, Shrestha, Smith, Strolger, Szalai, Taggart, Terreran, Terwel, Tinyanont, Valenti, Vinkó, Wheeler, Yang, Zheng, Ashall, DerKacy, Galbany, Hoeflich, de~Jaeger, Lu, Maund, Medler, Morell, Shappee, Stritzinger, Suntzeff, Tucker, \& Wang}]{kwok_ground-based_2024}
Kwok, L.~A., Siebert, M.~R., Johansson, J., {et~al.} 2024{\natexlab{b}}, The Astrophysical Journal, 966, 135, \dodoi{10.3847/1538-4357/ad2c0d}

\bibitem[{Kwok {et~al.}(2025{\natexlab{a}})Kwok, Singh, Jha, Blondin, Dastidar, Larison, Miller, Andrews, Andrews, Anupama, Auchettl, Bánhidi, Barna, Bostroem, Brink, Cartier, Chen, Christy, Coulter, Covarrubias, Davis, Dickinson, Dong, Farah, Filippenko, Flörs, Foley, Franz, Fremling, Galbany, Gangopadhyay, Garg, Garnavich, Gates, Graur, Gordon, Hiramatsu, Hoang, Howell, Hsu, Johansson, Joshi, Kahinga, Kaur, Kumar, Kumnurdmanee, Kuncarayakti, LeBaron, Liu, Maeda, Maguire, McCully, Mehta, Menotti, Metevier, Misra, Tanner~Murphey, Newsome, Padilla~Gonzalez, Patra, Pearson, Piro, Polin, Ravi, Rest, Rehemtulla, Meza~Retamal, Robinson, Rojas-Bravo, Sahu, Sand, Schmidt, Schulze, Schwab, Shrestha, Siebert, Simha, Smith, Sollerman, Subrayan, Szalai, Taggart, Singh~Teja, Temim, Terwel, Tinyanont, Valenti, Anais~Vilchez, Vinkó, Westerling, Yang, \& Zheng}]{kwok_jwst_2025}
Kwok, L.~A., Singh, M., Jha, S.~W., {et~al.} 2025{\natexlab{a}}, {JWST} and {Ground}-based {Observations} of the {Type} {Iax} {Supernovae} {SN} 2024pxl and {SN} 2024vjm: {Evidence} for {Weak} {Deflagration} {Explosions},  arXiv, \dodoi{10.48550/arXiv.2505.02944}

\bibitem[{Kwok {et~al.}(2025{\natexlab{b}})Kwok, Liu, Jha, Blondin, Larison, Miller, Dai, Foley, Filippenko, Andrews, Andrews, Auchettl, Badenes, Brink, Davis, Flörs, Galbany, Graur, Howell, Kumar, Könyves-Tóth, LeBaron, Macrie, Maeda, Maguire, McCully, Meza-Retamal, Padilla~Gonzalez, Pakmor, Pearson, Piro, Polin, Rehemtulla, Rojas-Bravo, Sand, Sangkachan, Schwab, Sears, Singh, Subrayan, Taggart, Temim, Terwel, Tinyanont, Vinkó, Wang, Wheeler, Yang, \& Zheng}]{kwok_jwst_2025-1}
Kwok, L.~A., Liu, C., Jha, S.~W., {et~al.} 2025{\natexlab{b}}, {JWST} {Spectroscopy} of {SN} {Ia} 2022aaiq and 2024gy: {Evidence} for {Enhanced} {Central} {Stable} {Ni} {Abundance} and a {Deflagration}-to-{Detonation} {Transition},  arXiv, \dodoi{10.48550/arXiv.2510.09760}

\bibitem[{Labrie {et~al.}(2023)Labrie, Simpson, Cardenes, Turner, Soraisam, Quint, Oberdorf, Placco, Berke, Smirnova, Conseil, Vacca, \& Thomas-Osip}]{labrie_dragonsquick_2023}
Labrie, K., Simpson, C., Cardenes, R., {et~al.} 2023, Research Notes of the AAS, 7, 214, \dodoi{10.3847/2515-5172/ad0044}

\bibitem[{Landolt(1992)}]{landolt_ubvri_1992}
Landolt, A.~U. 1992, The Astronomical Journal, 104, 340, \dodoi{10.1086/116242}

\bibitem[{Landolt(2009)}]{landolt_ubvri_2009}
---. 2009, The Astronomical Journal, 137, 4186, \dodoi{10.1088/0004-6256/137/5/4186}

\bibitem[{Li \& McCray(1992)}]{li_o_1992}
Li, H., \& McCray, R. 1992, The Astrophysical Journal, 387, 309, \dodoi{10.1086/171082}

\bibitem[{Liu {et~al.}(2023)Liu, Miller, Polin, Nugent, De, Nugent, Schulze, Gal-Yam, Fremling, Anand, Andreoni, Blanchard, Brink, Dhawan, Filippenko, Maguire, Schweyer, Sears, Sharma, Graham, Groom, Hale, Kasliwal, Masci, Purdum, Racine, Sollerman, \& Kulkarni}]{liu_sn_2023-1}
Liu, C., Miller, A.~A., Polin, A., {et~al.} 2023, The Astrophysical Journal, 946, 83, \dodoi{10.3847/1538-4357/acbb5e}

\bibitem[{Lunnan {et~al.}(2016)Lunnan, Chornock, Berger, Milisavljevic, Jones, Rest, Fong, Fransson, Margutti, Drout, Blanchard, Challis, Cowperthwaite, Foley, Kirshner, Morrell, Riess, Roth, Scolnic, Smartt, Smith, Villar, Chambers, Draper, Huber, Kaiser, Kudritzki, Magnier, Metcalfe, \& Waters}]{lunnan_ps1-14bj_2016}
Lunnan, R., Chornock, R., Berger, E., {et~al.} 2016, The Astrophysical Journal, 831, 144, \dodoi{10.3847/0004-637X/831/2/144}

\bibitem[{Lunnan {et~al.}(2017)Lunnan, Kasliwal, Cao, Hangard, Yaron, Parrent, McCully, Gal-Yam, Mulchaey, Ben-Ami, Filippenko, Fremling, Fruchter, Howell, Koda, Kupfer, Kulkarni, Laher, Masci, Nugent, Ofek, Yagi, \& Yan}]{lunnan_two_2017}
Lunnan, R., Kasliwal, M.~M., Cao, Y., {et~al.} 2017, The Astrophysical Journal, 836, 60, \dodoi{10.3847/1538-4357/836/1/60}

\bibitem[{Lyman {et~al.}(2014)Lyman, Levan, Church, Davies, \& Tanvir}]{lyman_progenitors_2014}
Lyman, J.~D., Levan, A.~J., Church, R.~P., Davies, M.~B., \& Tanvir, N.~R. 2014, Monthly Notices of the Royal Astronomical Society, 444, 2157, \dodoi{10.1093/mnras/stu1574}

\bibitem[{Medler {et~al.}(2023)Medler, Mazzali, Ashall, Teffs, Shahbandeh, \& Shappee}]{medler_flat-topped_2023}
Medler, K., Mazzali, P.~A., Ashall, C., {et~al.} 2023, Monthly Notices of the Royal Astronomical Society, 518, L40, \dodoi{10.1093/mnrasl/slac127}

\bibitem[{Medler {et~al.}(2025)Medler, Ashall, Hoeflich, Baron, DerKacy, Shahbandeh, Mera, Pfeffer, Hoogendam, Jones, Shiber, Fereidouni, Fox, Jencson, Galbany, Hinkle, Tucker, Shappee, Huber, Auchettl, Angus, Desai, Do, Payne, Shi, Kong, Romagnoli, Syncatto, Clayton, Dulude, Engesser, Filippenko, Gomez, Hsiao, de~Jaeger, Johansson, Krisciunas, Kumar, Lu, Matsuura, Mazzali, Milisavljevic, Morrell, O'Steen, Park, Phillips, Ravi, Rest, Rho, Suntzeff, Sarangi, Smith, Stritzinger, Strolger, Szalai, Temim, Tinyanont, Van~Dyk, Wang, Wang, Wesson, Yang, \& Zsiros}]{medler_jwst_2025}
Medler, K., Ashall, C., Hoeflich, P., {et~al.} 2025, {JWST} {Observations} of {SN} 2023ixf {II}: {The} {Panchromatic} {Evolution} {Between} 250 and 720 {Days} {After} the {Explosion},  arXiv, \dodoi{10.48550/arXiv.2507.19727}

\bibitem[{Mera {et~al.}(2026)Mera, Ashall, Hoeflich, Medler, Shahbandeh, Burns, Baron, DerKacy, Morrell, Lu, Hinkle, Mazzali, Fereidouni, Pfeffer, Shiber, Temim, Galbany, Coulter, Ferrari, Hoogendam, Hsiao, Phillips, \& Shappee}]{mera_jwst_2026}
Mera, T., Ashall, C., Hoeflich, P., {et~al.} 2026, The Astrophysical Journal, 997, 330, \dodoi{10.3847/1538-4357/ae317e}

\bibitem[{Milisavljevic {et~al.}(2010)Milisavljevic, Fesen, Gerardy, Kirshner, \& Challis}]{milisavljevic_doublets_2010}
Milisavljevic, D., Fesen, R.~A., Gerardy, C.~L., Kirshner, R.~P., \& Challis, P. 2010, The Astrophysical Journal, 709, 1343, \dodoi{10.1088/0004-637X/709/2/1343}

\bibitem[{Milisavljevic {et~al.}(2017)Milisavljevic, Patnaude, Raymond, Drout, Margutti, Kamble, Chornock, Guillochon, Sanders, Parrent, Lovisari, Chilingarian, Challis, Kirshner, Penny, Itagaki, Eldridge, \& Moriya}]{milisavljevic_iPTF15eqv_2017}
Milisavljevic, D., Patnaude, D.~J., Raymond, J.~C., {et~al.} 2017, The Astrophysical Journal, 846, 50, \dodoi{10.3847/1538-4357/aa7d9f}

\bibitem[{Miller {et~al.}(2020)Miller, Magee, Polin, Maguire, Zimmerman, Yao, Sollerman, Schulze, Perley, Kromer, Dhawan, Bulla, Andreoni, Bellm, De, Dekany, Delacroix, Fremling, Gal-Yam, Goldstein, Golkhou, Goobar, Graham, Irani, Kasliwal, Kaye, Kim, Laher, Mahabal, Masci, Nugent, Ofek, Phinney, Prentice, Riddle, Rigault, Rusholme, Schweyer, Shupe, Soumagnac, Terreran, Walters, Yan, Zolkower, \& Kulkarni}]{miller_spectacular_2020}
Miller, A.~A., Magee, M.~R., Polin, A., {et~al.} 2020, The Astrophysical Journal, 898, 56, \dodoi{10.3847/1538-4357/ab9e05}

\bibitem[{Modjaz {et~al.}(2008{\natexlab{a}})Modjaz, Kirshner, Blondin, Challis, \& Matheson}]{modjaz_double-peaked_2008}
Modjaz, M., Kirshner, R.~P., Blondin, S., Challis, P., \& Matheson, T. 2008{\natexlab{a}}, The Astrophysical Journal, 687, L9, \dodoi{10.1086/593135}

\bibitem[{Modjaz {et~al.}(2008{\natexlab{b}})Modjaz, Kewley, Kirshner, Stanek, Challis, Garnavich, Greene, Kelly, \& Prieto}]{modjaz_measured_2008}
Modjaz, M., Kewley, L., Kirshner, R.~P., {et~al.} 2008{\natexlab{b}}, The Astronomical Journal, 135, 1136, \dodoi{10.1088/0004-6256/135/4/1136}

\bibitem[{Morán-Fraile {et~al.}(2024)Morán-Fraile, Holas, Röpke, Pakmor, \& Schneider}]{moran-fraile_faint_2024}
Morán-Fraile, J., Holas, A., Röpke, F.~K., Pakmor, R., \& Schneider, F. R.~N. 2024, Astronomy and Astrophysics, 683, A44, \dodoi{10.1051/0004-6361/202347769}

\bibitem[{Mould {et~al.}(2000)Mould, Huchra, Freedman, Kennicutt, Ferrarese, Ford, Gibson, Graham, Hughes, Illingworth, Kelson, Macri, Madore, Sakai, Sebo, Silbermann, \& Stetson}]{mould_hubble_2000}
Mould, J.~R., Huchra, J.~P., Freedman, W.~L., {et~al.} 2000, The Astrophysical Journal, 529, 786, \dodoi{10.1086/308304}

\bibitem[{Nakaoka {et~al.}(2021)Nakaoka, Maeda, Yamanaka, Tanaka, Kawabata, Moriya, Kawabata, Tominaga, Takagi, Imazato, Morokuma, Sako, Ohsawa, Nagao, Jiang, Burgaz, Taguchi, Uemura, Akitaya, Sasada, Isogai, Otsuka, \& Maehara}]{nakaoka_calcium-rich_2021}
Nakaoka, T., Maeda, K., Yamanaka, M., {et~al.} 2021, The Astrophysical Journal, 912, 30, \dodoi{10.3847/1538-4357/abe765}

\bibitem[{Nicholl {et~al.}(2019)Nicholl, Short, Angus, Wiseman, Pursiainen, Taubenberger, Gromadzki, Frohmaier, {Chen. T.-W.}, Anderson, Inserra, Yaron, Young, Manulis, Smartt, Smith, Young, Wright, Chambers, Flewelling, Huber, Lowe, Magnier, Schultz, Waters, Wainscoat, \& Willman}]{nicholl_epessto_2019}
Nicholl, M., Short, P., Angus, C.~R., {et~al.} 2019, The Astronomer's Telegram, 12697, 1.
\newblock \url{https://ui.adsabs.harvard.edu/abs/2019ATel12697....1N}

\bibitem[{Oke {et~al.}(1995)Oke, Cohen, Carr, Cromer, Dingizian, Harris, Labrecque, Lucinio, Schaal, Epps, \& Miller}]{oke_keck_1995}
Oke, J.~B., Cohen, J.~G., Carr, M., {et~al.} 1995, \dodoi{10.1086/133562}

\bibitem[{Pakmor {et~al.}(2012)Pakmor, Kromer, Taubenberger, Sim, Röpke, \& Hillebrandt}]{pakmor_normal_2012}
Pakmor, R., Kromer, M., Taubenberger, S., {et~al.} 2012, The Astrophysical Journal Letters, 747, L10, \dodoi{10.1088/2041-8205/747/1/L10}

\bibitem[{Pakmor {et~al.}(2026)Pakmor, Shen, Bhat, Rajamuthukumar, Collins, O’Donnell, Bauer, Callan, Röpke, Pollin, Maguire, Kwok, Seth, Taubenberger, \& Justham}]{pakmor_violent_2026}
Pakmor, R., Shen, K.~J., Bhat, A., {et~al.} 2026, Astronomy \& Astrophysics, 706, A239.
\newblock \url{https://www.aanda.org/articles/aa/full_html/2026/02/aa57670-25/aa57670-25.html}

\bibitem[{Perets {et~al.}(2010)Perets, Gal-Yam, Mazzali, Arnett, Kagan, Filippenko, Li, Arcavi, Cenko, Fox, Leonard, Moon, Sand, Soderberg, Anderson, James, Foley, Ganeshalingam, Ofek, Bildsten, Nelemans, Shen, Weinberg, Metzger, Piro, Quataert, Kiewe, \& Poznanski}]{perets_faint_2010}
Perets, H.~B., Gal-Yam, A., Mazzali, P.~A., {et~al.} 2010, Nature, 465, 322, \dodoi{10.1038/nature09056}

\bibitem[{Perley(2019)}]{perley_fully_2019}
Perley, D.~A. 2019, Publications of the Astronomical Society of the Pacific, 131, 084503, \dodoi{10.1088/1538-3873/ab215d}

\bibitem[{Petrushevska {et~al.}(2024)Petrushevska, Ihanec, Gromadzki, \& Yaron}]{petrushevska_epessto_2024}
Petrushevska, T., Ihanec, N., Gromadzki, M., \& Yaron, O. 2024, Transient Name Server Classification Report, 2024-110, 1.
\newblock \url{https://ui.adsabs.harvard.edu/abs/2024TNSCR.110....1P}

\bibitem[{{Piro} {et~al.}(2021){Piro}, {Haynie}, \& {Yao}}]{Piro21}
{Piro}, A.~L., {Haynie}, A., \& {Yao}, Y. 2021, \apj, 909, 209, \dodoi{10.3847/1538-4357/abe2b1}

\bibitem[{Polin {et~al.}(2019)Polin, Nugent, \& Kasen}]{polin_observational_2019}
Polin, A., Nugent, P., \& Kasen, D. 2019, The Astrophysical Journal, 873, 84, \dodoi{10.3847/1538-4357/aafb6a}

\bibitem[{Polin {et~al.}(2021)Polin, Nugent, \& Kasen}]{polin_nebular_2021}
---. 2021, The Astrophysical Journal, 906, 65, \dodoi{10.3847/1538-4357/abcccc}

\bibitem[{Prentice {et~al.}(2020)Prentice, Maguire, Flörs, Taubenberger, Inserra, Frohmaier, Chen, Anderson, Ashall, Clark, Fraser, Galbany, Gal-Yam, Gromadzki, Gutiérrez, James, Jonker, Kankare, Leloudas, Magee, Mazzali, Nicholl, Pursiainen, Skillen, Smartt, Smith, Vogl, \& Young}]{prentice_rise_2020}
Prentice, S.~J., Maguire, K., Flörs, A., {et~al.} 2020, Astronomy and Astrophysics, 635, A186, \dodoi{10.1051/0004-6361/201936515}

\bibitem[{Ravi {et~al.}(2026)Ravi, Kumar, Baer-Way, Valenti, Modjaz, Baal, Jerkstrand, Dong, Kwok, Pearson, Sand, Hiramatsu, Filippenko, Andrews, Andrews, Arunachalam, Bostroem, Brink, Christy, Chen, Davis, Esamdin, Farah, Foley, Hoang, Hosseinzadeh, Howell, Hsu, Huang, Iskander, Janzen, Jha, Kaur, Lundquist, McCully, Mehta, Ni, Retamal, Patra, Ransome, Shrestha, Smith, Subrayan, Taggart, Wang, Wynn, Yang, Yan, Zheng, \& Coe}]{ravi_double-peaked_2026}
Ravi, A.~P., Kumar, S., Baer-Way, R., {et~al.} 2026, The {Double}-{Peaked} {Calcium}-{Strong} {SN} 2025coe: {Progenitor} {Constraints} from {Early} {Interaction} and {Ejecta} {Asymmetries},  arXiv, \dodoi{10.48550/arXiv.2601.00415}

\bibitem[{Rieke {et~al.}(2015)Rieke, Ressler, Morrison, Bergeron, Bouchet, García-Marín, Greene, Regan, Sukhatme, \& Walker}]{rieke_mid-infrared_2015}
Rieke, G.~H., Ressler, M.~E., Morrison, J.~E., {et~al.} 2015, \dodoi{10.1086/682257}

\bibitem[{Rigby {et~al.}(2023)Rigby, Perrin, McElwain, Kimble, Friedman, Lallo, Doyon, Feinberg, Ferruit, Glasse, Rieke, Rieke, Wright, Willott, Colon, Milam, Neff, Stark, Valenti, Abell, Abney, Abul-Huda, Acton, Adams, Adler, Aguilar, Ahmed, Albert, Alberts, Aldridge, Allen, Altenburg, Marquez, Oliveira, Andersen, Anderson, Anderson, Argyriou, Armstrong, Arribas, Artigau, Arvai, Atkinson, Bacon, Bair, Banks, Barrientes, Barringer, Bartosik, Bast, Baudoz, Beatty, Bechtold, Beck, Bergeron, Bergkoetter, Bhatawdekar, Birkmann, Blazek, Blome, Boccaletti, Boeker, Boia, Bonaventura, Bond, Bosley, Boucarut, Bourque, Bouwman, Bower, Bowers, Boyer, Bradley, Brady, Braun, Breda, Bresnahan, Bright, Britt, Bromenschenkel, Brooks, Brooks, Brown, Brown, Brown, Bunker, Burger, Bushouse, Cale, Cameron, Cameron, Canipe, Caplinger, Caputo, Cara, Carey, Carniani, Carrasquilla, Carruthers, Case, Catherine, Chance, Chapman, Charlot, Charlow, Chayer, Chen, Cherinka, Chichester, Chilton, Chonis, Clampin, Clark, Clark, Coe, Coleman,
  Comber, Comeau, Connolly, Cooper, Cooper, Coppock, Correnti, Cossou, Coulais, Coyle, Cracraft, Curti, Cuturic, Davis, Davis, Dean, DeLisa, deMeester, Dencheva, Dencheva, DePasquale, Deschenes, Detre, Diaz, Dicken, DiFelice, Dillman, Dixon, Doggett, Donaldson, Douglas, DuPrie, Dupuis, Durning, Easmin, Eck, Edeani, Egami, Ehrenwinkler, Eisenhamer, Eisenhower, Elie, Elliott, Elliott, Ellis, Engesser, Espinoza, Etienne, Etxaluze, Falini, Feeney, Ferry, Filippazzo, Fincham, Fix, Flagey, Florian, Flynn, Fontanella, Ford, Forshay, Fox, Franz, Fu, Fullerton, Galkin, Galyer, Marin, Gardner, Gardner, Garland, Garrett, Gasman, Gaspar, Gaudreau, Gauthier, Geers, Geithner, Gennaro, Giardino, Girard, Giuliano, Glassmire, Glauser, Glazer, Godfrey, Golimowski, Gollnitz, Gong, Gonzaga, Gordon, Gordon, Goudfrooij, Greene, Greenhouse, Grimaldi, Groebner, Grundy, Guillard, Gutman, Ha, Haderlein, Hagedorn, Hainline, Haley, Hami, Hamilton, Hammel, Hansen, Harkins, Harr, Hart, Hart, Hartig, Hashimoto, Haskins, Hathaway, Havey,
  Hayden, Hecht, Heller-Boyer, Henriques, Henry, Hermann, Hernandez, Hesman, Hicks, Hilbert, Hines, Hoffman, Holfeltz, Holler, Hoppa, Hott, Howard, Howard, Hunter, Hunter, Hurst, Husemann, Hustak, Ignat, Illingworth, Irish, Jackson, Jahromi, Jakobsen, James, James, Januszewski, Jenkins, Jirdeh, Johnson, Johnson, Jones, Jones, Jones, Jones, Jordan, Jordan, Jurczyk, Jurling, Kaleida, Kalmanson, Kammerer, Kang, Kao, Karakla, Kavanagh, Kelly, Kendrew, Kennedy, Kenny, Keski-kuha, Keyes, Kidwell, Kinzel, Kirk, Kirkpatrick, Kirshenblat, Klaassen, Knapp, Knight, Knollenberg, Koehler, Koekemoer, Kovacs, Kulp, Kumari, Kyprianou, Massa, Labador, Ortega, Lagage, Lajoie, Lallo, Lam, Lamb, Lambros, Lampenfield, Langston, Larson, Law, Lawrence, Lee, Leisenring, Lepo, Leveille, Levenson, Levine, Levy, Lewis, Lewis, Libralato, Lightsey, Link, Liu, Lo, Lockwood, Logue, Long, Long, Loomis, Lopez-Caniego, Alvarez, Love-Pruitt, Lucy, Luetzgendorf, Maghami, Maiolino, Major, Malla, Malumuth, Manjavacas, Mannfolk, Marrione, Marston,
  Martel, Maschmann, Masci, Masciarelli, Maszkiewicz, Mather, McKenzie, McLean, McMaster, Melbourne, Meléndez, Menzel, Merz, Meyett, Meza, Miskey, Misselt, Moller, Morrison, Morse, Moseley, Mosier, Mountain, Mueckay, Mueller, Mullally, Murphy, Murray, Murray, Mustelier, Muzerolle, Mycroft, Myers, Myrick, Nanavati, Nance, Nayak, Naylor, Nelan, Nickson, Nielson, Nieto-Santisteban, Nikolov, Noriega-Crespo, O'Shaughnessy, O'Sullivan, Ochs, Ogle, Oleszczuk, Olmsted, Osborne, Ottens, Owens, Pacifici, Pagan, Page, Park, Parrish, Patapis, Paul, Pauly, Pavlovsky, Pedder, Peek, Pena-Guerrero, Pennanen, Perez, Perna, Perriello, Phillips, Pietraszkiewicz, Pinaud, Pirzkal, Pitman, Piwowar, Platais, Player, Plesha, Pollizi, Polster, Pontoppidan, Porterfield, Proffitt, Pueyo, Pulliam, Quirt, Neira, Alarcon, Ramsay, Rapp, Rapp, Rauscher, Ravindranath, Rawle, Regan, Reichard, Reis, Ressler, Rest, Reynolds, Rhue, Richon, Rickman, Ridgaway, Ritchie, Rix, Robberto, Robinson, Robinson, Robinson, Rock, Rodriguez, Pino, Roellig,
  Rohrbach, Roman, Romelfanger, Rose, Roteliuk, Roth, Rothwell, Rowlands, Roy, Royer, Royle, Rui, Rumler, Runnels, Russ, Rustamkulov, Ryden, Ryer, Sabata, Sabatke, Sabbi, Samuelson, Sapp, Sappington, Sargent, Sauer, Scheithauer, Schlawin, Schlitz, Schmitz, Schneider, Schreiber, Schulze, Schwab, Scott, Sembach, Shanahan, Shaughnessy, Shaw, Shawger, Shay, Sheehan, Shen, Sherman, Shiao, Shih, Shivaei, Sienkiewicz, Sing, Sirianni, Sivaramakrishnan, Skipper, Sloan, Slocum, Slowinski, Smith, Smith, Smith, Smith, Snyder, Soh, Sohn, Soto, Spencer, Stallcup, Stansberry, Starr, Starr, Stewart, Stiavelli, Straughn, Strickland, Stys, Summers, Sun, Sunnquist, Swade, Swam, Swaters, Swoish, Taylor, Taylor, Plate, Tea, Teague, Telfer, Temim, Thatte, Thompson, Thompson, Thomson, Tikkanen, Tippet, Todd, Toolan, Tran, Trejo, Truong, Tsukamoto, Tustain, Tyra, Ubeda, Underwood, Uzzo, Campen, Vandal, Vandenbussche, Vila, Volk, Wahlgren, Waldman, Walker, Wander, Warfield, Warner, Wasiak, Watkins, Weaver, Weilert, Weiser, Weiss,
  Weissman, Welty, West, Wheate, Wheatley, Wheeler, White, Whiteaker, Whitehouse, Whiteleather, Whitman, Williams, Willmer, Willoughby, Wilson, Wirth, Wislowski, Wolf, Wolfe, Wolff, Workman, Wright, Wu, Wu, Wymer, Yates, Yeager, Yeates, Yerger, Yoon, Young, Yu, Zak, Zeidler, Zhou, Zielinski, Zincke, \& Zonak}]{rigby_science_2023}
Rigby, J., Perrin, M., McElwain, M., {et~al.} 2023, Publications of the Astronomical Society of the Pacific, 135, 048001, \dodoi{10.1088/1538-3873/acb293}

\bibitem[{Roelofs {et~al.}(2007)Roelofs, Groot, Benedict, McArthur, Steeghs, Morales-Rueda, Marsh, \& Nelemans}]{roelofs_hubble_2007}
Roelofs, G. H.~A., Groot, P.~J., Benedict, G.~F., {et~al.} 2007, The Astrophysical Journal, 666, 1174, \dodoi{10.1086/520491}

\bibitem[{Roming {et~al.}(2005)Roming, Kennedy, Mason, Nousek, Ahr, Bingham, Broos, Carter, Hancock, Huckle, Hunsberger, Kawakami, Killough, Koch, Mclelland, Smith, Smith, Soto, Boyd, Breeveld, Holland, Ivanushkina, Pryzby, Still, \& Stock}]{roming_swift_2005}
Roming, P. W.~A., Kennedy, T.~E., Mason, K.~O., {et~al.} 2005, Space Science Reviews, 120, 95, \dodoi{10.1007/s11214-005-5095-4}

\bibitem[{Schulze {et~al.}(2024)Schulze, Fransson, Kozyreva, Chen, Yaron, Jerkstrand, Gal-Yam, Sollerman, Yan, Kangas, Leloudas, Omand, Smartt, Yang, Nicholl, Sarin, Yao, Brink, Sharon, Rossi, Chen, Chen, Cikota, De, Drake, Filippenko, Fremling, Fréour, Fynbo, Ho, Inserra, Irani, Kuncarayakti, Lunnan, Mazzali, Ofek, Palazzi, Perley, Pursiainen, Rothberg, Shingles, Smith, Taggart, Tartaglia, Zheng, Anderson, Cassara, Christensen, George~Djorgovski, Galbany, Gkini, Graham, Gromadzki, Groom, Hiramatsu, Andrew~Howell, Kasliwal, McCully, Müller-Bravo, Paiano, Paraskeva, Pessi, Polishook, Rau, Rigault, \& Rusholme}]{schulze_1100_2024}
Schulze, S., Fransson, C., Kozyreva, A., {et~al.} 2024, Astronomy and Astrophysics, 683, A223, \dodoi{10.1051/0004-6361/202346855}

\bibitem[{Shen {et~al.}(2019)Shen, Quataert, \& Pakmor}]{shen_progenitors_2019}
Shen, K.~J., Quataert, E., \& Pakmor, R. 2019, The Astrophysical Journal, 887, 180, \dodoi{10.3847/1538-4357/ab5370}

\bibitem[{Simpson {et~al.}(2026)Simpson, Labrie, Hirst, Turner, Smirnova, Rawlings, Vacca, \& Berke}]{simpson_dragons_2026}
Simpson, C., Labrie, K., Hirst, P., {et~al.} 2026, {DRAGONS},  Zenodo, \dodoi{10.5281/zenodo.19373829}

\bibitem[{Smartt {et~al.}(2015)Smartt, Valenti, Fraser, Inserra, Young, Sullivan, Pastorello, Benetti, Gal-Yam, Knapic, Molinaro, Smareglia, Smith, Taubenberger, Yaron, Anderson, Ashall, Balland, Baltay, Barbarino, Bauer, Baumont, Bersier, Blagorodnova, Bongard, Botticella, Bufano, Bulla, Cappellaro, Campbell, Cellier-Holzem, Chen, Childress, Clocchiatti, Contreras, Dall'Ora, Danziger, de~Jaeger, De~Cia, Della~Valle, Dennefeld, Elias-Rosa, Elman, Feindt, Fleury, Gall, Gonzalez-Gaitan, Galbany, Morales~Garoffolo, Greggio, Guillou, Hachinger, Hadjiyska, Hage, Hillebrandt, Hodgkin, Hsiao, James, Jerkstrand, Kangas, Kankare, Kotak, Kromer, Kuncarayakti, Leloudas, Lundqvist, Lyman, Hook, Maguire, Manulis, Margheim, Mattila, Maund, Mazzali, McCrum, McKinnon, Moreno-Raya, Nicholl, Nugent, Pain, Pignata, Phillips, Polshaw, Pumo, Rabinowitz, Reilly, Romero-Cañizales, Scalzo, Schmidt, Schulze, Sim, Sollerman, Taddia, Tartaglia, Terreran, Tomasella, Turatto, Walker, Walton, Wyrzykowski, Yuan, \&
  Zampieri}]{smartt_pessto_2015}
Smartt, S.~J., Valenti, S., Fraser, M., {et~al.} 2015, Astronomy and Astrophysics, 579, A40, \dodoi{10.1051/0004-6361/201425237}

\bibitem[{Speagle(2020)}]{speagle_dynesty_2020}
Speagle, J.~S. 2020, {\textbackslash}mnras, 493, 3132, \dodoi{10.1093/mnras/staa278}

\bibitem[{Srivastav {et~al.}(2023)Srivastav, Smartt, Huber, Dimitriadis, Chambers, Fulton, Moore, Callan, Gillanders, Maguire, Nicholl, Shingles, Sim, Smith, Anderson, de~Boer, Chen, Gao, \& Young}]{srivastav_luminous_2023}
Srivastav, S., Smartt, S.~J., Huber, M.~E., {et~al.} 2023, The Astrophysical Journal Letters, 943, L20, \dodoi{10.3847/2041-8213/acb2ce}

\bibitem[{Tonry {et~al.}(2024)Tonry, Denneau, Weiland, Lawrence, Siverd, Erasmus, Koorts, Jordan, Suc, Smartt, Smith, Young, Nicholl, Fulton, McCollum, Moore, Weston, Sheng, Ramsden, Aamer, Shingles, Srivastav, Ramaiya, Gillanders, Rhodes, Andersson, Stevance, Rest, Chen, Stubbs, \& Sommer}]{tonry_atlas_2024}
Tonry, J., Denneau, L., Weiland, H., {et~al.} 2024, Transient Name Server Discovery Report, 2024-85, 1.
\newblock \url{https://ui.adsabs.harvard.edu/abs/2024TNSTR..85....1T}

\bibitem[{Tonry {et~al.}(2018)Tonry, Denneau, Heinze, Stalder, Smith, Smartt, Stubbs, Weiland, \& Rest}]{tonry_atlas_2018}
Tonry, J.~L., Denneau, L., Heinze, A.~N., {et~al.} 2018, Publications of the Astronomical Society of the Pacific, 130, 064505, \dodoi{10.1088/1538-3873/aabadf}

\bibitem[{Touchard-Paxton {et~al.}(2025)Touchard-Paxton, Frohmaier, Pursiainen, Sullivan, Polin, Dimitriadis, Galbany, Killestein, Kumar, \& Lyman}]{touchard-paxton_sn_2025}
Touchard-Paxton, C.~G., Frohmaier, C., Pursiainen, M., {et~al.} 2025, Monthly Notices of the Royal Astronomical Society, 537, 1015, \dodoi{10.1093/mnras/staf069}

\bibitem[{{Valenti} {et~al.}(2008){Valenti}, {Benetti}, {Cappellaro}, {Patat}, {Mazzali}, {Turatto}, {Hurley}, {Maeda}, {Gal-Yam}, {Foley}, {Filippenko}, {Pastorello}, {Challis}, {Frontera}, {Harutyunyan}, {Iye}, {Kawabata}, {Kirshner}, {Li}, {Lipkin}, {Matheson}, {Nomoto}, {Ofek}, {Ohyama}, {Pian}, {Poznanski}, {Salvo}, {Sauer}, {Schmidt}, {Soderberg}, \& {Zampieri}}]{Valenti08}
{Valenti}, S., {Benetti}, S., {Cappellaro}, E., {et~al.} 2008, \mnras, 383, 1485, \dodoi{10.1111/j.1365-2966.2007.12647.x}

\bibitem[{{Valenti} {et~al.}(2016){Valenti}, {Howell}, {Stritzinger}, {Graham}, {Hosseinzadeh}, {Arcavi}, {Bildsten}, {Jerkstrand}, {McCully}, {Pastorello}, {Piro}, {Sand}, {Smartt}, {Terreran}, {Baltay}, {Benetti}, {Brown}, {Filippenko}, {Fraser}, {Rabinowitz}, {Sullivan}, \& {Yuan}}]{Valenti16}
{Valenti}, S., {Howell}, D.~A., {Stritzinger}, M.~D., {et~al.} 2016, \mnras, 459, 3939, \dodoi{10.1093/mnras/stw870}

\bibitem[{Virtanen {et~al.}(2020)Virtanen, Gommers, Oliphant, Haberland, Reddy, Cournapeau, Burovski, Peterson, Weckesser, Bright, van~der Walt, Brett, Wilson, Millman, Mayorov, Nelson, Jones, Kern, Larson, Carey, Polat, Feng, Moore, VanderPlas, Laxalde, Perktold, Cimrman, Henriksen, Quintero, Harris, Archibald, Ribeiro, Pedregosa, van Mulbregt, \& {SciPy 1. 0 Contributors}}]{virtanen_scipy_2020}
Virtanen, P., Gommers, R., Oliphant, T.~E., {et~al.} 2020, Nature Methods, 17, 261, \dodoi{10.1038/s41592-019-0686-2}

\bibitem[{Waldman {et~al.}(2011)Waldman, Sauer, Livne, Perets, Glasner, Mazzali, Truran, \& Gal-Yam}]{waldman_helium_2011}
Waldman, R., Sauer, D., Livne, E., {et~al.} 2011, The Astrophysical Journal, 738, 21, \dodoi{10.1088/0004-637X/738/1/21}

\bibitem[{Williams(1994)}]{williams_extinction_1994}
Williams, R.~E. 1994, The Astrophysical Journal, 426, 279, \dodoi{10.1086/174062}

\bibitem[{Wong \& Bildsten(2025)}]{wong_mass_2025}
Wong, T. L.~S., \& Bildsten, L. 2025, Mass {Loss} and {Subsequent} {Thermal} {Evolution} of {Surviving} {Helium} {White} {Dwarfs} {Shocked} by {Thermonuclear} {Supernovae},  arXiv, \dodoi{10.48550/arXiv.2508.12529}

\bibitem[{Woudt {et~al.}(2009)Woudt, Steeghs, Karovska, Warner, Groot, Nelemans, Roelofs, Marsh, Nagayama, Smits, \& O’Brien}]{woudt_expanding_2009}
Woudt, P.~A., Steeghs, D., Karovska, M., {et~al.} 2009, The Astrophysical Journal, 706, 738, \dodoi{10.1088/0004-637X/706/1/738}

\bibitem[{Wright {et~al.}(2015)Wright, Wright, Goodson, Rieke, Aitink-Kroes, Amiaux, Aricha-Yanguas, Azzollini, Banks, Barrado-Navascues, Belenguer-Davila, Bloemmart, Bouchet, Brandl, Colina, Detre, Diaz-Catala, Eccleston, Friedman, García-Marín, Güdel, Glasse, Glauser, Greene, Groezinger, Grundy, Hastings, Henning, Hofferbert, Hunter, Jessen, Justtanont, Karnik, Khorrami, Krause, Labiano, Lagage, Langer, Lemke, Lim, Lorenzo-Alvarez, Mazy, McGowan, Meixner, Morris, Morrison, Müller, rgaard Nielson, Olofsson, O’Sullivan, Pel, Penanen, Petach, Pye, Ray, Renotte, Renouf, Ressler, Samara-Ratna, Scheithauer, Schneider, Shaughnessy, Stevenson, Sukhatme, Swinyard, Sykes, Thatcher, Tikkanen, Dishoeck, Waelkens, Walker, Wells, \& Zhender}]{wright_mid-infrared_2015}
Wright, G.~S., Wright, D., Goodson, G.~B., {et~al.} 2015, \dodoi{10.1086/682253}

\bibitem[{Xi {et~al.}(2024)Xi, Wang, Li, Liu, Yan, Lin, Zhao, Filippenko, Zheng, Brink, Yang, Ehgamberdiev, Mirzaqulov, Reguitti, Pastorello, Tomasella, Cai, Zhang, Li, Zhang, Sai, Chen, Liu, Ma, \& Xiang}]{xi_sn_2024}
Xi, G., Wang, X., Li, G., {et~al.} 2024, Monthly Notices of the Royal Astronomical Society, 527, 9957, \dodoi{10.1093/mnras/stad3691}

\bibitem[{Zenati {et~al.}(2023)Zenati, Perets, Dessart, Jacobson-Galán, Toonen, \& Rest}]{zenati_origins_2023}
Zenati, Y., Perets, H.~B., Dessart, L., {et~al.} 2023, The Astrophysical Journal, 944, 22, \dodoi{10.3847/1538-4357/acaf65}

\end{thebibliography}
\end{document}